\documentclass[twocolumn,bibyear]{aa}
\usepackage[varg]{txfonts}

\usepackage{natbib}
\usepackage{float}
\usepackage{fleqn}
\usepackage{graphicx}
\usepackage{color,soul}
\usepackage{color,amssymb}
\usepackage{hyperref}
\usepackage{url}
\usepackage{multirow}
\usepackage{makecell}
\usepackage{lscape}
\usepackage{pdflscape}
\usepackage{ae,aecompl}
\usepackage[table]{xcolor}
\usepackage{tikz,xcolor,hyperref}

\definecolor{lime}{HTML}{A6CE39}
\DeclareRobustCommand{\orcidicon}{
	\begin{tikzpicture}
	\draw[lime, fill=lime] (0,0) 
	circle [radius=0.16] 
	node[white] {{\fontfamily{qag}\selectfont \tiny ID}};
	\draw[white, fill=white] (-0.0625,0.095) 
	circle [radius=0.007];
	\end{tikzpicture}
	\hspace{-2mm}
}
\foreach \x in {A, ..., Z}{\expandafter\xdef\csname orcid\x\endcsname{\noexpand\href{https://orcid.org/\csname orcidauthor\x\endcsname}
			{\noexpand\orcidicon}}
}

\begin{document}

\title{Investigating the detectability of hydrocarbons in exoplanet atmospheres with JWST}

\titlerunning{Hydrocarbons in exoplanet atmospheres with JWST}

\author{\orcidA{}Danny Gasman\inst{1,2}\thanks{danny.gasman@kuleuven.be}, \orcidB{}Michiel Min\inst{2}, and \orcidC{}Katy L. Chubb\inst{2,3}\thanks{klc20@st-andrews.ac.uk}}

\authorrunning{Gasman, Min, and Chubb}

\institute{Institute of Astronomy, KU Leuven, Celestijnenlaan 200D, 3001, Leuven, Belgium \and SRON Netherlands Institute for Space Research, Sorbonnelaan 2, 3584 CA, Utrecht, Netherlands\and Centre for Exoplanet Science, University of St Andrews, Nort Haugh, St Andrews, KY169SS, UK}

\abstract {} {We investigate at what abundances various hydrocarbon molecules (e.g. acetylene (C$_2$H$_2$), ethylene (C$_2$H$_4$), and methane (CH$_4$)) become detectable when observing the atmospheres of various planets using the \textit{James Webb} Space Telescope (JWST).} 
{We focused on atmospheric models based on the parameters of a small sample of planets: HD~189733b, HD 209458b (hot Jupiters orbiting bright stars); HD 97658b (a sub-Neptune/super-Earth orbiting a bright star); and Kepler-30c (a warm Jupiter orbiting a faint star).
We computed model transmission spectra, assuming equilibrium chemistry and clear atmospheres for all planets apart from HD~189733b, where we also computed spectra with a moderate cloud layer included. We used the Bayesian retrieval package ARCiS for the model atmospheres, and simulated observed spectra from different instruments that will be on board JWST using the PandExo package.
We subsequently ran retrievals on these spectra to determine whether the parameters input into the forward models, with a focus on molecular abundances, can be accurately retrieved from these simulated spectra.}
{We find that generally we can detect and retrieve abundances of the hydrocarbon species as long as they have a volume mixing ratio (VMR) above approximately 1$\times10^{-7}$--1$\times10^{-6}$, at least for the brighter targets.
There are variations based on planet type and instrument(s) used, and these limits will likely change depending on the abundance of other strong absorbers. We also find scenarios where the presence of one hydrocarbon is confused with another, particularly when  a small wavelength region is covered; this is often improved when two instruments are combined.}
{The molecules C$_2$H$_2$, CH$_4$, and C$_2$H$_4$ will all be detectable with JWST, provided they are present in high enough abundances, and that the optimal instruments are chosen for the exoplanet system being observed. Our results indicate that generally a combination of two instruments, either NIRSpec G395M and MIRI LRS, or NIRCam F322W2 and MIRI LRS, are best for observing these hydrocarbons in bright exoplanet systems with planets of various sizes, with NIRSpec G395M and MIRI LRS the best option for the HD~189733b-like atmosphere with clouds included. The use of NIRSpec Prism is tentatively found to be best for fainter targets, potentially in combination with the MIRI LRS slit mode, although the target we test is too faint to draw any strong conclusions. 
Instrument sensitivity, noise, and wavelength range are all thought to play a role in being able to distinguish spectral features.} 

\maketitle

\section{Introduction}\label{sec:intro}

The \textit{James Webb} Space Telescope (JWST) will allow new wavelength ranges to be probed for the signatures of a rich variety of molecular species in exoplanet atmospheres, including many major carbon- and oxygen-based species, particularly when considering favourable targets such as hot gas giant exoplanets orbiting bright stars. 
Gas-phase hydrocarbons such as acetylene (C$_2$H$_2$), ethylene (C$_2$H$_4$), and methane 
(CH$_4$) have been observed in and play an important role in Solar System giant planets~\citep{19SiMoHu.jupiter, 06BuOrCl.uranus, 00MoBeLe.saturn}; in Titan~\citep{10ClCuBa.titan, 19LoNiAc.titan}, with potential links to life~\citep{19RiSh.exo,16Mc.titan}; and in carbon stars~\citep{75TaWi.exo}. Only CH$_4$~\citep{ref:15MaGrBa,ref:19GuSoBr} and, very recently, C$_2$H$_2$~\citep{21GiBrGa} have been detected in an exoplanet's atmosphere. Others have been considered there theoretically~\citep{ref:14HuSe,ref:20ZiMiMo,14RiHeBi},  in early Earth’s atmosphere~\citep{19RiSh.exo}, and in the context of hydrocarbon hazes and clouds~\citep{19KaReIk.exo, 18HoYoUg.exo, ref:20GaThLe}. On Earth, methane is the most common hydrocarbon and is formed by geological or biological processes. Studies like \cite{19RiFeWa} explore the potential for enhanced abundances of hydrocarbons (including C$_2$H$_2$) on Earth-like exoplanets, due to bombardment processes. 
The formation of heavy hydrocarbons (with two or more carbon atoms) such as C$_2$H$_4$, C$_2$H$_6$, and C$_2$H$_8$ in planets <1000K are thought to be driven by the photodissociation of methane. These are able to polymerise to build complex polyaromatic hydrocarbons (PAHs) and/or aerosols, and eventually amino acids~\citep{08Ti.exo,19RiSh.exo}. Their detection in exoplanet atmospheres would allow us to infer important information about a planet’s atmospheric dynamics, formation, and evolution~\citep{11ObMuBe.exo,17BrFiMa.exo,18EiWaDi}. Simultaneous detections of several molecules will help give constraints on elemental abundances such as the C/O ratio~\citep{21GiBrGa}, for which hydrocarbons act as a particularly good probe~\citep{19DrCaHe.exo}, and metallicity.

Various groups have conducted studies to predict the transit spectroscopy performance of the JWST instruments from various perspectives. Different aspects, such as brightness limits and stitching, of the different instruments have been examined:  NIRCam's capabilities \citep{ref:07GrBeEi}, MIRI's capabilities \citep{ref:15BaAiIr}, and NIRSpec's capabilities \citep{ref:15BaAiIr}. The ability to detect and characterise super-Earths and Earth-like exoplanets has been examined, which may take up to 100 days or 25 transits to characterise \citep[e.g.][]{ref:15BaAiIr,ref:09DeSeWi,ref:09KaTr,ref:15BaKaLu,ref:16BaAiIr}, although well over 30 transits are expected to be required for some molecules in some planet atmospheres, such as O$_3$ in Trappist-1~e~\citep{20WuScGo,21LiMaKa}. Additionally, the deterioration of retrieval accuracy due to stellar activity has been studied \citep{ref:15BaAiIr}. \citet{ref:16GrLiMo} examine retrieval performances for temperature and H$_2$O, CO, CO$_2$, CH$_4$, and NH$_3$ abundances. For clear and solar composition atmospheres, they are able to retrieve these species by analysing spectra in the 1-2.5 $\mu$m wavelength range, but cloudy and dense atmospheres require data observed in the range 1-11 $\mu$m to yield accurate retrievals. \citet{ref:17MoBoBo} and \citet{ref:19MaLi} study the influence of clouds and hazes on emission and transmission spectra observed with various instruments of JWST. A direct comparison of instrument modes is performed in \citet{ref:17HoBuDe}, \citet{ref:17BaLi}, and \citet{ref:20GuKiFi} (NIRSpec only). Additionally, the expected improvement in S/N of JWST observations may produce a bias when assuming isothermal \citep{ref:16RoWaVe} or one-dimensional \citep{ref:20MaGoLe} forward models in the retrieval of transmission spectra. Such biases have also been studied in the context of emission spectra~\citep{16FeLiFo.exo,20TaPaIr.exo}. On the topic of hydrocarbons, \citet{ref:20ZiMiMo} include the three hydrocarbons in their forward models, but only examine the presence of features in NIRCam and MIRI LRS observations in the atmosphere of a super-Earth (55 Cnc e);  \citet{ref:20GuKiFi} only consider  NIRSpec, and do  not include C$_2$H$_4$ in their model and machine-learning analysis of warm Neptunes. Therefore, a study focusing on the retrieval outcomes of CH$_4$, C$_2$H$_2$, and C$_2$H$_4$ combined in transmission spectra using a Bayesian retrieval framework and including a wider variety of the JWST instruments has not been performed. Considering the occurrence of false detections \citep[see e.g. the review of][]{ref:17DeSe}, it is essential to provide estimates of the limitations of the instruments in addition to recommendations for their use cases.

The mid-infrared wavelengths reachable by MIRI could prove to be vital for detecting higher-order  hydrocarbons, especially in hazy planets \citep{ref:17BaLi}. In this work we simulated a range of exoplanet atmospheres with selected JWST instruments in order to investigate the detectability of three hydrocarbons: CH$_4$, C$_2$H$_2$, and C$_2$H$_4$. The paper is structured as follows. We  provide some background information on the instruments and planets considered in this study in Sect.~\ref{sec:background}. We outline the methodology used in Sect.~\ref{sec:method}, and discuss our results in Sect.~\ref{sec:results}. We link our findings to expectations of atmospheres in Sect.~\ref{sec:discussion}. A summary of our conclusions and a discussion of possible future work are given in Sect.~\ref{sec:conclusion}. 

\section{Background}
\label{sec:background}

\subsection{The instruments of JWST}\label{sec:JWST}
\begin{figure*}[h!]
    \centering
    \includegraphics[width=\textwidth]{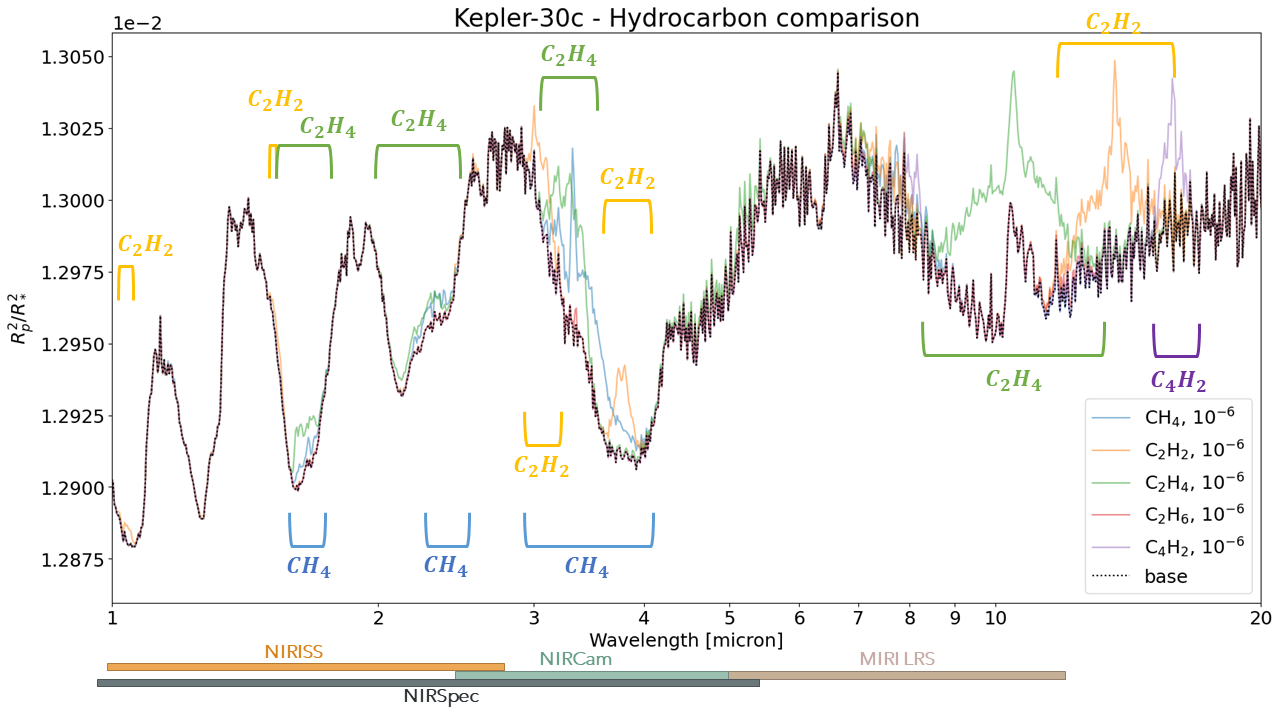}
    \caption{Comparison of transit spectra including increased hydrocarbon abundances, with respect to a base of equilibrium chemistry abundances for the main molecular species with the hydrocarbons removed. The spectra shown are for a planet modelled on Kepler-30c, with an equilibrium temperature of 769~K.}
    \label{fig:all}
\end{figure*}

JWST carries four instruments that can be used for transit spectroscopy of exoplanet atmospheres. The resolutions and wavelength ranges of these instruments offer great improvements with respect to their past or current counterparts, such as the instruments of the \textit{Hubble} Space Telescope (HST). The specifications and other relevant information about the JWST instruments MIRI, NIRSpec, NIRCam, and NIRISS are summarised in Table~\ref{t:JWST_instruments}, and more details are given below. 
In the wavelength region covered by these instruments, several prominent molecular absorption features exist for C$_2$H$_2$, CH$_4$, and C$_2$H$_4$, as illustrated by Fig.~\ref{fig:all}, along with H$_2$O, NH$_3$, HCN, CO, CO$_2$, H$_2$S, PH$_3$, and many others \citep{ref:18TeYu}.

\begin{table}[h!]
        \caption{Selected details of the instruments of JWST relevant to the present study. The particular modes focused on in this study are highlighted in bold.}
        \label{t:JWST_instruments} 
        \centering 
        \resizebox{\columnwidth}{!}{%
        \begin{tabular}{llll}
                \hline\hline
                \hline
                \rule{0pt}{3ex}Instrument & Mode/Filter & R=$\frac{\lambda}{\Delta\lambda}$ & $\lambda$ range ($\mu$m)              \vspace{1.5ex}\\
                \hline\hline
        \rule{0pt}{3ex}\textbf{MIRI} & \textbf{LRS} & \textbf{$\sim$100}  & \textbf{5 - 12}\\
                \rule{0pt}{3ex}MIRI  & MRS& 1500 - 3500 & 4.9 - 28.3\\
                \rule{0pt}{3ex}NIRSpec & G140M/F070LP & $\sim$1000 & 0.70 – 1.27 \\
                \rule{0pt}{3ex}NIRSpec & G140M/F100LP & $\sim$1000 & 0.97 – 1.84 \\
                \rule{0pt}{3ex}NIRSpec & G235M & $\sim$1000 & 1.7 – 3.1 \\
                \rule{0pt}{3ex}\textbf{NIRSpec} & \textbf{G395M} & \rule{0pt}{3ex}\textbf{$\sim$1000} & \textbf{2.9 - 5.1} \\
                \rule{0pt}{3ex}\textbf{NIRSpec} & \textbf{Prism} & \rule{0pt}{3ex}\textbf{$\sim$100} & \textbf{0.6 – 5.3} \\
                \rule{0pt}{3ex}\textbf{NIRCam} & \textbf{F322W2} & \rule{0pt}{3ex}\textbf{$\sim$1600} & \textbf{2.4 - 4.2} \\
                \rule{0pt}{3ex}NIRCam & F444W & $\sim$1600 & 3.8 - 5.1 \\
                \rule{0pt}{3ex}\textbf{NIRISS} & \textbf{SOSS} & \textbf{$\sim$700} & \textbf{0.6 - 2.8} \\
                \hline\hline
        \end{tabular}%
        }
\end{table}

\paragraph{MIRI} The Mid-Infrared Instrument (MIRI) contains a spectrograph that is able to detect wavelengths between 4.9 and 28.3 microns in its Medium Resolution Spectroscopy mode (MIRI MRS)\footnote{\url{https://jwst-docs.stsci.edu/jwst-mid-infrared-instrument/miri-observing-modes/miri-medium-resolution-spectroscopy}},
and 5-12 microns in its Low Resolution Spectroscopy mode (MIRI LRS)\footnote{\url{https://jwst-docs.stsci.edu/jwst-mid-infrared-instrument/miri-observing-modes/miri-low-resolution-spectroscopy}}. The latter provides a choice between slitless and slit spectroscopy, where its slitless option has been optimised for observations of bright compact sources such as exoplanet host stars, and is considered here. It should be noted that for fainter objects the slit mode may be more attractive, but \citet{ref:17HoBuDe} found the slitless mode to yield a higher information content for exoplanet transit spectroscopy targets. The LRS mode offers a resolving power of about 100 at 7.5 microns, whereas the MRS mode has a resolving power of about 1500 to 3500. 
The MRS, however,
requires three separate exposures to cover the full range, making it less attractive to consider for transit spectroscopy proposals \citep{ref:17HoBuDe}. We do not consider MIRI MRS in this study. 

\paragraph{NIRSpec} The Near InfraRed Spectrograph (NIRSpec) is able to observe the wavelength range of 0.6 to 5.3 microns, divided over a range of grisms.\footnote{\url{https://jwst-docs.stsci.edu/jwst-near-infrared-spectrograph}} It has four different observing modes, where the bright object time-series (BOTS) mode is the most relevant for exoplanet transmission spectroscopy. The resolving power is 1000 and 2700 for the medium- and high-resolution gratings, respectively. The Prism mode,  useful for observing faint objects \citep{ref:17HoBuDe},   has a resolving power of about 100.

\paragraph{NIRCam} The Near Infrared Camera (NIRCam) operates in the wavelength range of 0.6 to 5.0 microns, and offers various modes of operation.\footnote{\url{https://jwst-docs.stsci.edu/jwst-near-infrared-camera}} The most relevant of these is the Grism Time Series mode, which has a resolving power of 1600 at 4 microns, and is able to observe the 2.4-5.0 micron range. Different parts of this wavelength range can be observed using different grisms, where F322W2 and F444W are included in PandExo \citep{ref:17BaMaKl}, the JWST instrument simulation package used in this work. F322W2 covers approximately 2.4 to 4.2 microns, whereas F444W covers around 3.8 to 5.1 microns.

\paragraph{NIRISS} The Near Infrared Imager and Slitless Spectrograph (NIRISS) offers the possibility to observe the 0.6 to 5.0 micron range, where the most relevant mode, single object slitless spectroscopy (SOSS), covers 0.6 to 2.8 microns.\footnote{\url{https://jwst-docs.stsci.edu/jwst-near-infrared-imager-and-slitless-spectrograph}} There are various options for subarray use, and the resolving power is 700 at 1.4 microns.

\subsection{Molecular data}

\begin{figure}[h!]
    \centering
    \includegraphics[width = \columnwidth]{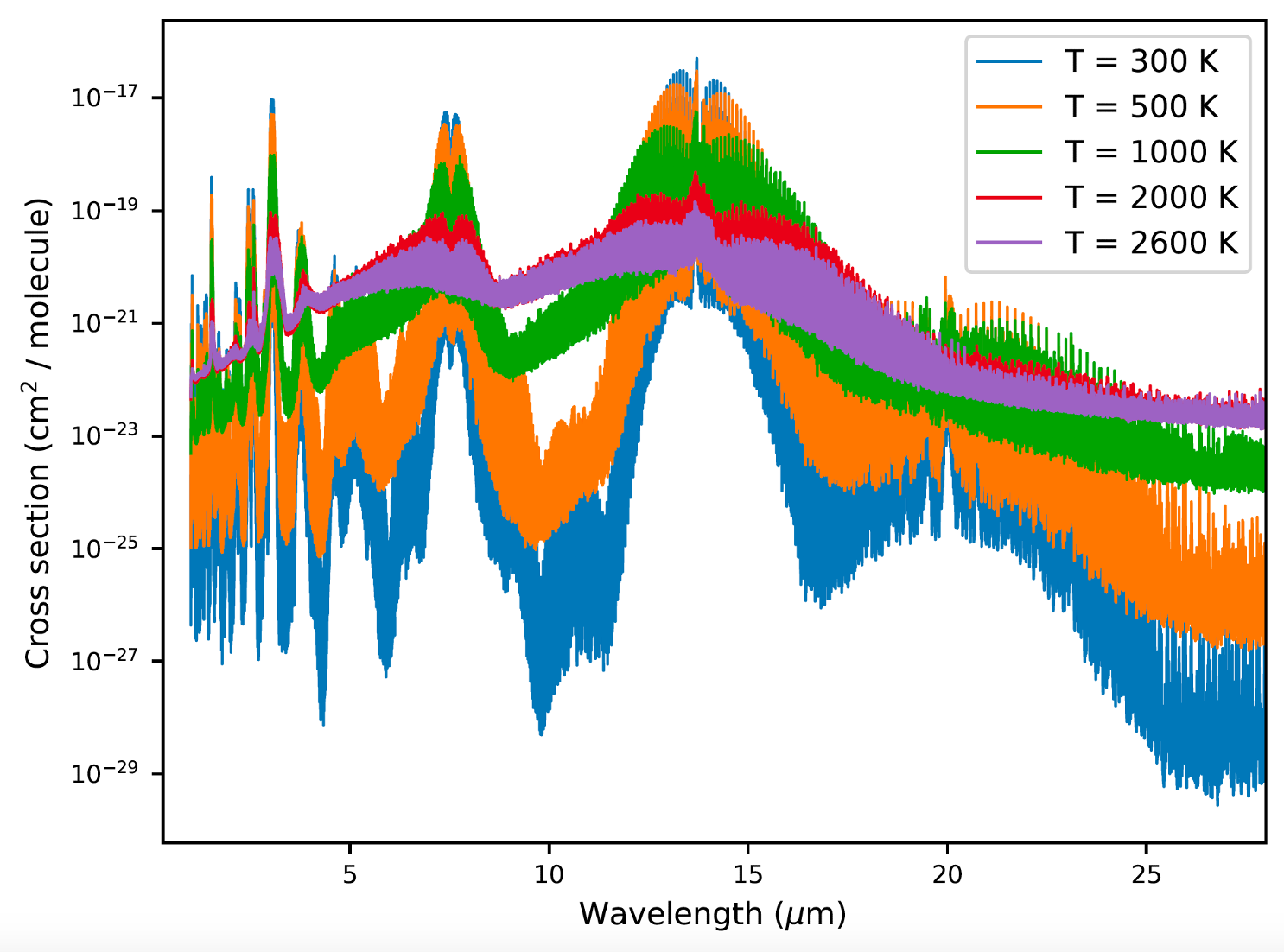}
    \caption{Cross sections of C$_2$H$_2$ computed using the ExoMol aCeTY line list~\citep{ref:20ChTeYu} for a range of temperatures at a pressure of 0.1 bar.}
    \label{fig:c2h2_Tcomp}
\end{figure}

The bulk of the analysis uses molecular data from the ExoMol database, with theoretical line lists largely designed to be suitable up to high temperatures \citep{ref:20TeYu}. These line lists were formatted into opacities by \cite{20ChRoAl.exo} as part of the ExoMolOP database\footnote{\url{http://www.exomol.com/data/data-types/opacity/}}. These theoretical determinations typically become increasingly computationally demanding as molecules gain more atoms and electrons. The molecules included in this database are thus limited, and the only hydrocarbons currently available are CH$_4$~\citep{jt698}, C$_2$H$_2$~\citep{ref:20ChTeYu}, and C$_2$H$_4$~\citep{ref:18MaYaTe}. Other hydrocarbons, namely C$_2$H$_6$ and C$_4$H$_2$, are included as part of HITRAN~\citep{ref:17Go}, a database of largely experimental data that has been measured at room temperature. For this reason, although HITRAN data is often very accurate (more accurate than purely theoretically calculated data), it is not considered complete  as many of the weaker lines in particular are missing; it is only designed for temperatures in the region of 296 K or lower, with a few exceptions.
In the Pacific Northwest National Laboratory (PNNL) experimental database, cross sections for C$_3$H$_8$ are also available, but this is again only valid for room temperature environments \citep{ref:PNNL}. Molecular spectra can change drastically for higher temperatures, making the experimental data invalid at the high-temperature environments mainly considered in this paper, due to hotter planets generally being better characterised than colder smaller planets. An example of this temperature variation for C$_2$H$_2$ can be found in \autoref{fig:c2h2_Tcomp}, which shows the flattening of the spectra at higher temperatures. The line lists used in this study for CH$_4$ and C$_2$H$_2$ are complete up to 2000~K and 2200~K, respectively. The line list for C$_2$H$_4$ is only complete up to 700~K, and so we expect some opacity to be missing at higher temperatures, namely  for the atmospheres of HD~189733b and HD~209458b. Therefore, we could be underestimating the detectability of C$_2$H$_4$ in these cases. It also has a lower wavelength cutoff of 1.4~$\mu$m, due to computational demands for computing higher-energy--lower-wavelength transitions. This means that the region covered by the NIRISS SOSS and NIRISS Prism is partly not covered by the opacity of C$_2$H$_4$. Nevertheless, the ExoMol MaYTY line list~\citep{ref:18MaYaTe} is the most complete line list currently available for the main isotopologue of C$_2$H$_4$. Accurate molecular absorption data suitable for characterising atmospheres up to high temperatures are likely to become available for other hydrocarbon species such as C$_4$H$_2$ and C$_2$H$_6$ in the near future~\citep{ref:20TeYu,HITEMP,TheoReTS,MOLLIST}. It is known from room temperature laboratory spectra that prominent absorption features for these species are expected within the infrared wavelength region accessible with JWST~\citep{ref:17Go}. In particular, C$_2$H$_6$ is expected to be present in the upper atmospheres of certain carbon-rich gas giant exoplanets and brown dwarfs \citep{ref:13BiRiHe}.
We therefore only focus on CH$_4$, C$_2$H$_2$, and C$_2$H$_4$ for the current work, but hope to be able to expand to these other heavier hydrocarbon species in the future. 

\subsection{Planets considered in this study as the bases of our models}

In the current study we focus on a small sample of planets
that offer a range of physical properties based on the following: HD~189733b, HD~209458b, HD~97658b, and Kepler-30c.
HD~189733b and HD~209458b are hot Jupiters orbiting bright stars, HD~97658b is a sub-Neptune orbiting a bright star, and Kepler-30c is a warm Jupiter orbiting a faint star. 
\autoref{t:planet_pars} gives some of the derived physical properties of the exoplanetary systems considered in this study. More details of the planets and previous detections are also given below. The model spectra we use in this study are from atmospheric models produced using the physical parameters of those planets listed here, but we do not use any observed data to produce them. Any clouds, for example, or atmospheric species that have potentially been detected (as detailed below) are thus not included. HD~189733b is an exception where we consider the effect of adding a cloud layer, as detailed in Sect.~\ref{sec:pandexo}. We assume equilibrium chemistry based on the assumed equilibrium temperature of the planet for all species apart from the hydrocarbons of interest (C$_2$H$_2$, CH$_4$, C$_2$H$_4$), which we add in at various abundances.

\begin{table*}[h!]
        \caption{Physical properties of the exoplanetary systems considered in this study: effective temperature of the host star $T_{\rm eff(s)}$, solar radius $R_s$, planet equilibrium temperature $T_{\rm eq(p)}$, planet radius $R_p$, and star-planet separation $a$.}
        \label{t:planet_pars} 
        \centering  
        \begin{tabular}{llllllll}
                \hline
                \hline
                \rule{0pt}{3ex}Planet & $T_{\rm eff(s)}$ (K) & $R_s$($R_{\rm Sun}$) & $T_{\rm eq(p)}$ (K)  & $R_p$($R_{\rm Jup}$) & $a$ (AU)  \vspace{1.5ex}\\
                \hline
                \rule{0pt}{3ex}HD209458 b & 6092 $\pm$ 103$^b$ &1.38 $\pm$ 0.02 $^g$ & $\sim$1400 &1.380 $\pm$ 0.018$^i$ & 0.04747 $\pm$ 0.00055$^i$ \\
                \rule{0pt}{3ex}HD189733 b &4875 $\pm$ 43$^a$ & 0.805 $\pm$ 0.016$^f$  & $\sim$1200&1.138 $\pm$ 0.077$^h$ & 0.03142 $\pm$ 0.00052$^i$ \\
                \rule{0pt}{3ex}HD97658 b & 5170 $\pm$ 44  $^c$ &0.73 $\pm$ 0.02$^c$ &$\sim$ 730 &  0.201 $\pm$ 0.009$^j$ & 0.08  $_{-0.018} ^{+0.0017}$ $^j$ \\
                \rule{0pt}{3ex}Kepler-30c  & 5498 $\pm$ 54 $^d$  &0.95 $\pm$ 0.12$^d$  &$\sim$ 430 & 1.069  $\pm$ 0.036$^k$ & 0.3$^d$  \\
                \hline
                \hline
        \end{tabular}
                        \mbox{}\\

{\flushleft
$^a$: \cite{08Belle.exo}; $^b$ \cite{11ScFlCu.exo}; $^c$: \cite{11HoAsMa.exo}; $^d$: \cite{12FaFoSt.exo}; $^e$: \cite{18SoTrPi.exo}; $^f$: \cite{08BaMcBr.exo}; $^g$: \cite{14BoBrFe.exo}; $^h$: \cite{08ToWiHo.exo}; $^i$: \cite{10Sou.exo}; $^j$: \cite{14VaGiVa.exo}; $^k$: \cite{12SaFaWi.exo}; $^l$: \cite{06McStVa}\\
}
\end{table*}

\subsubsection{HD 189733b}
HD~189733b is a hot Jupiter orbiting a K1-K2 type star approximately 19.3 parsec  from Earth~\citep{05BoUdMa}. The star's flux in the near-infrared (J) band is 6.07 \citep{ref:star_flux}.
The emission spectra of HD~189733b will be observed using NIRCam's slitless grism mode as part of the JWST Guaranteed Time Observations (GTO) programme  (GTO 1274; P.I. Jonathan I. Lunine). In addition, a number of general observer (GO) programmes will observe HD~189733b in Cycle 1, including in transmission using NIRCam and MIRI MRS (GO 1633; P.I. Drake Deming) and MIRI LRS (GO 2001; P.I. Michiel Min), and an eclipse map also using MIRI LRS (GO 2021; P.I. Brian Kilpatrick).
Evidence for CH$_4$ in the transmission spectra of HD 189733b was first claimed to have been observed by \cite{08SwVaTi.exo}. CO and H$_2$O have been detected on the dayside in high resolution \citep{13KoBrSn,13BiKoBr}, along with sodium \citep{15WyEhLo} and potentially potassium \citep{19KeMaEs} in the planet's terminator region.
H$_2$O, CO, and HCN were all detected on the planet's dayside at high resolution by \cite{18CaMaHa}.

\subsubsection{HD 209458b}
HD~209458b is a hot Jupiter orbiting a G0 V-type  star approximately 47.0 parsec  from Earth. The J-band flux of the star is 6.591 \citep{ref:star_flux}, making its brightness when observed from Earth similar to that of HD 189733. As a consequence this planet is also very well studied.
H$_2$O was first claimed to be found by \citet{ref:07Ba}, and has been detected in the atmospheric transmission \citep{ref:13DeWiMc} and emission \citep{16LiStBe} spectra of HD~209458b since then. CO and H$_2$O, and potentially HCN, were detected in high resolution by \cite{18HaMaCa}. The presence of multiple molecules (H$_2$O, CO, HCN, C$_2$H$_2$, CH$_4$, and NH$_3$) were recently detected in the atmosphere of HD~209458b by \cite{21GiBrGa} using high-resolution spectra and cross-correlation methods. A high C/O for the planet's atmosphere was inferred. Observations that indicate patchy clouds  present at the terminator region of HD~209458b have been made by \cite{17MaMa} and \cite{20Barstow}. 
The emission spectra of HD~209458b will be observed using NIRCam's slitless grism mode as part of the JWST Guaranteed Time Observations (GTO) programme (GTO 1274; P.I. Jonathan I. Lunine), and the transmission spectra with MIRI LRS as part of the GO programme (GO 2667; P.I. Hannah Wakeford).

\subsubsection{HD 97658b}
HD 97658b~\citep{11HoAsMa.exo,14KnDrKr} is much smaller than the hot Jupiters discussed previously, and is classified as a sub-Neptune or super-Earth. Its host star is of the K1V spectral type, and is approximately 21.1 parsec from Earth. Its J-band flux is 6.203 \citep{ref:star_flux}.
Not much is known about the atmosphere's composition, although
\cite{20GuCoDr.exo} recently found that atmospheres with high C/O ratios ($\geq$ 0.8) are favoured when compared to observed transmission data for HD~97658b. \cite{20GuCoDr.exo} also found HD~97658b to be a high-priority target for atmospheric characterisation when assessing sub-Neptune exoplanets cooler than 1000~K. The planet's relatively low temperature makes it a potentially good target for observing hydrocarbons using JWST.

\subsubsection{Kepler-30c}
Kepler-30c, despite being classified as a warm Jupiter, orbits a much fainter star than the targets mentioned previously, making it a more challenging object to observe. With a flux in the near-infrared of 14.00 magnitudes \citep{ref:star_flux}, it is used to examine what instrument settings are best suited to observing faint targets. We note that this is an extremely faint target and that even instruments such as NIRSpec Prism have a J-band flux limit of 9.82~\citep{16NiFeGi}. It is thus included as an extreme case of a faint target, and we would expect better results from a similar but slightly brighter target. Not much is currently known about the planet's atmosphere, most likely due to the host star being faint.

\section{Methodology}
\label{sec:method}

\subsection{Forward models using ARCiS}\label{sec:ARCiS}

The ARtful modeling Code for exoplanet Science (ARCiS) package performs  atmospheric modelling and Bayesian retrieval (see \citealt{ref:20MiOrCh} and \citealt{ref:19OrMi} for full details).
The code consists of a forward modelling routine based on k-distribution molecular opacities from the ExoMolOP database \citep{20ChRoAl.exo} and cloud opacities using Mie and distribution of hollow spheres \citep[DHS; see][]{05MiHoKo.arcis} computations. With ARCiS one can compute cloud formation~\citep{ref:19OrMi} and chemistry \citep{ref:18WoHeHu} from physical and chemical principles. The code was benchmarked against petitCODE \citep{ref:15MoBoDu,ref:17MoBoBo} by \cite{ref:19OrMi}. For the retrieval part the {\sc Multinest} algorithm \citep{08FeHo.multi,09FeGaHo.multi,13FeHoCa.multi} was employed to sample the specified parameter space for the region of maximum likelihood. Benchmarks for the retrieval were   performed in the framework of the ARIEL mission~\citep{18PaBeBa.ARIEL}, showing excellent agreement with multiple other retrieval codes.

\begin{figure}[h!]
  \centering
  \includegraphics[width=\columnwidth]{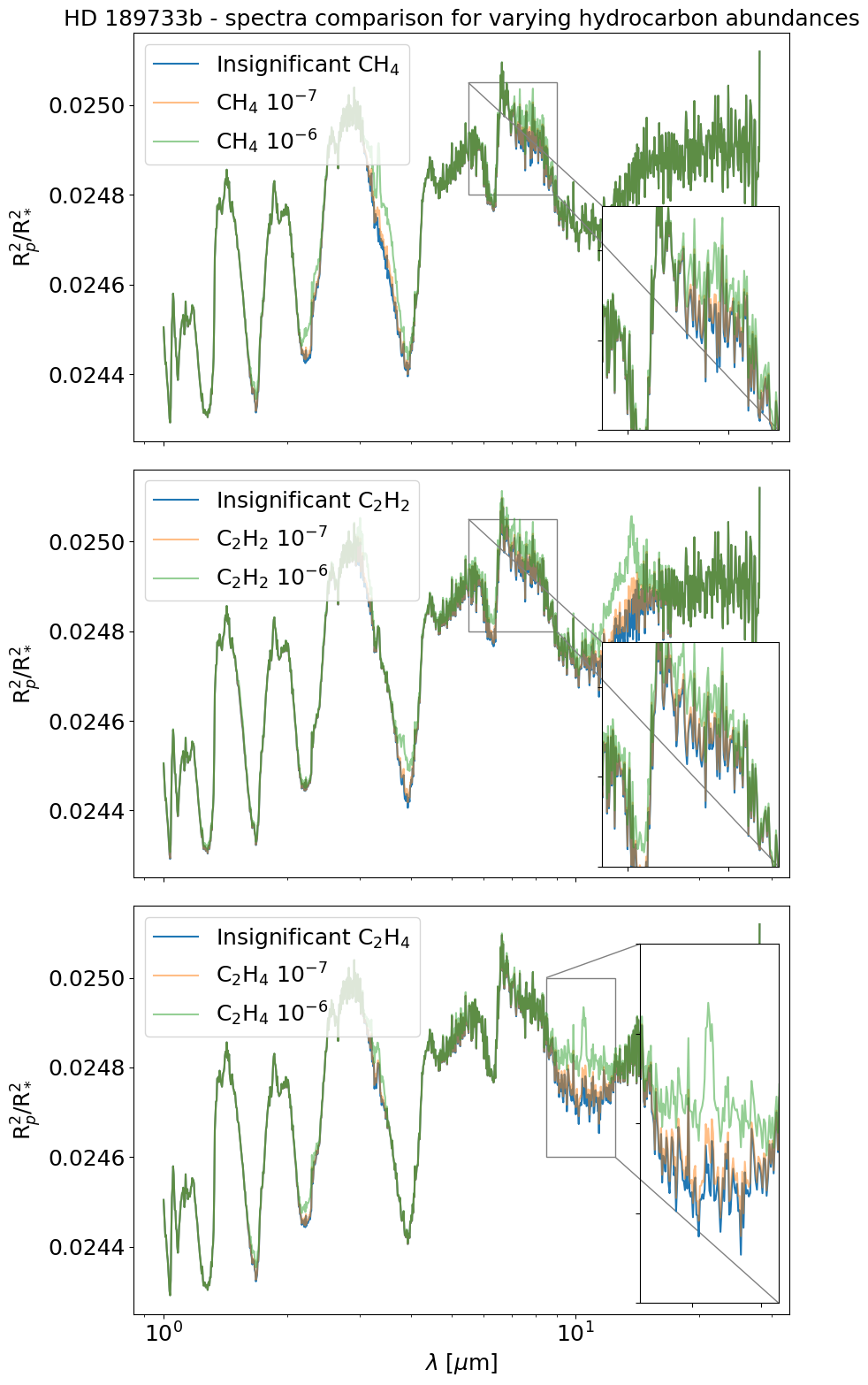}
  \caption{Comparison of transmission spectra of a planet modelled on HD~189733b with varying hydrocarbon abundances. All species become visible at a VMR starting around $10^{-7}$.}
  \label{fig:chab}
\end{figure}

We first computed forward model atmospheres for each of our selected targets using ARCiS coupled with GG$_{CHEM}$ \citep{ref:18WoHeHu} in order to obtain assumed equilibrium chemistry abundances of molecules and to compute the pressure-temperature structure of the atmosphere. 
These model atmospheres included the following molecules: H$_2$O~\citep{jt734}, CO$_2$~\citep{20YuMeFr.co2}, CO~\citep{15LiGoRo.CO}, CH$_4$~\citep{jt698}, NH$_3$~\citep{jt771}, C$_2$H$_2$~\citep{ref:20ChTeYu}, TiO~\citep{ExoMol_TiO}, and VO~\citep{jt644}; the atmospheres were  filled with H$_2$ and He (at a ratio of 0.85:0.15, respectively). We included the most common constituents of atmospheres (H$_2$O, CO$_2$, CO, NH$_3$), the three hydrocarbons considered in this study (CH$_4$, C$_2$H$_2$, and C$_2$H$_4$), and two strong optical absorbers (TiO and VO). 
The temperature structure and molecular abundances were computed self-consistently using ARCiS~\citep{ref:20MiOrCh} coupled with GG$_{CHEM}$ \citep{ref:18WoHeHu} for the equilibrium chemistry computations. \autoref{tab:plan_gen} gives the molecular abundances and temperatures for the upper bound of the pressure layers (lower bound of altitude) that we would typically probe using transmission spectroscopy, as illustrated by Fig.~\ref{fig:contrib_funcs}.
These are the `base' atmospheric compositions of each planet that we adopt, assuming an isothermal temperature structure and constant molecular abundances, which is not an unreasonable assumption when probing transmission spectra in this wavelength region~\citep{ref:19ZhChKe}, although we note below this may lead to some potential biases. We subsequently added varying abundances of the hydrocarbons we are focusing on in this study (C$_2$H$_2$, CH$_4$, C$_2$H$_4$) to the computed forward model atmospheres (at R~=~$\frac{\lambda}{\Delta \lambda}$~=~300) using these base compositions.
It can be seen from Fig. \ref{fig:chab} that the difference between forward model spectra can be observed at a volume mixing ratio (VMR) of around 1~$\times$~$10^{-7}$~-~1~$\times$~$10^{-6}$. These example spectra are for a clear planet atmosphere with similar planetary and stellar properties to those of  the HD~189733b system. 
As mentioned above, we computed our base models based on a layer of the atmosphere at approximately the upper limit of pressure that contributes to the observed transmission spectra for all planets. If we had chosen to model a lower-pressure layer, which has a greater contribution to the base spectra, there would be some effect on our initial model spectra.
The main difference would be for the cooler planets, HD 97658b and Kepler-30c, where there is a greater variation in the temperature structure across the pressure layer that we expect to be probed by transmission spectroscopy, as can be seen in Fig.~\ref{fig:PT_profs} and Fig.~\ref{fig:contrib_funcs}. Here we could be overestimating the temperature of our base models by a few hundred Kelvin. The main difference to the equilibrium chemistry would then be a lower CO abundance and higher CH$_4$ abundance. Ideally, we would use a non-isothermal pressure-temperature parametrisation in our retrievals in order to limit any potential biases~\citep{ref:16RoWaVe,ref:20MaGoLe}.

\begin{table*}[h!]
        \centering
        \caption{
                Abundances of various species computed assuming equilibrium chemistry and solar C/O and metallicity using GG$_{CHEM}$ \citep{ref:18WoHeHu} at $\sim$1.8$\times$10$^{-1}$~bar. The temperatures given are those assumed from equilibrium chemistry forward models using ARCiS, with a self-consistently computed temperature structure, coupled with GG$_{CHEM}$ at this altitude of $\sim$1.8$\times$10$^{-1}$~bar, an upper bound of a typical pressure layer able to be probed with transmission spectroscopy.}
        \label{tab:plan_gen}
        \begin{tabular}{lcccc}
                & \textbf{HD 189733b} & \textbf{HD 209458b} & \textbf{HD 97658b} & \textbf{Kepler-30c} \\ \hline
                \rule{0pt}{2.5ex}T (at $\sim$1.8e-1 bar) [K]               & 1353           & 1709            & 1077           & 769    \vspace{0.5ex} \\ \hline
                \rule{0pt}{2.5ex}H$_2$O [log VMR]        & -3.5            & -3.5            & -3.7          & -3.1   \vspace{0.5ex}     \\ \hline
                \rule{0pt}{2.5ex}CO$_2$ [log VMR]        & -7.1            & -7.3           & -7.2           & -8.4     \vspace{0.5ex} \\ \hline
                \rule{0pt}{2.5ex}CO [log VMR]            & -3.4            & -3.3           & -3.6          & -6.1    \vspace{0.5ex}  \\ \hline
                \rule{0pt}{2.5ex}CH$_4$ [log VMR]        & -6.6           & -8.4            & -4.5           & -3.2    \vspace{0.5ex} \\ \hline
                \rule{0pt}{2.5ex}NH$_3$ [log VMR]        & -7.2            & -7.7            & -6.8         & -5.5    \vspace{0.5ex} \\ \hline
                \rule{0pt}{2.5ex}C$_2$H$_2$ [log VMR]    & -12.3           & -12.6            & -12.0          & -17.2   \vspace{0.5ex}  \\ \hline
                \rule{0pt}{2.5ex}TiO [log VMR]           & -7.1            & -6.8           & -7.9          & -10.1  \vspace{0.5ex} \\ \hline
                \rule{0pt}{2.5ex}VO [log VMR]            & -8.3            & -8.0            & -9.3         & -11.6  \vspace{0.5ex} \\ \hline
                                \rule{0pt}{2.5ex}C$_2$H$_4$ [log VMR]           & -12.6          & -14.3            & -10.7          & -13.4  \vspace{0.5ex} \\ \hline
        \end{tabular}
\end{table*}

\subsection{Simulating JWST spectra using PandExo}
\label{sec:pandexo}
\begin{figure*}[h!]
    \centering
   \includegraphics[width=\textwidth]{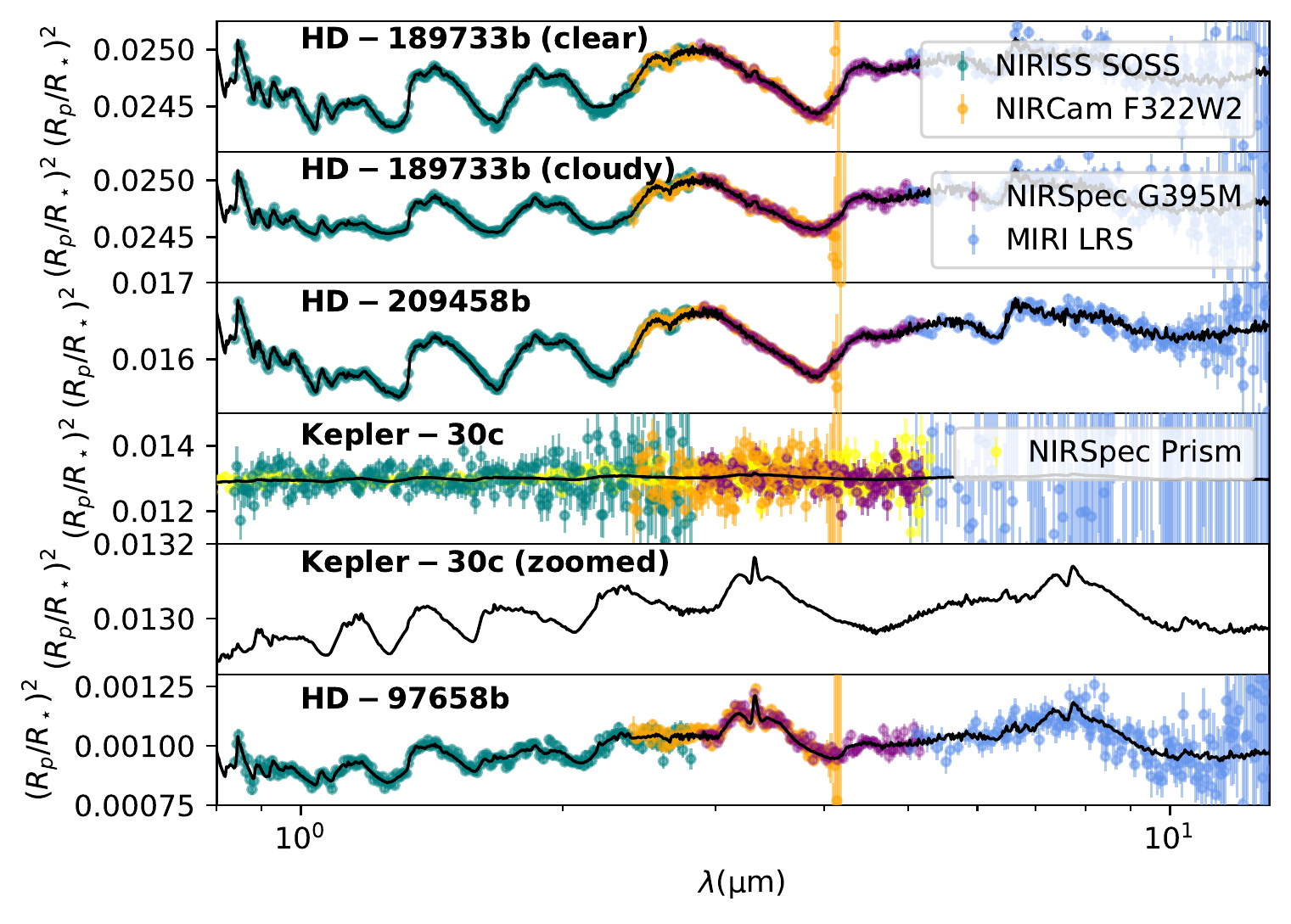}
    \caption{Base spectra (computed from equilibrium chemistry) of each of the four planet types used in this study, as detailed in each panel. Both the clear and the cloudy models for HD 189733b are included. Simulated JWST spectra and the corresponding error bars using PandExo are also plotted for each instrument used in this study (see  figure legend). }
    \label{fig:trans_base}
\end{figure*}

Once the forward model spectra were computed with ARCiS, noise and other error sources were added to the spectrum in order to simulate observations of our targets by various JWST instruments, as described in Sect.~\ref{sec:JWST}. This was done using PandExo, a \textit{Python} package that is capable of simulating JWST measurements of transiting exoplanets \citep{ref:17BaMaKl}, and that can retrieve planetary and stellar properties from the online exoplanet database ExoMast\footnote{\url{https://exo.mast.stsci.edu/}}. Using this package, the data outputs of the instruments in various modes and using various subarrays can be simulated\footnote{\url{https://natashabatalha.github.io/PandExo/tutorialjwst.html}}. PandExo requires  a noise floor and saturation limit that are instrument dependent. 
Both of these quantities are uncertain and will not be available until after the start of operation. However, the JWST User Guide  gives approximate saturation limits for each instrument mode:\footnote{\url{https://jwst-docs.stsci.edu/jwst-exposure-time-calculator-overview/jwst-etc-calculations-page-overview/jwst-etc-saturation}} 78\% of full-well depth for MIRI; 70\% of full-well depth for NIRCam time series and grism time series; 76\% of full-well depth for NIRISS SOSS;  70\% of full-well depth for NIRSpec's full subarray; and 84\% for the other subarrays. These values are used here. Furthermore, the noise floor level is even more unpredictable. Attempts have been made to approximate the noise floor levels by comparison with currently operational missions, resulting in 20ppm, 30ppm, and 50ppm for NIRISS \citep{ref:14KrBeDe}, NIRCam \citep{ref:14KrBeDe}, and MIRI \citep{ref:09KnChCo}, respectively. \citet{ref:20FoDr} assume a noise floor level of 25 ppm for NIRSpec. They argue that this is a conservative assumption, but it is in line with the noise floors of the other instruments and thus is used here.
PandExo outputs a wide range of information, but the relevant parts for the observation file to be used for retrieval are the wavelengths, the output spectrum with random noise, and the error including the noise floor. The final part of the observation file is the spectral resolution per bin. We binned the data to a constant resolution of 100 in order to resize the data to a more manageable size for retrieval. PandExo also includes a binning tool, which was used for this purpose. The spectral resolution per bin is given by
\begin{equation}
    R = \frac{\lambda}{\Delta \lambda}.
    \label{eq:bin}
\end{equation}
This observation file, with error bars added by PandExo, is then used for the free molecular retrievals (i.e. where we do not constrain molecular abundances based on equilibrium chemistry and allow them to vary freely) using ARCiS~\citep{ref:20MiOrCh,ref:19OrMi} (see Sect.~\ref{sec:ARCiS}). An illustration of the base equilibrium chemistry spectra for each of the four planets in this study is given in Fig.~\ref{fig:trans_base}, along with the simulated spectra for the various JWST instruments used for each planet. We included both a clear and a cloudy model for HD 189733b. For the cloudy model we added in a fully opaque grey cloud layer at $P_{top}=0.01$ bar, and above that a Rayleigh scattering haze with a total opacity of $f_{mix}$~=~1~$\times$~10$^2$~$\times$~H$_2$.
It can be seen that the cloudy model is similar to the clear model, but with muted molecular absorption features. The base spectrum of Kepler-30c is shown with and without the simulated JWST spectra, with the zoomed-in version included to make the main absorption features clearer.

\subsection{Retrieval of JWST-simulated spectra using ARCiS}
Using our simulated spectra for various JWST instruments (those highlighted in bold in Table~\ref{t:JWST_instruments} and illustrated in Fig.~\ref{fig:trans_base}), we ran a series of retrievals on each spectrum. A summary of the set-ups for the models and retrievals of the four planets considered in this study can be found in Table~\ref{t:ret_setup}, including the JWST instruments and modes used for each planet. These retrievals  test how accurately we can retrieve  what was input into the model atmospheres; see Sect.~\ref{sec:bayes} for details on how this is quantified. In particular, we focus on whether the VMRs of the hydrocarbons of interest (C$_2$H$_2$, C$_2$H$_4$, and CH$_4$) can be determined accurately. For this we allow free molecular retrievals with no conditions on the chemistry enforced. The free parameters and their ranges and priors are given in Table~\ref{t:priors}, and the planetary and stellar parameters used are given in Table~\ref{t:planet_pars}. 
Although self-consistent temperature structures were computed for the forward models (see Fig.~\ref{fig:PT_profs}), the retrieval assumes an isothermal atmosphere. This may result in an overestimation of the abundances and certainties for JWST's capabilities \citep{ref:16RoWaVe}. 
We assumed a population size of 500 in the {\rm Multinest} algorithm used for the retrievals.

\begin{table*}[h!]
        \caption{Instrument and abundance set-ups for the models and retrievals of the four planets considered in this study. The molecular abundances for the species listed in the first column were set to the values specified, with all other molecules abundances based on equilibrium chemistry as given in Table~\ref{tab:plan_gen}.  }
        \label{t:ret_setup}
        \centering
        \begin{tabular}{ll}
                \hline\hline
                \rule{0pt}{3ex}    Molecules and abundances & Instrument/mode   \\
                \hline
                \multicolumn{2}{l}{\rule{0pt}{3ex}\textbf{HD~189733b and HD~209458b}}\\
                        \makecell{\rule{0pt}{3ex}C$_2$H$_2$=1$\times$10$^{-4}$, 1$\times$10$^{-6}$, 1$\times$10$^{-7}$ \\ C$_2$H$_4$=1$\times$10$^{-4}$, 1$\times$10$^{-6}$, 1$\times$10$^{-7}$ \\
                        CH$_4$=1$\times$10$^{-4}$, 1$\times$10$^{-6}$, 1$\times$10$^{-7}$ } & \makecell{\rule{0pt}{3ex}MIRI LRS \\ NIRISS SOSS \\ NIRCam F322W2\\ NIRSpec G395M  \\ NIRISS SOSS + NIRSpec G395M \\ NIRSpec G395M + MIRI LRS \\ NIRCam F322W2 + MIRI LRS } \\
                                \multicolumn{2}{l}{\rule{0pt}{3ex}\textbf{HD~97658b}}\\
                \makecell{\rule{0pt}{3ex}C$_2$H$_2$=1$\times$10$^{-4}$, 1$\times$10$^{-6}$, 1$\times$10$^{-7}$ \\ C$_2$H$_4$=1$\times$10$^{-4}$, 1$\times$10$^{-6}$, 1$\times$10$^{-7}$ \\
                        CH$_4$=1$\times$10$^{-4}$, 1$\times$10$^{-6}$, 1$\times$10$^{-7}$ } & \makecell{\rule{0pt}{3ex}MIRI LRS (+ 25 transits) \\ NIRISS SOSS (+ 25 transits)\\ NIRCam F322W2 (+ 25 transits)\\ NIRSpec G395M (+ 25 transits) \\ NIRISS SOSS + NIRSpec G395M \\ NIRSpec G395M + MIRI LRS \\ NIRCam F322W2 + MIRI LRS } \\
                                        \multicolumn{2}{l}{\rule{0pt}{3ex}\textbf{Kepler-30c}}\\
                \makecell{\rule{0pt}{3ex}C$_2$H$_2$=1$\times$10$^{-4}$, 1$\times$10$^{-6}$ \\ C$_2$H$_4$=1$\times$10$^{-4}$, 1$\times$10$^{-6}$\\
                        CH$_4$=1$\times$10$^{-4}$, 1$\times$10$^{-6}$ } & \makecell{\rule{0pt}{3ex}MIRI LRS (slit/slitless)\\ NIRISS SOSS \\NIRCam F322W2\\ NIRSpec G395M  \\NIRSpec Prism \\ NIRISS SOSS + NIRSpec G395M \\ NIRSpec Prism + MIRI LRS (slit/slitless) } \\
                \hline \hline
        \end{tabular}
        \rule{0pt}{0.2ex}
\end{table*}

\begin{table*}[h!]
        \caption{Parameters and the corresponding ranges and priors used in the retrievals in this study. $P_{\rm top}$ and $f_{\rm mix}$ are the cloud pressure and Rayleigh scattering relating to the cloud deck, respectively. }
        \label{t:priors}
        \centering
        \begin{tabular}{lll}
                        \hline\hline
                \rule{0pt}{3ex}Parameter       &       Range & Prior   \\
                                        \hline
                                        \multicolumn{3}{l}{\rule{0pt}{3ex}\textbf{Parameters included in all set-ups}}\\
                \rule{0pt}{3ex}Planet radius, $R_p$ ($R_J$)     & 5$\sigma$ around literature value       &  Flat linear       \\
                \rule{0pt}{3ex}Atmospheric temperature, $T_p$ (K) & 800-2800; 300-1300$^{a}$ & Flat linear \\
                \rule{0pt}{3ex}Molecular abundances (VMR)$^{b}$ & 10$^{-12}$ \dots  1 & Flat log\\
                \multicolumn{3}{l}{\rule{0pt}{3ex}\textbf{Parameters included in cloud set-up only}}\\
                \rule{0pt}{3ex}P$_{\rm top}$ (Pa)  & -3 \dots 3  & Flat log\\
                        \rule{0pt}{3ex}f$_{\rm mix}$ ($\times$ H$_2$ scattering)  & 10$^{-3}$ \dots $10^6$ & Flat log\\
\hline \hline
        \end{tabular}
                        \rule{0pt}{0.2ex}
\flushleft{\textit{$^a$: The first temperature range is for HD~189733b, HD~209458b, and HD~97658b. The second temperature range is for Kepler-30c. }}
\flushleft{\textit{$^b$: The molecules included in all retrievals were H$_2$O, C$_2$H$_2$, CH$_4$, C$_2$H$_4$, CO$_2$, CO, NH$_3$, TiO, and VO. }}
\end{table*}

\subsection{Detection evidence from the Bayes factor}\label{sec:bayes}
In order to determine the detectability of a particular hydrocarbon at a given VMR, we used the Nested Sampling Global Log-Evidence (log($E$)) as a metric. This is given as an output from the {\rm Multinest} algorithm \citep{08FeHo.multi,09FeGaHo.multi,13FeHoCa.multi}. 
This Bayesian log-evidence is then used to find the Bayes factor ($B_{01}$) (see e.g.  \citealt{13BeSe,15WaTiRo.taurex,20Br.exo}), which is a measure to assess the significance of one model against another and, given identical model priors, is reduced to the ratio of evidence ($E$). Here, `model' refers to the set of free parameters included in the retrieval, in particular if we include the hydrocarbon of interest (whose VMR  we are varying) or not. If the Bayes factor  $B_{01}$~$>$~5, then according to \cite{08Trotta.stats} the model can be considered significant with respect to the `base' model: $B_{01}$~$>$~5 corresponds to $>$~3.6~$\sigma$ detection over the base model; $B_{01}$~$>$~11 corresponds to $>$~5~$\sigma$ detection over the base model; $B_{01}$~$<$~1 indicates that the evidence is not significant; 1~$<$~$B_{01}$~$<$~2.5 indicates weak evidence; 2.5~$<$~$B_{01}$~$<$~5 indicates moderate evidence; $B_{01}$~$>$~5 indicates strong evidence; and $B_{01}$~$>$~11 indicates very strong evidence.

\section{Results}
\label{sec:results}

We    attempted to quantify the minimum values of  the VMR that are observable for each molecule, planet, and instrument (or instrument combination).
In Tables~\ref{t:stats_HD209}~-~\ref{t:stats_Kepler30c} we specify the Bayes factor, $B_{01}$, for each retrieval set-up, comparing the retrieval with the specified molecule (C$_2$H$_2$, C$_2$H$_4$, CH$_4$) included versus not included. In all cases the specified molecule was included in the initial forward atmospheric simulation, at the log(VMR) specified. Based on \cite{08Trotta.stats}, we consider all cases with $B_{01}$~$\geq$~5 to be significant.
More details are provided below for each planet, with notes on the differences between the detectability of the three species (C$_2$H$_2$, CH$_4$, C$_2$H$_4$) for different instruments.

        \begin{table*}[h!]
        \caption{Bayes factor ($B_{01}$) for various retrievals for HD~209458b. $B_{01}$~$>$~11 indicates a very strong detection, $B_{01}$~$>$~5 indicates a strong detection, $B_{01}$~$>$~2.5 indicates a moderate detection, $B_{01}$~$>$~1 indicates a weak detection, $B_{01}$~$<$~1 indicates no detection. 
        }       
        \label{t:stats_HD209} 
        \centering  
        \begin{tabular}{|c|p{0.5cm}|p{0.5cm}|p{0.5cm}||p{0.5cm}|p{0.5cm}|p{0.5cm}||p{0.5cm}|p{0.5cm}|p{0.5cm}|}
                \hline
                \hline
                \rule{0pt}{2.5ex}       &       \multicolumn{9}{|c|}{HD 209458b} \\
                \hline
                \rule{0pt}{2.5ex}Molecule &     \multicolumn{3}{|c|}{C$_2$H$_2$}        &       \multicolumn{3}{|c|}{C$_2$H$_4$}        &       \multicolumn{3}{|c|}{CH$_4$} \\
                \hline
                \rule{0pt}{2.5ex}log(VMR) &     -4      &       -6      &       -7      &       -4      &       -6      &       -7      &       -4      &       -6      &       -7      \\
                \hline
                \rule{0pt}{2.5ex}MIRI LRS                                                                & \cellcolor{blue!20}$>$11 & \cellcolor{blue!10}$>$2.5 & $<$1 &        \cellcolor{blue!20}$>$11  & $<$1 & $<$1 &       \cellcolor{blue!20}$>$11  & \cellcolor{blue!10}$>$2.5& $<$1 \\
                \rule{0pt}{2.5ex}NIRISS SOSS                                                             & \cellcolor{blue!20}$>$11  & $<$1 & $<$1 &       \cellcolor{blue!20}$>$11  & $<$1 & $<$1 &       \cellcolor{blue!20}$>$11  & $<$1 & $<$1 \\
                \rule{0pt}{2.5ex}NIRCam F322W2                                                           & \cellcolor{blue!20}$>$11 & \cellcolor{blue!15}$>$5 & $<$1 &        \cellcolor{blue!20}$>$11  & \cellcolor{blue!20}$>$11& $<$1 &       \cellcolor{blue!20}$>$11 & \cellcolor{blue!20}$>$11 & $<$1 \\
                \rule{0pt}{2.5ex}NIRSpec G395M                                                           & \cellcolor{blue!20}$>$11& \cellcolor{blue!10}$>$2.5  & $<$1 &       \cellcolor{blue!20}$>$11 & \cellcolor{blue!15}$>$5 & $<$1 &       \cellcolor{blue!20}$>$11 & \cellcolor{blue!20}$>$11 & $<$1 \\
                \makecell{\rule{0pt}{2.5ex}NIRISS SOSS + NIRSpec G395M} & \cellcolor{blue!20}$>$11& \cellcolor{blue!20}$>$11 & $<$1 &     \cellcolor{blue!20}$>$11 & \cellcolor{blue!20}$>$11 & $<$1 &     \cellcolor{blue!20}$>$11 & \cellcolor{blue!20}$>$11 & $<$1 \\
                \makecell{\rule{0pt}{2.5ex}NIRSpec G395m + MIRI LRS}    & \cellcolor{blue!20}$>$11& \cellcolor{blue!20}$>$11 & $<$1 &     \cellcolor{blue!20}$>$11 & \cellcolor{blue!20}$>$11 & $<$1 &     \cellcolor{blue!20}$>$11 & \cellcolor{blue!20}$>$11 & $<$1 \\
                \makecell{\rule{0pt}{2.5ex}NIRCam F322W2 + MIRI LRS}    & \cellcolor{blue!20}$>$11& \cellcolor{blue!20}$>$11 & $<$1 &     \cellcolor{blue!20}$>$11 & \cellcolor{blue!20}$>$11 & $<$1 &     \cellcolor{blue!20}$>$11 & \cellcolor{blue!20}$>$11 & $<$1 \\
                \hline
                \hline
        \end{tabular}
\end{table*}

        \begin{table*}[h!]
        \caption{Bayes factor ($B_{01}$) for various retrievals for HD~189733b. $B_{01}$~$>$~11 indicates a very strong detection, $B_{01}$~$>$~5 indicates a strong detection, $B_{01}$~$>$~2.5 indicates a moderate detection, $B_{01}$~$>$~1 indicates a weak detection, $B_{01}$~$<$~1 indicates no detection.}
        \label{t:stats_HD189} 
        \centering  
        \begin{tabular}{|c|p{0.5cm}|p{0.5cm}|p{0.5cm}||p{0.5cm}|p{0.5cm}|p{0.5cm}||p{0.5cm}|p{0.5cm}|p{0.5cm}|}
                \hline
                \hline
        \rule{0pt}{2.5ex}       &       \multicolumn{9}{|c|}{HD 189733b} \\
                                        \hline
\rule{0pt}{2.5ex}Molecule &     \multicolumn{3}{|c|}{C$_2$H$_2$}        &       \multicolumn{3}{|c|}{C$_2$H$_4$}        &       \multicolumn{3}{|c|}{CH$_4$} \\
\hline
\rule{0pt}{2.5ex}log(VMR) &     -4      &       -6      &       -7      &       -4      &       -6      &       -7      &       -4      &       -6      &       -7      \\
\hline
        \rule{0pt}{2.5ex}MIRI LRS                                                                & \cellcolor{blue!20}$>$11 & \cellcolor{blue!10}$>$2.5 & $<$1 &        \cellcolor{blue!20}$>$11  & $<$1 & $<$1 &       \cellcolor{blue!20}$>$11  & \cellcolor{blue!5}$>$1& $<$1 \\

        \rule{0pt}{2.5ex}NIRISS SOSS                                                             & \cellcolor{blue!20}$>$11  & $<$1 & $<$1 &     \cellcolor{blue!20}$>$11  & $<$1 & $<$1 &        \cellcolor{blue!20}$>$11  & $<$1 & $<$1 \\

        \rule{0pt}{2.5ex}NIRCam F322W2                                                           & \cellcolor{blue!20}$>$11 & \cellcolor{blue!20}$>$11 & $<$1 &        \cellcolor{blue!20}$>$11  & \cellcolor{blue!15}$>$5 & $<$1 &       \cellcolor{blue!20}$>$11 & \cellcolor{blue!20}$>$11 & $<$1 \\

        \rule{0pt}{2.5ex}NIRSpec G395M                                                           & \cellcolor{blue!20}$>$11& \cellcolor{blue!20}$>$11 & $<$1 &        \cellcolor{blue!20}$>$11 & \cellcolor{blue!20}$>$11 & $<$1 &       \cellcolor{blue!20}$>$11 & \cellcolor{blue!20}$>$11 & $<$1 \\

        \makecell{\rule{0pt}{2.5ex}NIRISS SOSS + NIRSpec G395M} & \cellcolor{blue!20}$>$11& \cellcolor{blue!20}$>$11 & $<$1 &       \cellcolor{blue!20}$>$11 & \cellcolor{blue!20}$>$11 & $<$1 &        \cellcolor{blue!20}$>$11 & \cellcolor{blue!20}$>$11 & $<$1 \\

        \makecell{\rule{0pt}{2.5ex}NIRSpec G395m + MIRI LRS}    & \cellcolor{blue!20}$>$11& \cellcolor{blue!20}$>$11 & $<$1 &       \cellcolor{blue!20}$>$11 & \cellcolor{blue!20}$>$11 & $<$1 &        \cellcolor{blue!20}$>$11 & \cellcolor{blue!20}$>$11 & $<$1 \\

        \makecell{\rule{0pt}{2.5ex}NIRCam F322W2 + MIRI LRS}    & \cellcolor{blue!20}$>$11& \cellcolor{blue!20}$>$11 & $<$1 &       \cellcolor{blue!20}$>$11 & \cellcolor{blue!20}$>$11 & $<$1 &        \cellcolor{blue!20}$>$11 & \cellcolor{blue!20}$>$11 & $<$1 \\

                \hline
                \hline
        \end{tabular}
        \rule{0pt}{0.2ex}

\end{table*}

        \begin{table*}[h!]
        \caption{Bayes factor ($B_{01}$) for various retrievals for HD~189733b, which include a cloud deck at $P_{\rm top}$~=~0.01~bar with a scattering slope of $f_{\rm mix}$=1$\times$10$^{2}$~$\times$~H$_2$ scattering. $B_{01}$~$>$~11 indicates a very strong detection, $B_{01}$~$>$~5 indicates a strong detection, $B_{01}$~$>$~2.5 indicates a moderate detection, $B_{01}$~$>$~1 indicates a weak detection, $B_{01}$~$<$~1 indicates no detection. }
        \label{t:stats_HD189_clouds} 
        \centering  
        \begin{tabular}{|c|p{0.5cm}|p{0.5cm}|p{0.5cm}||p{0.5cm}|p{0.5cm}|p{0.5cm}||p{0.5cm}|p{0.5cm}|p{0.5cm}|}

                \hline
                \hline
                \rule{0pt}{2.5ex}       &       \multicolumn{9}{|c|}{HD 189733b} \\
                \hline
                \rule{0pt}{2.5ex}Molecule &     \multicolumn{3}{|c|}{C$_2$H$_2$}        &       \multicolumn{3}{|c|}{C$_2$H$_4$}        &       \multicolumn{3}{|c|}{CH$_4$} \\
                \hline
                \rule{0pt}{2.5ex}log(VMR) &     -4      &       -6      &       -7      &       -4      &       -6      &       -7      &       -4      &       -6      &       -7      \\
                \hline
                \rule{0pt}{2.5ex}MIRI LRS                                                                & \cellcolor{blue!20}$>$11 & $<$1& $<$1 &       \cellcolor{blue!20}$>$11  & $<$1 & $<$1 &        \cellcolor{blue!20}$>$11  & \cellcolor{blue!5}$>$1& $<$1 \\
                \rule{0pt}{2.5ex}NIRISS SOSS                                                             & \cellcolor{blue!20}$>$11  & $<$1 & $<$1 &       \cellcolor{blue!20}$>$11  & $<$1& $<$1 &        \cellcolor{blue!20}$>$11  & $<$1 & $<$1 \\
                \rule{0pt}{2.5ex}NIRCam F322W2                                                           & \cellcolor{blue!20}$>$11 & \cellcolor{blue!10}$>$2.5 & $<$1 &        \cellcolor{blue!20}$>$11  & \cellcolor{blue!10}$>$2.5 & $<$1 &       \cellcolor{blue!20}$>$11 & \cellcolor{blue!20}$>$11 & $<$1 \\
                \rule{0pt}{2.5ex}NIRSpec G395M                                                           & \cellcolor{blue!20}$>$11& \cellcolor{blue!20}$>$11 & $<$1 &        \cellcolor{blue!20}$>$11 & \cellcolor{blue!20}$>$11 & $<$1 &       \cellcolor{blue!20}$>$11 & \cellcolor{blue!20}$>$11 & $<$1 \\
                \makecell{\rule{0pt}{2.5ex}NIRISS SOSS + NIRSpec G395M} & \cellcolor{blue!20}$>$11& \cellcolor{blue!20}$>$11 & $<$1 &     \cellcolor{blue!20}$>$11 & \cellcolor{blue!20}$>$11 & $<$1 &     \cellcolor{blue!20}$>$11 & \cellcolor{blue!20}$>$11 & $<$1 \\
                \makecell{\rule{0pt}{2.5ex}NIRSpec G395m + MIRI LRS}    & \cellcolor{blue!20}$>$11& \cellcolor{blue!20}$>$11 & $<$1 &     \cellcolor{blue!20}$>$11 & \cellcolor{blue!20}$>$11 & $<$1 &     \cellcolor{blue!20}$>$11 & \cellcolor{blue!20}$>$11 & $<$1 \\
                \makecell{\rule{0pt}{2.5ex}NIRCam F322W2 + MIRI LRS}    & \cellcolor{blue!20}$>$11& \cellcolor{blue!15}$>$5  & $<$1 &     \cellcolor{blue!20}$>$11 & \cellcolor{blue!15}$>$5 & $<$1 &      \cellcolor{blue!20}$>$11 & \cellcolor{blue!20}$>$11 & $<$1\\
                \hline
                \hline
        \end{tabular}
        \rule{0pt}{0.2ex}
\end{table*}

        \begin{table*}[h!]
        \caption{Bayes factor ($B_{01}$) for various retrievals for HD~97658b. $B_{01}$~$>$~11 indicates a very strong detection, $B_{01}$~$>$~5 indicates a strong detection, $B_{01}$~$>$~2.5 indicates a moderate detection, $B_{01}$~$>$~1 indicates a weak detection, $B_{01}$~$<$~1 indicates no detection. }
        \label{t:stats_HD976} 
        \centering  
        \begin{tabular}{|c|p{0.5cm}|p{0.5cm}|p{0.5cm}||p{0.5cm}|p{0.5cm}|p{0.5cm}||p{0.5cm}|p{0.5cm}|p{0.5cm}|}
                \hline
                \hline
                \rule{0pt}{2.5ex}       &       \multicolumn{9}{|c|}{HD 97658b} \\
                \hline
                \rule{0pt}{2.5ex}Molecule &     \multicolumn{3}{|c|}{C$_2$H$_2$}        &       \multicolumn{3}{|c|}{C$_2$H$_4$}        &       \multicolumn{3}{|c|}{CH$_4$} \\
                \hline
                \rule{0pt}{2.5ex}log(VMR) &     -4      &       -6      &       -7      &       -4      &       -6      &       -7      &       -4      &       -6      &       -7      \\
                \hline
                \rule{0pt}{2.5ex}MIRI LRS                                                                & \cellcolor{blue!20}$>$11 & $<$1 & $<$1 &       \cellcolor{blue!20}$>$11  & $<$1 & $<$1 &       \cellcolor{blue!20}$>$11  & $<$1 & $<$1 \\
                                \rule{0pt}{2.5ex}MIRI LRS (25 transits)  & \cellcolor{blue!20}$>$11 & \cellcolor{blue!10}$>$2.5 & $<$1 & \cellcolor{blue!20}$>$11  & $<$1 & $<$1 &        \cellcolor{blue!20}$>$11  & \cellcolor{blue!5}$>$1& $<$1 \\
                \rule{0pt}{2.5ex}NIRISS SOSS                                                             & \cellcolor{blue!20}$>$11  & $<$1 & $<$1 &       \cellcolor{blue!20}$>$11  & $<$1 & $<$1 &       \cellcolor{blue!20}$>$11  & $<$1 & $<$1 \\
                \rule{0pt}{2.5ex}NIRISS SOSS (25 transits)                                                       & \cellcolor{blue!20}$>$11  & \cellcolor{blue!10}$>$2.5 & $<$1 &        \cellcolor{blue!20}$>$11  & \cellcolor{blue!5}$>$1& $<$1 &       \cellcolor{blue!20}$>$11  & \cellcolor{blue!20}$>$11 & $<$1 \\
                \rule{0pt}{2.5ex}NIRCam F322W2                                                           & \cellcolor{blue!20}$>$11 & $<$1 & $<$1 &       \cellcolor{blue!20}$>$11  & $<$1& $<$1 &        \cellcolor{blue!20}$>$11 & \cellcolor{blue!15}$>$5 & $<$1 \\
                \rule{0pt}{2.5ex}NIRCam F322W2 (25 transits)                                                     & \cellcolor{blue!20}$>$11 & \cellcolor{blue!15}$>$5  & $<$1 &       \cellcolor{blue!20}$>$11  & $<$1& $<$1 &        \cellcolor{blue!20}$>$11 & \cellcolor{blue!20}$>$11 & $<$1 \\
                \rule{0pt}{2.5ex}NIRSpec G395M                                                           & \cellcolor{blue!20}$>$11& $<$1  & $<$1 &       \cellcolor{blue!20}$>$11 & $<$1& $<$1 & \cellcolor{blue!20}$>$11 & \cellcolor{blue!20}$>$11 & $<$1 \\
                \rule{0pt}{2.5ex}NIRSpec G395M (25 transits)                                                             & \cellcolor{blue!20}$>$11& \cellcolor{blue!15}$>$5  & $<$1 &       \cellcolor{blue!20}$>$11 & $<$1& $<$1 &  \cellcolor{blue!20}$>$11 & \cellcolor{blue!20}$>$11 & $<$1 \\
                \makecell{\rule{0pt}{2.5ex}NIRISS SOSS + NIRSpec G395M} & \cellcolor{blue!20}$>$11& $<$1 & $<$1 & \cellcolor{blue!20}$>$11 & \cellcolor{blue!20}$>$11 & $<$1 &        \cellcolor{blue!20}$>$11 & \cellcolor{blue!20}$>$11 & $<$1 \\
                \makecell{\rule{0pt}{2.5ex}NIRSpec G395m + MIRI LRS}    & \cellcolor{blue!20}$>$11& $<$1 & $<$1 & \cellcolor{blue!20}$>$11 & \cellcolor{blue!5}$>$1& $<$1 &  \cellcolor{blue!20}$>$11 & \cellcolor{blue!20}$>$11 & $<$1 \\
                \makecell{\rule{0pt}{2.5ex}NIRCam F322W2 + MIRI LRS}    & \cellcolor{blue!20}$>$11&  \cellcolor{blue!15}$>$5 & $<$1 &     \cellcolor{blue!20}$>$11 & \cellcolor{blue!20}$>$11 & $<$1 &     \cellcolor{blue!20}$>$11 & \cellcolor{blue!20}$>$11 & $<$1 \\
                \hline
                \hline
        \end{tabular}
\end{table*}

        \begin{table*}[h!]
        \caption{Bayes factor ($B_{01}$) for various retrievals for Kepler-30c. $B_{01}$~$>$~11 indicates a very strong detection, $B_{01}$~$>$~5 indicates a strong detection, $B_{01}$~$>$~2.5 indicates a moderate detection, $B_{01}$~$>$~1 indicates a weak detection, $B_{01}$~$<$~1 indicates no detection. }
        \label{t:stats_Kepler30c} 
        \centering  
        \begin{tabular}{|c|p{0.5cm}|p{0.5cm}||p{0.5cm}|p{0.5cm}||p{0.5cm}|p{0.5cm}|}
                \hline
                \hline
                \rule{0pt}{2.5ex}       &       \multicolumn{6}{|c|}{Kepler-30c} \\
                \hline
                \rule{0pt}{2.5ex}Molecule &     \multicolumn{2}{|c|}{C$_2$H$_2$}        &       \multicolumn{2}{|c|}{C$_2$H$_4$}        &       \multicolumn{2}{|c|}{CH$_4$} \\
                \hline
                \rule{0pt}{2.5ex}log(VMR) &     -4      &       -6      &       -4      &       -6      &               -4      &       -6              \\
                \hline
                                \rule{0pt}{2.5ex}MIRI LRS (slit)                                                         & $<$1 & $<$1 &         $<$1  & $<$1 &       $<$1  & $<$1 \\
                                                                \rule{0pt}{2.5ex}MIRI LRS (slitless)                                                          & $<$1 & $<$1 &   $<$1  & $<$1 &  $<$1  & $<$1 \\
                                                \rule{0pt}{2.5ex}NIRISS SOSS                                                                 & $<$1 & $<$1 &  $<$1  & $<$1 &  $<$1  & $<$1\\
                \rule{0pt}{2.5ex}NIRCam F322W2                                                           & $<$1 & $<$1 &         $<$1  & $<$1 &  $<$1  & $<$1\\
                \rule{0pt}{2.5ex}NIRSpec G395M                                                           & $<$1 & $<$1 &         $<$1  & $<$1 &  $<$1  & $<$1\\
                \rule{0pt}{2.5ex}NIRSpec Prism                                                                   & $<$1 & $<$1  &        $<$1  & $<$1  &      \cellcolor{blue!5}$>$1  & $<$1  \\
                \rule{0pt}{2.5ex}NIRISS SOSS + NIRSpec G395M                                                             & $<$1 & $<$1 &         $<$1  & $<$1 &       \cellcolor{blue!5}$>$1   & $<$1  \\
                \rule{0pt}{2.5ex}NIRSpec Prism + MIRI LRS (slit)                                                                 & $<$1 & $<$1 &         $<$1  & $<$1 &       \cellcolor{blue!10}$>$2.5  & $<$1 \\
                                \rule{0pt}{2.5ex}NIRSpec Prism + MIRI LRS (slitless)                                                              & $<$1 & $<$1 &   $<$1  & $<$1 &  \cellcolor{blue!5}$>$1  & $<$1 \\
                \hline
                \hline
        \end{tabular}
\end{table*}

\subsection{HD 189733b \& HD 209458b}
\begin{figure*}[h!]
    \centering
    \includegraphics[width=\textwidth]{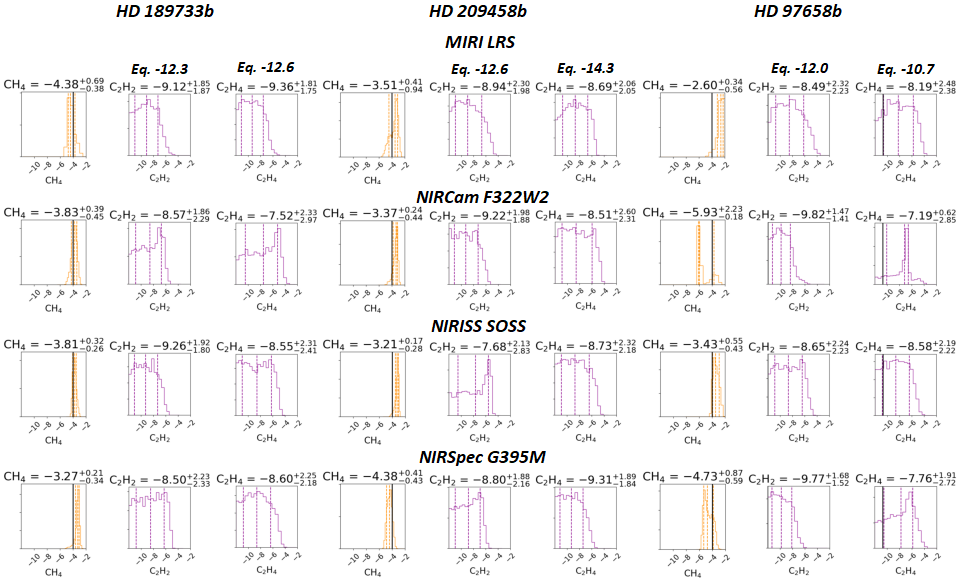}
    \caption{Posterior distributions from retrievals of model atmospheres of HD~189733b, HD~209458b, and HD~97658b for an input CH$_4$ abundance of $10^{-4}$ (orange) and C$_2$H$_2$ and C$_2$H$_4$ in their respective equilibrium abundances (purple). The results using a variety of instruments are given.
    The model input abundance is indicated by a black line, which is not shown when lower than the presented range.}
    \label{fig:single_ch4-4}
\end{figure*}

\begin{figure*}[h!]
    \centering
    \includegraphics[width=\textwidth]{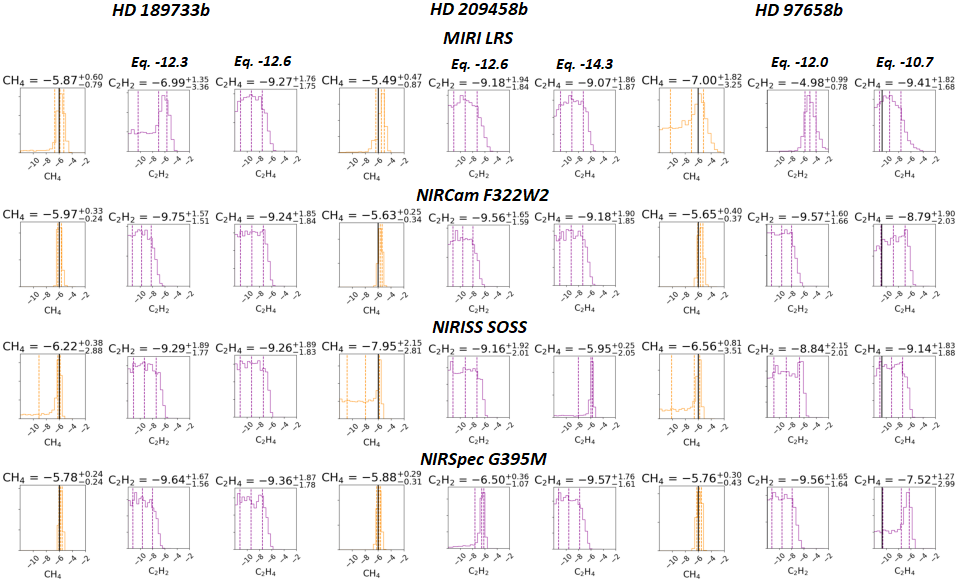}
    \caption{Posterior distributions from retrievals of model atmospheres of HD~189733b, HD~209458b, and HD~97658b for an input CH$_4$ abundance of $10^{-6}$ (orange) and C$_2$H$_2$ and C$_2$H$_4$ in their respective equilibrium abundances (purple). The results using a variety of instruments are given.
    The model input abundance is indicated by a black line, which is not shown when lower than the presented range.}
    \label{fig:single_ch4-6}
\end{figure*}

\begin{figure*}[h!]
    \centering
    \includegraphics[width=\textwidth]{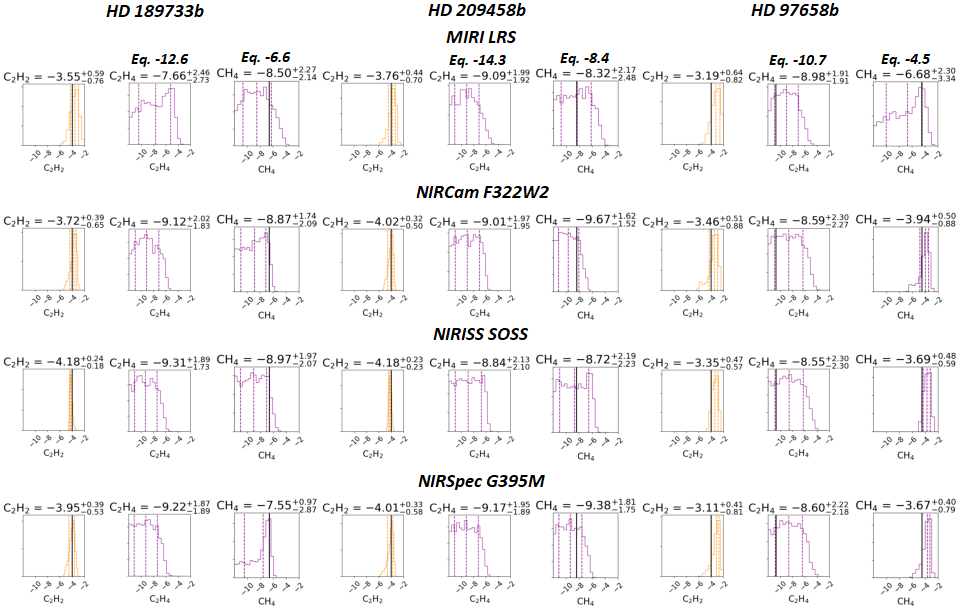}
    \caption{Posterior distributions from retrievals of model atmospheres of HD~189733b, HD~209458b, and HD~97658b for an input C$_2$H$_2$ abundance of $10^{-4}$ (orange) and C$_2$H$_4$ and CH$_4$ in their respective equilibrium abundances (purple). The results using a variety of instruments are given.
    The model input abundance is indicated by a black line, which is not shown when lower than the presented range.}
    \label{fig:single_c2h2-4}
\end{figure*}

\begin{figure*}[h!]
    \centering
    \includegraphics[width=\textwidth]{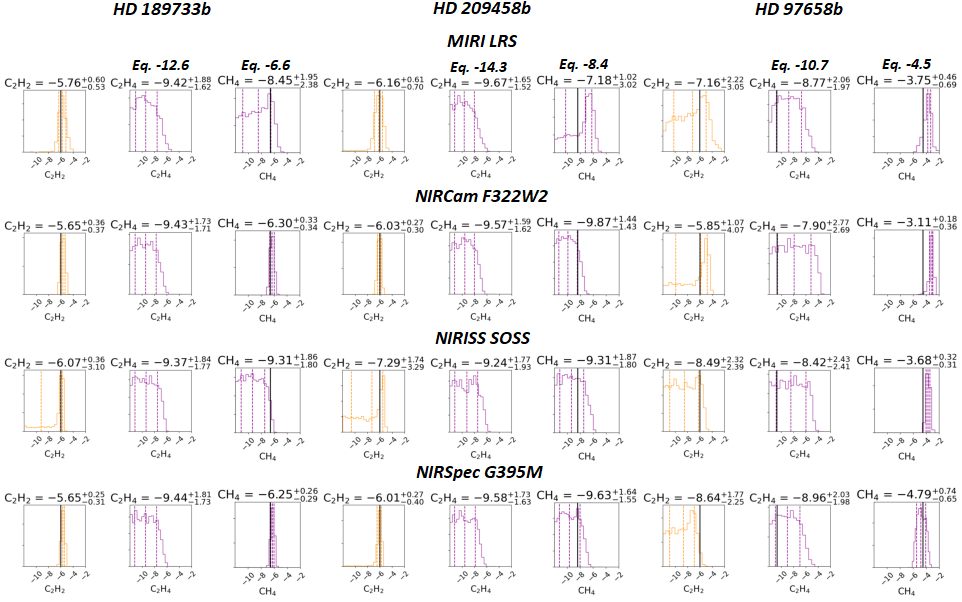}
    \caption{Posterior distributions from retrievals of model atmospheres of HD~189733b, HD~209458b, and HD~97658b for an input C$_2$H$_2$ abundance of $10^{-6}$ (orange) and C$_2$H$_4$ and CH$_4$ in their respective equilibrium abundances (purple). The results using a variety of instruments are given.
    The model input abundance is indicated by a black line, which is not shown when lower than the presented range.}
    \label{fig:single_c2h2-6}
\end{figure*}

\begin{figure*}[h!]
    \centering
    \includegraphics[width=\textwidth]{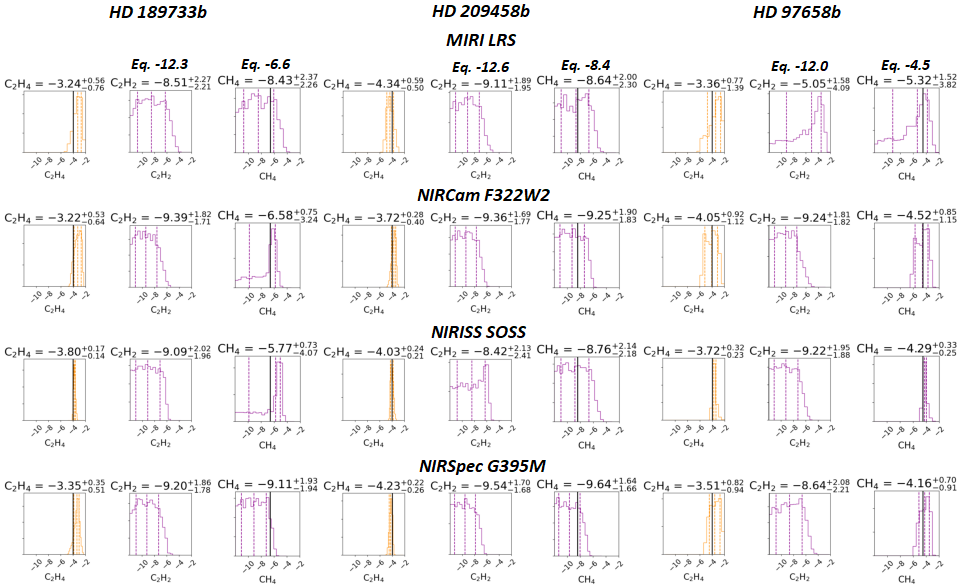}
    \caption{Posterior distributions from retrievals of model atmospheres of HD~189733b, HD~209458b, and HD~97658b for an input C$_2$H$_4$ abundance of $10^{-4}$ (orange) and C$_2$H$_2$ and CH$_4$ in their respective equilibrium abundances (purple). The results using a variety of instruments are given.
    The model input abundance is indicated by a black line, which is not shown when lower than the presented range.}
    \label{fig:single_c2h4-4}
\end{figure*}

\begin{figure*}[h!]
\centering
\begin{minipage}{\textwidth}
    \centering
    \includegraphics[width=\textwidth]{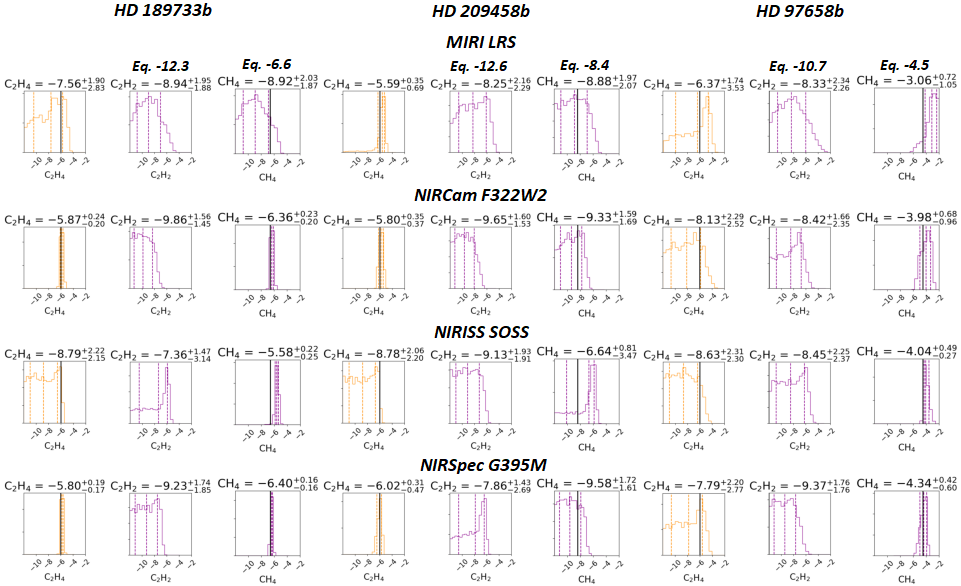}
    \caption{Posterior distributions from retrievals of model atmospheres of HD~189733b, HD~209458b, and HD~97658b for an input C$_2$H$_4$ abundance of $10^{-6}$ (orange) and C$_2$H$_2$ and CH$_4$ in their respective equilibrium abundances (purple). The results using a variety of instruments are given.
    The model input abundance is indicated by a black line, which is not shown when lower than the presented range.}
    \label{fig:single_c2h4-6}
\end{minipage}\vspace{1mm}
\begin{minipage}{\textwidth}
    \centering
    \includegraphics[width=\textwidth]{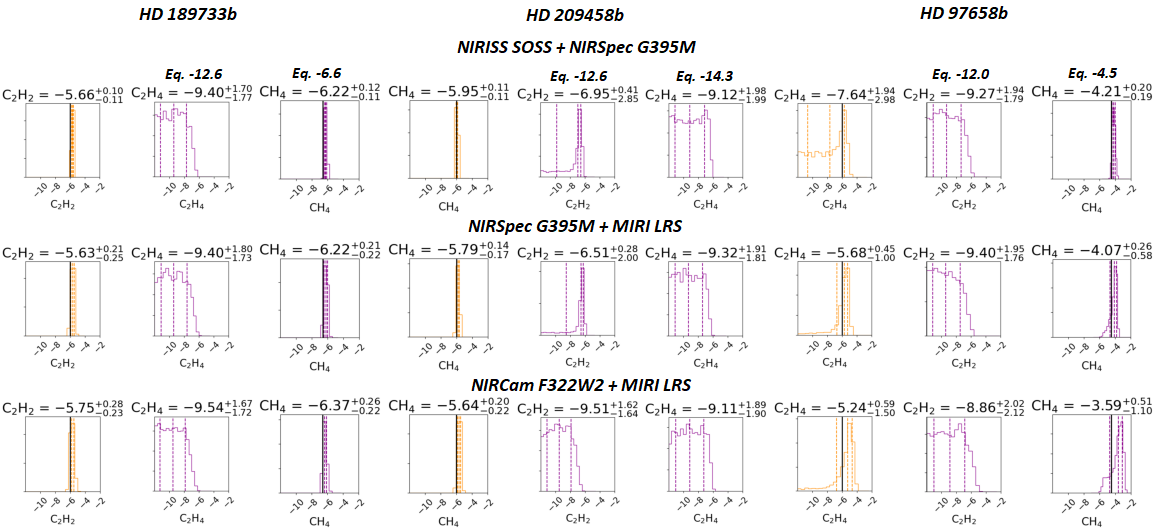}
    \caption{Posterior distributions from retrievals using selected combinations of instruments of model atmospheres of HD~189733b, HD~209458b, and HD~97658b for an input abundance of $10^{-4}$ for the main hydrocarbon specified (orange) and other species in their respective equilibrium abundances (purple). 
    The model input abundance is indicated by a black line, which is not shown when lower than the presented range.}
    \label{fig:combined}
\end{minipage}
\end{figure*}

The posterior distributions of all hydrocarbon VMRs retrieved for various abundances (1~$\times$~10$^{-4}$ or 1~$\times$~10$^{-6}$ for CH$_4$, C$_2$H$_2$, and C$_2$H$_4$) are collected in Figs.~\ref{fig:single_ch4-4}-\ref{fig:single_c2h4-6} for the single instrument set-ups of HD 189733b, HD 209458b, and HD 97658b. The posterior distributions for the VMRs of the other hydrocarbons are also given in each case. We give a short summary for each instrument or instrument combination for these two planets below.

\subsubsection{HD 189733b \&\ HD 209458b: MIRI LRS}
MIRI LRS is able to detect the 5-12 micron range. The outcome is similar for both planets: all species are retrieved with high evidence when in an abundance of $10^{-4}$, but difficulties occur around the $10^{-6}$ mark. 
The Bayes factors in Tables~\ref{t:stats_HD209} and~\ref{t:stats_HD189} become <1 at this point. Interestingly, for the $10^{-6}$ cases, C$_2$H$_2$ is moderately detectable in both planets, while there is no significant evidence for C$_2$H$_4$ at all. Observations using MIRI LRS may be relatively noisy, especially at wavelengths   longer than 10 $\mu$m. As shown in Figs.~\ref{fig:all} and \ref{fig:chab}, the most distinct features for C$_2$H$_4$ occur in this range, while at the shorter wavelengths of MIRI's range C$_2$H$_2$ and CH$_4$ have overlapping features. Therefore, in lower abundances C$_2$H$_4$ may be overshadowed by noise, while features of CH$_4$ could be attributed to C$_2$H$_2$ and vice versa. This is evident in the posterior distribution plot for C$_2$H$_2$ of Fig.~\ref{fig:single_ch4-6} in the case of HD~189733b, where VMR(CH$_4$)~=~1~$\times$~10$^{-6}$ (and C$_2$H$_2$ should be in negligible abundance). Similarly, in the case of HD~209458b using MIRI LRS, Fig.~\ref{fig:single_c2h2-6} shows that a peak becomes visible in the posterior distribution of CH$_4$ at a higher abundance than it should be, for the case where VMR(C$_2$H$_2$)~=~1~$\times$~10$^{-6}$.

\subsubsection{HD 189733b \&\ HD 209458b: NIRISS SOSS}

NIRISS SOSS  operates in the 0.6 to 2.8 micron range, where, as shown in \autoref{fig:all}, the strongest signatures are from CH$_4$ and C$_2$H$_4$. C$_2$H$_2$  also shows signatures here, but they are not significant and could easily be overshadowed by the other hydrocarbons when these are sufficiently abundant. Similarly to MIRI LRS, detection becomes more uncertain for an abundance of 1~$\times$~10$^{-6}$. Contrary to MIRI LRS, however, all Bayes factors are $<$1 in these cases. The cause for this is likely of the same nature; Fig.~\ref{fig:all} demonstrates that CH$_4$ and C$_2$H$_4$ overlap in its wavelength range, while the features of C$_2$H$_2$ are narrow and weak. The corresponding posterior of C$_2$H$_4$ in Fig.~\ref{fig:single_ch4-6} (for HD~209458b where VMR(CH$_4$)~=~10$^{-6}$) shows that C$_2$H$_4$ is retrieved as abundant when it should be negligible.  
Figure~\ref{fig:single_c2h4-6} (where VMR(C$_2$H$_4$)~=~10$^{-6}$) shows a higher retrieved abundance for CH$_4$ than it should, in particular for HD~209458b.
These cases indicate that removing the species of interest can still result in a good fit with the other hydrocarbon present in a higher VMR, which leads to a low value of B$_{01}$. We therefore find that for lower abundances NIRISS SOSS is likely capable of finding that {a} hydrocarbon is  present, but seems unable to ascertain which one.

\subsubsection{HD 189733b \& HD 209458b: NIRCam F322W2}
The detectable wavelength range of NIRCam F322W spans approximately 2.4 to 4.2 microns, where the hydrocarbon signatures are strong and unique. The Bayes factors (Tables~\ref{t:stats_HD209} and \ref{t:stats_HD189} for HD~209458b and HD~189733b, respectively) are larger than 5, and in many cases larger than 11, for all three hydrocarbons, even for VMRs of 10$^{-6}$.
The equilibrium abundance of CH$_4$ in HD 189733b is sufficient to be detectable in the absence or lower abundance of other species, evidenced by the distributions in Fig.~\ref{fig:single_c2h2-6} for VMR(C$_2$H$_2$)~=~10$^{-6}$ and in Fig.~\ref{fig:single_c2h4-6} for VMR(C$_2$H$_4$)~=~10$^{-6}$, but 1$\cdot$10$^{-7}$ is not. NIRCam F322W2 has very few problems distinguishing between the hydrocarbons, as opposed to NIRISS SOSS. The only notable example is the inability to find CH$_4$ when C$_2$H$_2$ is increased to 10$^{-4}$ in HD~189733b (see Fig.~\ref{fig:single_c2h2-4}). Although NIRCam F322W2 is still able to strongly detect CH$_4$ at a VMR of 10$^{-6}$ for the cloudy atmosphere case for HD~189733b, only moderate detections are found for C$_2$H$_2$ or C$_2$H$_4$ at the same VMR (see Table~\ref{t:stats_HD189_clouds}).

\subsubsection{HD 189733b \& HD 209458b: NIRSpec G395M}

Tables~\ref{t:stats_HD209} and \ref{t:stats_HD189} (for HD~209458b and HD~189733b, respectively) demonstrate that the performance of NIRSpec G395M is very similar to NIRCam F322W2,
with both set-ups operating in approximately the same wavelength range. Even when clouds are included in our model atmospheres and retrievals, NIRSpec G395M gives strong evidence for the presence of all hydrocarbons down to a VMR of 10$^{-6}$ (see the Bayes factors in Table~\ref{t:stats_HD189_clouds}). This is an improvement in comparison to using NIRCam F322W2 for the cloudy HD~189733b atmospheres.
NIRSpec, just like NIRCam, is able to detect the equilibrium abundance of CH$_4$ in HD 189733b for the cases where VMR(C$_2$H$_4$) and VMR(C$_2$H$_2$) are 10$^{-6}$, but starts to have issues when these are increased to 10$^{-4}$, particularly for VMR(C$_2$H$_4$)~=~10$^{-4}$ (Fig.~\ref{fig:single_c2h4-4}).
There are a few more cases where hydrocarbons interfere with one another, namely for VMR(CH$_4$)~=~10$^{-6}$ in HD 209458b (Fig.~\ref{fig:single_ch4-6}), where C$_2$H$_2$ is retrieved when it should be at negligible abundance, but also VMR(C$_2$H$_4$)~=~10$^{-6}$ (Fig.~\ref{fig:single_c2h4-6}) where C$_2$H$_2$ is similarly retrieved when it should not be.

\subsubsection{HD 189733b \& HD 209458b: NIRISS SOSS and NIRSpec G395M}
\citet{ref:17BaLi} find that increasing the wavelength range is more effective than using the same instrument twice, as is done when observing multiple transits. We therefore test various instrument combinations. Furthermore, they find that the combination of NIRISS SOSS and NIRSpec G395M   yields the most information and the best results. We present a selection of the results for the combined instrument retrievals in Fig.~\ref{fig:combined}.
Combining these two instrument modes shows improvements for retrieving the abundances of C$_2$H$_2$ and C$_2$H$_4$ in HD~209458b in terms of the Bayes factors (Table~\ref{t:stats_HD209}). No improvements are observed for HD~189733b when compared to using NIRSpec G395M by itself, as NIRSpec G395M alone was already very accurate for HD~189733b. While the evidence for the other two hydrocarbons has improved, the strange shape of the C$_2$H$_2$ distribution remains for CH$_4$ increased to 10$^{-6}$ in HD~209458b in Fig.~\ref{fig:combined}.
Although this instrument combination appears to offer an accurate retrieval for abundances down to 10$^{-6}$, we note that the improvements are not drastic in comparison to using NIRSpec G395M alone. This is likely due to the performance of NIRISS SOSS alone, which does not cover a beneficial wavelength region for the detection of the hydrocarbons we are interested in.

\begin{figure*}[h!]
    \centering
    \includegraphics[width=\textwidth]{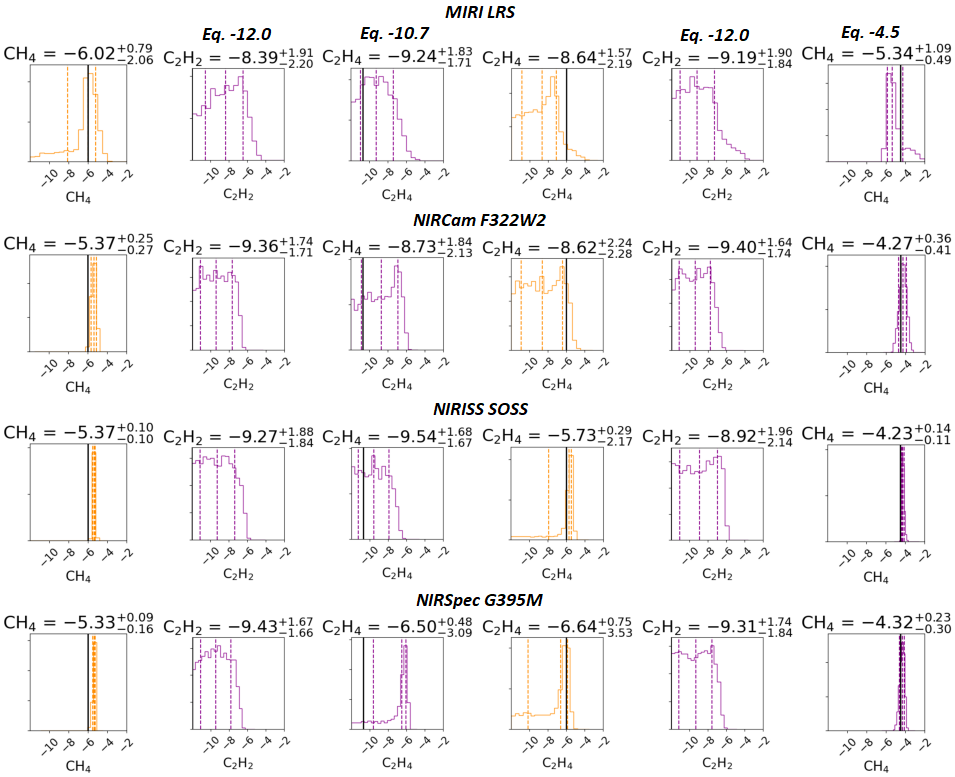}
    \caption{Posterior distributions from retrievals using 25 transits of model atmospheres
    of HD 97658b, for increased CH$_4$ and C$_2$H$_4$ abundances of $10^{-6}$ (orange) and equilibrium abundance of the remaining species (purple). The model input abundance is indicated by a black line, which is not shown when lower than the presented range.}
    \label{fig:transits}
\end{figure*}

\subsubsection{HD 189733b \& HD 209458b: NIRSpec G395M and MIRI LRS}
The combination of NIRSpec G395M and MIRI LRS covers a wide range of wavelengths, including the longer wavelengths expected to be minimally affected by hazes blocking the spectrum. In such cases, using a different combination of instruments could yield more information regarding molecular abundances than the combination treated previously of NIRISS SOSS and NIRSpec G395M, which operates in the shorter wavelengths. We set out to test whether there is an advantage to using this instrument combination as opposed to one of the single instruments alone. In particular, there were previous issues in accurately determining the VMR of C$_2$H$_2$ in HD 189733b down to low values for these instruments individually. 
Similarly to the combination discussed above, we find an improvement in evidence for C$_2$H$_2$ and C$_2$H$_4$ in HD~209458b in comparison to using either instrument alone (Table~\ref{t:stats_HD209}). For HD~189733b, the performance is just as good as NIRSpec G395M individually but there is improvement in comparison to using MIRI LRS alone (Table~\ref{t:stats_HD189}). We  find, however, that combining these instruments does not reduce the retrieved abundance of C$_2$H$_2$ in HD~209458b in the case where VMR(CH$_4$)~=~10$^{-6}$, which should be negligible (Fig.~\ref{fig:combined}).

\subsubsection{HD 189733b \& HD 209458b: NIRCam F322W2 and MIRI LRS}

This combination of instruments demonstrated an improvement in comparison to the individual instruments alone. Although NIRCam F322W2 alone offered medium to strong detections in all cases, this was improved to very strong detections in all cases when combined with MIRI LRS. This can be seen in the Bayes factors of Table~\ref{t:stats_HD209} (for HD~209458b) and Table~\ref{t:stats_HD189} (for HD~189733b). An illustration of the different spectra retrieved using this instrument combination, with and without each of the molecules (CH$_4$, C$_2$H$_2$, C$_2$H$_4$) included in the retrieval, is given in the right panels of Figs.~\ref{fig:comp1_HD189}~and~\ref{fig:HD189_zoomed_NIRSpec_NIRCam}. Here the retrieval with the molecule included (best-fit  spectra in purple) fits the observed data points better than the retrieval with the molecule removed (best-fit spectra in orange) for HD~189733b. Similar figures for HD~209458b can be found in the supplementary information document\footnote{\url{https://doi.org/10.5281/zenodo.5918500}} associated with this work.
Additionally, the false detection of C$_2$H$_2$ in HD~209458b that occurs in the NIRSpec results does not occur here, or for NIRCam F322W2 alone. This can be seen  in Fig.~\ref{fig:single_ch4-6}, where VMR(CH$_4$)~=~10$^{-6}$.

\subsection{Clouds}
\begin{figure*}[h!]
    \centering
    \includegraphics[width=\textwidth]{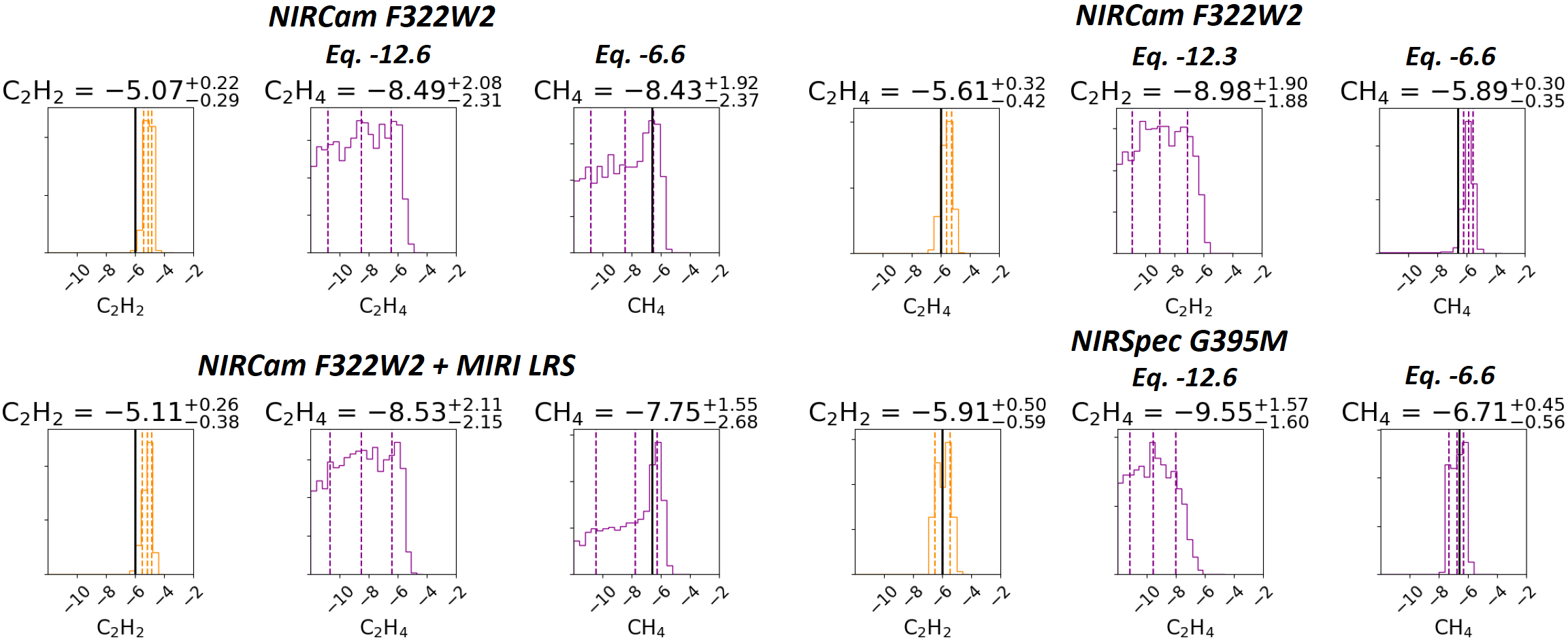}
    \caption{Retrieval results using NIRCam F322W2 and its combination, and NIRSpec G395M for HD~189733b when including clouds and a scattering slope, for increased C$_2$H$_2$ and C$_2$H$_4$ abundance to $10^{-6}$ (orange), respectively. The remaining species are in equilibrium abundance (purple) in each set of columns. The model input abundance is indicated by a black line, which is not shown when lower than the presented range.}
    \label{fig:post_clouds}
\end{figure*}

The presence of clouds and hazes can cause flattening of the spectrum, as can be seen for HD~189733b in Fig.~\ref{fig:trans_base}. They also cause a scattering slope, which largely affects spectra at shorter wavelengths than we consider in this study. These muted features may complicate the detection of molecular species in atmospheres. As such, we have included a case of HD~189733b with a cloud layer at $P_{top}$~=~0.01~bar, and a scattering slope with $f_{mix}=1\times$~10$^2$~$\times$~H$_2$ scattering. The corresponding Bayes factors are included in Table~\ref{t:stats_HD189_clouds}. Overall, the inclusion of clouds in this configuration does not seem to significantly reduce the evidence of the hydrocarbon detections; we mainly see a reduction in the results for MIRI LRS and NIRCam F322W2. In addition to the deterioration in evidence, both NIRCam F322W2 and its combination with MIRI LRS are now unable to confidently find CH$_4$ in equilibrium abundance when VMR(C$_2$H$_2$)~=~10$^{-6}$ (see Fig.~\ref{fig:post_clouds}), while this was previously possible (Figs.~\ref{fig:single_c2h2-6} and \ref{fig:combined}). On the other hand, this problem does not occur for NIRCam F322W2 when VMR(C$_2$H$_4$)~=~10$^{-6}$ (also shown in Fig.~\ref{fig:post_clouds}). The NIRSpec G395M observations are not affected by this, as shown in the bottom right set of the same plots. The full posterior plots, illustrating the clear versus cloudy case for HD~189733b for various instruments and combinations of instruments, can be found in the supplementary information document associated with this work, along with selected retrieved spectra for the cloudy case.

\subsubsection{HD 189733b \& HD 209458b: Summary}
The most accurate results were found using the combinations of either NIRCam F322W2 and MIRI LRS or NIRSpec G395M and MIRI LRS. Although the results from NIRCam F322W2 and NIRSpec G395M were similar, it seems particularly beneficial to use NIRSpec G395M over NIRCam F322W2 in the case of a cloudy atmosphere for HD~189733b.
In addition, we note that improvements in the retrieved abundances of other non-hydrocarbon species (although not detailed here) are also generally found by extending the wavelength region observed. We  include some posterior distributions including the full range of molecules in the supplementary information to this work. The input model abundances are given as red vertical lines in all plots. In general we note a few observations. Firstly, the combination of NIRISS SOSS and NIRSpec G395M, although not found to be optimal for detecting the hydrocarbon molecules we are interested in, gives reasonable and well-constrained abundances for TiO, VO, CO$_2$, and CO (for both HD~189733b and HD~209458b). In general,  H$_2$O is  retrieved reasonably for all instrument combinations, although in some cases it is found to be slightly more abundant than in the input models. It can also be observed for both planets that the abundances of CO and CO$_2$ are retrieved confidently around the input values for NIRSpec G395M and MIRI LRS, but not for NIRCam F322W2 and MIRI LRS. We note, however,  that in the case of HD~209458b that C$_2$H$_2$ is often retrieved in the cases where it should not be for NIRSpec G395M and MIRI LRS, but this issue does not occur for NIRCam F322W2 and MIRI LRS. Even so, we conclude that generally the best choice for observing HD~189733b and HD~209458b is probably NIRSpec G395M and MIRI LRS (combined), particularly if clouds are expected, based on the results of HD~189733b, with NIRCam F322W2 and MIRI LRS (combined) a close second.

\subsection{HD 97658b}

The posteriors of all hydrocarbons retrieved for various abundances (of 1~$\times$~10$^{-4}$ or 1~$\times$~10$^{-6}$ for CH$_4$, C$_2$H$_2$, and C$_2$H$_4$) are collected in Figs.~\ref{fig:single_ch4-4}--\ref{fig:single_c2h4-6} for the single instrument set-ups of HD 189733b, HD 209458b, and HD 97658b (with HD~97658b in the right-hand columns). The Bayes factors for HD~97658b can be found in Table~\ref{t:stats_HD976}.

\subsubsection{HD 97658b: MIRI LRS}
The smaller transit depth and larger relative error (see Fig.~\ref{fig:trans_base}) in the observed spectra of HD~97658b makes it more difficulte to  differentiate between hydrocarbon signatures.
As such, no certain detections occur for VMRs of 10$^{-6}$, which results in Bayes factors of $<$1 for all cases in Table~\ref{t:stats_HD976}. When in lower abundance, the features can be attributed to different hydrocarbons, resulting in a good fit whether the species of interest is included or not. The equilibrium abundance of CH$_4$ is significant enough that it should be able to be detected (see Table~\ref{t:planet_pars}), but in Fig.~\ref{fig:single_c2h2-4} (VMR(C$_2$H$_2$)~=~10$^{-4}$) and Fig.~\ref{fig:single_c2h4-4} (VMR(C$_2$H$_4$)~=~10$^{-4}$) it can be seen that it is not confidently retrieved when the other hydrocarbons are of significant abundances. The posteriors are more constrained for CH$_4$ in Figs.~\ref{fig:single_c2h2-6} and \ref{fig:single_c2h4-6}, where VMR(C$_2$H$_2$)~=~10$^{-6}$ and VMR(C$_2$H$_4$)~=~10$^{-6}$, respectively.

\subsubsection{HD 97658b: NIRISS SOSS}
Similar to the results for MIRI LRS, NIRISS SOSS is unable to retrieve when the abundance is 10$^{-6}$ and lower. However, the high equilibrium abundance of CH$_4$ is found here in combination with the other hydrocarbons even when they are significant (VMR of 10$^{-4}$, see Fig.~\ref{fig:single_c2h2-4} for C$_2$H$_4$ and Fig.~\ref{fig:single_c2h4-4} for C$_2$H$_2$).

\subsubsection{HD 97658b: NIRCam F322W2}
NIRCam F322W2 performs better than both MIRI LRS and NIRISS SOSS for HD~97658b, as was the case for HD~189733b and HD~209458b. It is able to accurately retrieve CH$_4$ down to an abundance of 10$^{-6}$, without influencing the distribution of C$_2$H$_2$ abundances (see Fig.~\ref{fig:single_ch4-6}). 
The posterior plot in Fig.~\ref{fig:single_c2h2-6} starts to become skewed for C$_2$H$_2$, and the Bayes factor indicates that there is no statistical evidence that C$_2$H$_2$ can be retrieved at this abundance using NIRCam F322W2 alone. The retrieved CH$_4$ abundance is reasonable but slightly higher than it should be when the VMR of C$_2$H$_2$ is 10$^{-6}$, possibly due to difficulties in differentiating between these hydrocarbon species. 
C$_2$H$_4$ is not retrievable at an abundance of 10$^{-6}$ (see Fig.~\ref{fig:single_c2h4-6}), probably due to the high equilibrium CH$_4$ abundance, which again in this case is retrieved reasonably well, but slightly higher than in the input models. Additionally, we note that the distributions in Fig.~\ref{fig:single_ch4-4} (where VMR(CH$_4$)~=~10$^{-4}$) look peculiar, with some likelihood of a C$_2$H$_4$ contribution being found. Overall, NIRCam F322W2 does not seem to be able to distinguish between the different hydrocarbons as well as it did for HD~189733b and HD~209458b. We therefore test cases where the number of transits is increased to 25 (see Sect.~\ref{sec:25_transits}), as well as the instrument combinations considered previously.

\subsubsection{HD 97658b: NIRSpec G395M}
The performance of NIRSpec G395M is again very similar to NIRCam F322W2, where C$_2$H$_4$ or C$_2$H$_2$ for abundances of 10$^{-6}$ are not detected, see Table~\ref{t:stats_HD976} and the corresponding posterior plots in Figs.~\ref{fig:single_ch4-4} to \ref{fig:single_c2h4-6}.

\subsubsection{HD 97658b: NIRISS SOSS and NIRSpec G395M}
We note that especially C$_2$H$_4$ in a VMR of 10$^{-6}$ in combination with the equilibrium abundance of CH$_4$ proved difficult to find using all single instruments tested. In an attempt to resolve these issues, we examine the same instrument combinations as for HD~189733b and HD~209458b, the resulting posterior distributions of which can be found in Fig.~\ref{fig:combined}. The Bayes factors for all other cases are again presented in Table~\ref{t:stats_HD976}.
Combining these two instrument modes shows great improvements in comparison to the individual instruments for most cases in HD~97658b, resulting in very confident retrievals for all cases apart from when VMR(C$_2$H$_2$)~=~10$^{-6}$. The Bayes factor corresponding to C$_2$H$_4$ with a VMR of 10$^{-6}$ is in excess of 11, indicating that it is now retrieved. Interestingly, the distribution is not a single narrow peak in this case, which indicates that the exact abundance is not as certain as appears based on the Bayes factor alone.

\subsubsection{HD 97658b: NIRSpec G395M and MIRI LRS}

We find the improvements when combining NIRSpec G395M and MIRI LRS to be minor in comparison to NIRSpec G395M alone when examining the Bayes factors of Table~\ref{t:stats_HD976}. The retrieved abundances based on Fig.~\ref{fig:combined} now seem to be well constrained.

\subsubsection{HD 97658b: NIRCam F322W2 and MIRI LRS}
There were some issues with these individual instruments; both were unable to retrieve C$_2$H$_4$ or C$_2$H$_2$ at a VMR of 10$^{-6}$ (see the Bayes factors in Table~\ref{t:stats_HD976}), and MIRI LRS could also not retrieve CH$_4$  at a VMR of 10$^{-6}$. Both MIRI LRS and NIRCam F322W2 had some trouble distinguishing between different hydrocarbon species. Combining the instruments shows a great improvement for all these issues, and is found to be far more accurate than using the single instruments alone (see, for example, the posterior distributions in  Fig.~\ref{fig:combined}). The factors in Table~\ref{t:stats_HD976} indicate this to be the best instrument combination to use for our selected small target.

\subsubsection{HD 97658b: Multiple transits}\label{sec:25_transits}

\citet{ref:15BaKaLu} find that observing multiple transits  shows advantages for smaller targets like super-Earths and other small objects. As such, this set-up is tested for HD~97658b. Based on the Bayes factors in Table~\ref{t:stats_HD976} and the results of selected species in Fig.~\ref{fig:transits}, we find in general that there are improvements in some cases of retrievals when the number of transits is increased to 25 in comparison to a single transit, as demonstrated by the increase in $B_{01}$ in Table~\ref{t:stats_HD976} for smaller abundances. However, the increase in observational time likely does not justify these small improvements, and we found similar or even better improvements when different instruments were combined to give a larger wavelength coverage. While the multiple transits only improve the detection limit of C$_2$H$_2$ to $10^{-6}$, a significant increase in $B_{01}$ is found for C$_2$H$_4$ abundances of $10^{-6}$ as well using the instrument combination
NIRCam F322W2 and MIRI LRS.

\subsubsection{HD 97658b: Summary}
The most accurate instrument combination for HD~97658b was found to be NIRCam F322W2 and MIRI LRS (combined). The improvements compared to the individual instruments alone was found to be much greater in the case of this planet than with the more favourable targets of HD~189733b and HD~209458b. Similar observations  for HD~189733b and HD~209458b can be made for HD~97658b regarding the posterior distributions of the other (non-hydrocarbons) molecules included in the input models, which can be found in the supplementary information document of this work. H$_2$O is retrieved reasonably well for all of the following instrument combinations. CO and CO$_2$ are are retrieved reasonably well for NIRSpec G395M and MIRI LRS (although the abundances are slightly too high in some cases), but not for NIRCam F322W2 and MIRI LRS. NIRISS SOSS and NIRSpec G395M generally gives confident and reasonably accurate retrievals of the abundances of CO, CO$_2$, VO, and TiO in comparison to their input values, which are given as red lines in all figures. While observing multiple transits does improve the detection limits and certainty, a significantly longer observational time may be required. We therefore find that a single transit using the combination of NIRCam F322W2 and MIRI LRS is preferable to using 25 transits for any individual instrument for HD~97658b.

\subsection{Kepler-30c}

\subsubsection{Kepler-30c: Standard instrument set-ups}
\begin{figure*}[h!]
\centering
\begin{minipage}{\textwidth}
    \centering
    \includegraphics[width=\textwidth]{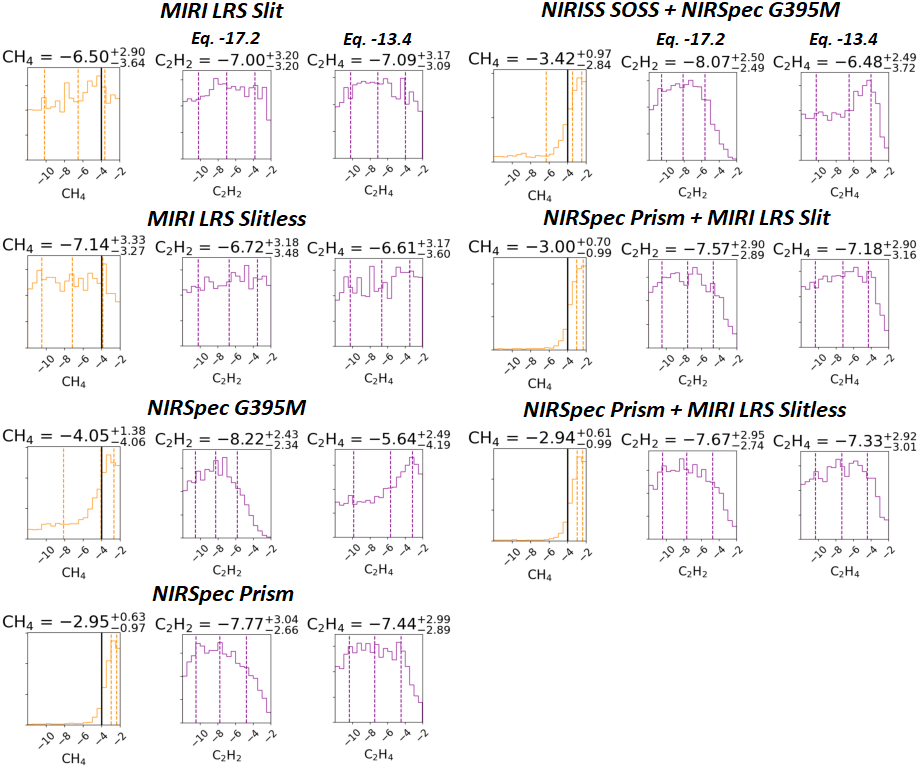}
    \caption{Posterior distributions from retrievals using a variety of instruments for Kepler-30c, for increased CH$_4$ abundance to $10^{-4}$ (orange) and equilibrium abundance for the remaining species (purple) in each set of columns. The model input abundance is indicated by a black line, which is not shown when lower than the presented range.}
    \label{fig:kepler_ch4}
\end{minipage}
\end{figure*}

\begin{figure*}
\begin{minipage}{\textwidth}
    \centering
    \includegraphics[width=\textwidth]{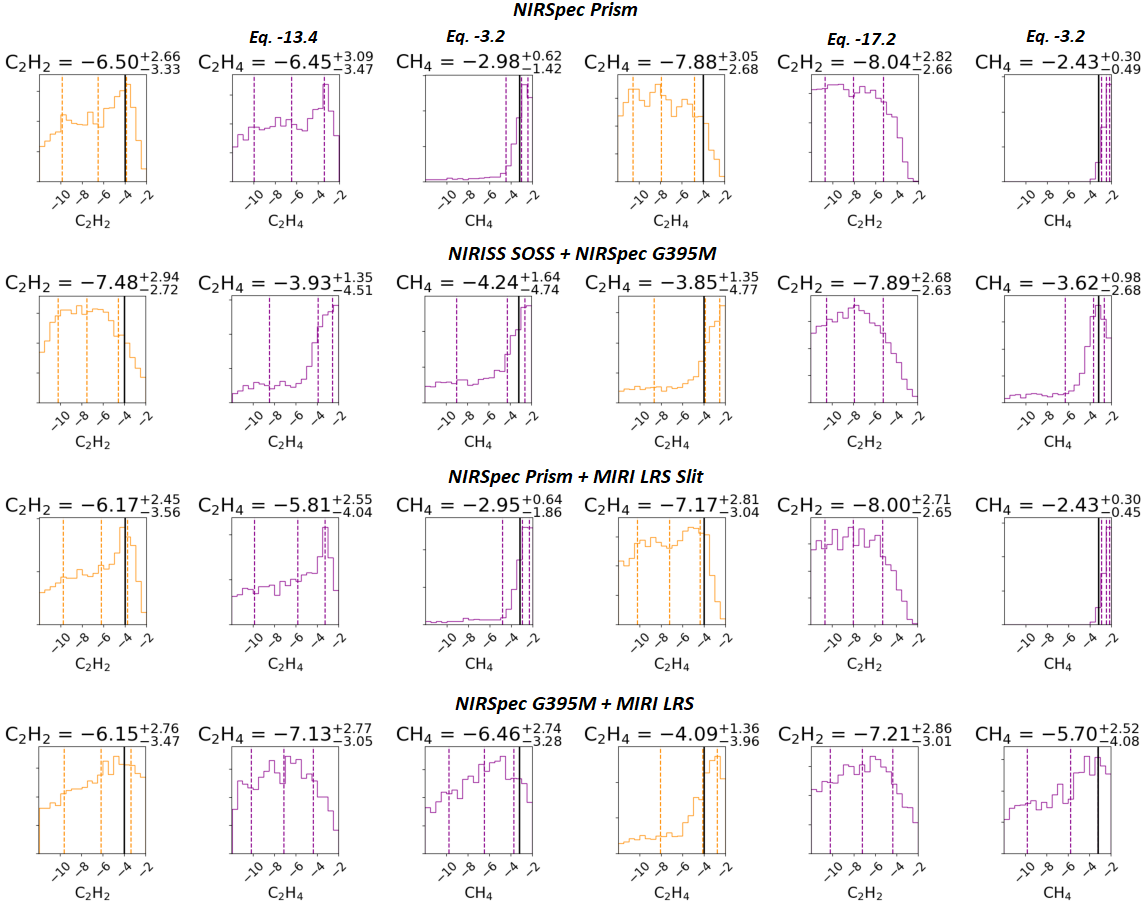}
    \caption{Posterior distributions from retrievals using  a variety of instruments and instrument combinations for Kepler-30c, for increased hydrocarbon abundances to $10^{-4}$ (orange) and equilibrium abundance for the remaining species (purple) in each set of columns. The model input abundance is indicated by a black line, which is not shown when lower than the presented range.}
    \label{fig:kepler_comb}
\end{minipage}
\end{figure*}

The Bayes factors and posterior plots for our selected faint target can be found in Table~\ref{t:stats_Kepler30c} and Figs.~\ref{fig:kepler_ch4} and \ref{fig:kepler_comb}. The standard instrument set-ups, which refer to MIRI LRS, NIRISS SOSS, NIRCam F322W2, and NIRSpec G395M, perform quite poorly overall for Kepler-30c. MIRI LRS, NIRISS SOSS, NIRSpec G395M, and NIRCam F322W2 are all unable to find any hydrocarbons, based on the Bayes factors of Table~\ref{t:stats_Kepler30c}. Of these standard single instrument set-ups, NIRSpec G395M looks the most reasonable based on the posterior distribution for the case where VMR(CH$_4$)~=~10$^{-4}$, as shown in Fig.~\ref{fig:kepler_ch4}. Both the slitless mode, which is  the most commonly used  in this study,  and the slit mode were tested for MIRI LRS. This was done since using the slit could have advantages for faint objects due to a reduction in background light, but our results confirm the findings of \citet{ref:17HoBuDe}, since no improvements were found when using the slit mode as a single instrument.


\subsubsection{Kepler-30c: NIRSpec Prism}
\citet{ref:17BaLi} suggest using NIRSpec Prism for detecting faint objects. The mode is easily saturated, but may be better for observing objects like Kepler-30c. NIRSpec Prism covers a relatively wide range of wavelengths,  approximately from 0.6 to 5.3 $\mu$m; the resolving power is comparable to MIRI LRS ($\sim$100 nominally). We found that NIRSpec Prism shows weak detection (>1 in Table~\ref{t:stats_Kepler30c}) of CH$_4$ down to the VMR~=~10$^{-4}$ region, but is unable to find hydrocarbons at VMR~=~10$^{-6}$. Furthermore, the equilibrium abundance of CH$_4$ is VMR~=~10$^{-3.16}$, which possibly inhibits detection of the other hydrocarbons. In fact, CH$_4$ is consistently retrieved at an abundance higher than the input abundance, most likely due to C$_2$H$_2$ or C$_2$H$_4$ being present and increasing the strength of the features (see the narrowly constrained distributions in Fig.~\ref{fig:kepler_comb}). Even for the case where VMR(CH$_4$)~=~10$^{-4}$, the abundance retrieved is too high, as shown in Fig.~\ref{fig:kepler_ch4}.
In the presence of abundant CH$_4$ it becomes incredibly difficult to find less abundant hydrocarbons in faint objects like Kepler-30c.

\subsubsection{Kepler-30c: NIRSpec Prism and MIRI LRS}

The combination NIRSpec Prism and MIRI LRS covers the widest wavelength range achievable with the instrument settings currently included in PandExo: 0.6-12 microns. Again, here we test both the slit and slitless modes of MIRI LRS. The performance including the slitless mode is approximately the same as using NIRSpec Prism by itself (see Table~\ref{t:stats_Kepler30c}), and thus there are no advantages that would warrant using this instrument combination for faint objects like Kepler-30c. However, minor improvements are found when combining NIRSpec Prism with MIRI LRS slit (see Fig.~\ref{fig:kepler_comb} and Table~\ref{t:stats_Kepler30c}).

\subsubsection{Kepler-30c: Summary}
In summary, Kepler-30c is not an easy target to accurately detect signatures of the hydrocarbons we are looking for using the instruments tested here. We note that the stellar brightness of Kepler-30 is below that recommended for all instruments tested, even NIRSpec Prism which is designed for fainter targets. In addition, due to the low equilibrium temperature of Kepler-30c, the equilibrium chemistry simulations yield quite a high CH$_4$ abundance, which can make distinguishing between different hydrocarbons more challenging. In the presence of abundant CH$_4$ it becomes incredibly difficult to find less abundant hydrocarbons in faint objects like Kepler-30c. We find limited benefits of combining NIRSpec Prism with MIRI LRS observations over observing with NIRSpec Prism alone. We find worse results when using NIRSpec G395M in comparison to observing with NIRSpec Prism, but find that there is some slight improvement when combining NIRISS SOSS and NIRSpec G395M in comparison to NIRSpec G395M alone. Based on the posterior plots alone (see Fig.~\ref{fig:kepler_ch4} and Fig.~\ref{fig:kepler_comb}), NIRISS SOSS and NIRSpec G395M appear to be better for retrieving C$_2$H$_4$ at a VMR of 10$^{-4}$. This is not mirrored by the Bayes factors in Table~\ref{t:stats_Kepler30c}, although we note that the difference between the two is not significant. The posteriors for NIRISS SOSS and NIRSpec G395M (combined) and NIRSpec Prism are similar for retrieving CH$_4$ at a VMR of 10$^{-4}$. NIRSpec Prism is probably the best choice in terms of performance and to avoid using more than one instrument unnecessarily or to observe multiple transits, although there is a very minor improvement when MIRI LRS slit is combined with NIRSpec Prism. Our findings support the guidance that the JWST instruments are not suited to targets with a J-band flux fainter than 9.82~\citep{16NiFeGi}, but they hint that  a slightly brighter but still faint target could benefit from the use of NIRSpec Prism and MIRI LRS slit (combined), or potentially NIRISS SOSS and NIRSpec G395M (combined).

\subsection{Distinguishing absorption features}\label{sec:ab_feat}
\begin{figure}[h!]
    \centering
    \includegraphics[width=\columnwidth]{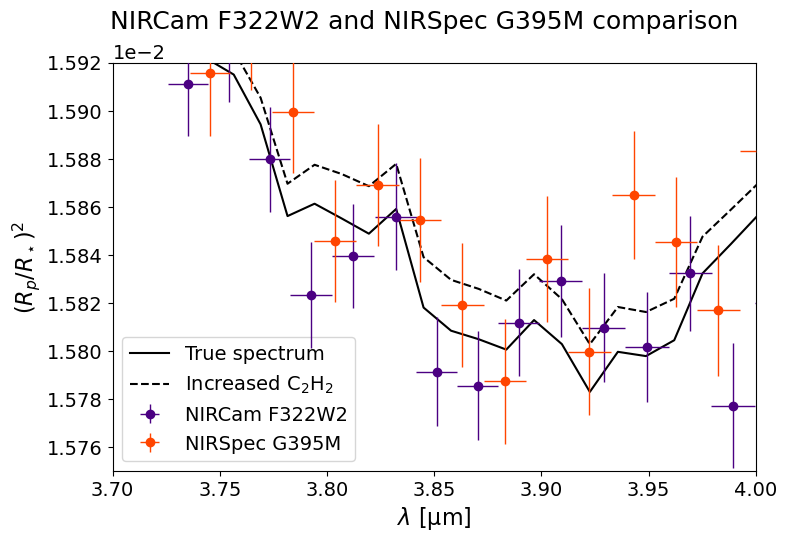}
    \caption{Comparison of the observations for NIRCam F322W2 and NIRSpec G395M, for HD 209458b with CH$_4$ increased to 10$^{-6}$ (true spectrum), and C$_2$H$_2$ increased to 3$\cdot$10$^{-7}$ (as found from NIRSpec G395M observations).}
    \label{fig:noise_comp}
\end{figure}
As discussed previously, it can sometimes be difficult to distinguish between different hydrocarbons, since some signatures are in similar wavelength regions. It was found that a large abundance of one hydrocarbon, here most often methane, results in other hydrocarbons not being detected or in large uncertainties in detection. This could be caused by random noise pushing the best fit to the wrong model spectrum or being somewhere in between. 
For example, NIRCam F322W2 and NIRSpec G396M observe partially overlapping parts of the spectrum, yet NIRSpec G395M incorrectly finds C$_2$H$_2$ in HD 209458b when CH$_4$ is increased to a VMR of 10$^{-6}$ (see Fig.~\ref{fig:single_ch4-6}), while NIRCam F322W2 does not. When examining the observations around the range of a prominent C$_2$H$_2$ feature (see Fig.~\ref{fig:all}) in Fig.~\ref{fig:noise_comp}, we see that the simulated spectra using NIRSpec G395M shows a deeper transit depth (i.e. more absorption) than the same region from NIRCam F322W2,
therefore causing the retrievals using the NIRSpec G395M simulated spectra to find C$_2$H$_2$ where it should be insignificant.
However, some wavelength ranges seem better suited to distinguishing between different hydrocarbons than others, due to the difference in nature of signatures between wavelength ranges. Both NIRSpec G395M and NIRCam F322W2 are best at differentiating between the hydrocarbons since they operate in a similar wavelength range where wider and more unique signatures can be found. MIRI LRS is not able to distinguish between the hydrocarbons as well as NIRSpec G395M and NIRCam F322W2, but performs slightly better than NIRISS SOSS, which performed worst. This is most likely due to the widest and most unique signatures being present in the mid-IR range. A large part of this is covered by MIRI LRS, but the majority of the major C$_2$H$_2$ signatures are at the end of the wavelength range where the error of MIRI LRS increases significantly. The larger error causes smaller changes, for example in the 10$^{-6}$ range, to be harder to detect. Furthermore, the noise floor of MIRI is the highest of all instruments, preventing it from reaching the same precision as NIRSpec G395M and NIRCam F322W2. We therefore find optimum results when different instruments are combined. In particular: 
NIRSpec G395M--MIRI LRS and
NIRCam F322W2--MIRI LRS. We note that the presence of hydrocarbon hazes (see Sect.~\ref{sec:hazes}) may mute absorption features in the shorter wavelength regions in particular, so instrument combinations that include MIRI LRS may be preferable.

\section{Discussion}
\label{sec:discussion}

\subsection{Feasibility of detectable abundances}

\subsubsection{Different chemistry models}

The importance of including hydrocarbons in chemical reaction schemes is discussed by \citet{ref:15VeHeAg}, who show that these extra reactions are especially relevant for cooler atmospheres with T~<~1000~K and for hotter atmospheres when C/O~>~1. A similar conclusion is reached by other works such as \citet{ref:12Ma}, who perform a study on the effects of different C/O ratios on the atmospheric composition of hot Jupiters when assuming equilibrium chemistry. The assumption of chemical equilibrium is generally thought to hold for hot planets since this class of planets is less affected by disequilibrium effects than those with $T_{\rm eq}$~<~1200 K. \cite{ref:12Ma} find that for C/O~>~1 a change in properties occurs for $T_{\rm eq}$~>~1200 K, where the atmosphere may contain abundant C$_2$H$_2$ and HCN in addition to CH$_4$, although CO remains the dominant carbon bearing species. 
\citet{ref:15MoBoDu} use a similar assumption in their model, where planets with $T_{\rm eff}\geq$~1500~K are assumed to be unaffected by disequilibrium processes, based on the findings of \citet{ref:14MiKa}.
Theoretical models such as these thus predict that it is difficult to find hydrocarbons at detectable abundances when considering equilibrium chemistry alone, unless C/O is very high.
In this case a combination of very high temperatures and high C/O (at least higher than unity) can result in highly abundant hydrocarbons, where heavier hydrocarbons can surpass CH$_4$ in abundance.
This is mainly applicable to C$_2$H$_2$, which is not destroyed by high temperatures like C$_2$H$_4$.

On the other hand, planets with an effective temperature of around 1500~K or less are expected to have atmospheres strongly affected by disequilibrium processes such as vertical mixing~\citep{ref:14MiKa,17TsLyGr}. Vertical mixing essentially allows us to probe a deeper layer of the atmosphere than we would otherwise be able to see, and tends to enhance the abundances of species such as CH$_4$, C$_2$H$_2$, and C$_2$H$_4$ for all C/O ratios, but particularly for C/O$\geq$1~\citep{12MoMaVi.exo,ref:13BiRiHe,17TsLyGr}. Methane and acetylene have been found to be amongst the most instructive species when it comes to diagnosing vertical mixing effects in an exoplanet's atmosphere~\citep{17TsLyGr}. Accurate abundance measurements of these species thus allow for evidence to be inferred not only of C/O, but also of atmospheric vertical mixing strength (quantified by the eddy diffusion coefficient, $K_{\rm zz}$). Photodissociation is also thought to play a key role in producing higher abundances of hydrocarbons~\citep{ref:15VeHeAg}. Detection of hydrocarbon species other than methane will therefore provide important information on both disequilibrium chemistry processes and C/O ratio.

\subsection{Effect of hydrocarbon hazes on detectability}\label{sec:hazes}

It is expected that haze formation will become efficient at temperatures below around 1000~K~\citep{ref:19KaIk}, and so we would expect it to particularly affect the atmospheres, and therefore transmission spectra, of Kepler-30c and HD~97658b out of the planets considered in this study. 
There are many uncertainties related to haze formation, and the models discussed previously rely on many assumptions regarding haze precursor types and particle size and shape, but also the fraction of haze precursors that in the end form a haze and mixing effects. Since these factors can affect the haze thickness, the resulting spectra can differ greatly depending on the assumptions. As such, it is essential to gain more information about the processes involved. 
The detection of the haze and molecules present in deeper layers of the atmosphere can be of use here. However, the haze can block large parts of the spectrum. Overall, lower C/O ratios and irradiation intensities combined with higher equilibrium temperatures result in clearer atmospheres (see e.g. Figs. 9 and 12 in \citealt{ref:19KaIk}), which are favourable when identifying the atmosphere's composition. However, these conditions are exactly the  opposite of what is required for complex hydrocarbon formation. As such, it is expected that atmospheres with abundant complex hydrocarbons could be greatly affected by hazes, thus making their identification more difficult. NIRISS SOSS will be most affected by hazes because its wavelength range is in the shorter wavelengths close to where a spectral slope occurs. Longer wavelengths, specifically in the MIRI region, will generally be less affected by haze effects, making it an interesting option for detecting molecular abundances. However, very thick hazes can result in flattening over the entire spectrum where wider wavelength ranges are better suited (see e.g. \citealt{ref:16GrLiMo}), likely making instrument combinations a better option. 

\section{Conclusion}\label{sec:conclusion}

In this work we have investigated the detectability of some major hydrocarbon species (C$_2$H$_2$, CH$_4$, and C$_2$H$_4$) in 
various types of exoplanets: two favourable hot Jupiter targets modelled on HD~189733b and HD~209458b; a smaller super-Earth/sub-Neptune modelled on HD~97658b; and a warm Jupiter orbiting a faint star modelled on Kepler-30c. We simulated the expected observed transit spectra of these planets using various instruments on board the JWST, using the PandExo package~\citep{ref:17BaMaKl}. We used the retrieval package ARCiS~\citep{ref:20MiOrCh,ref:19OrMi} with opacities from the ExoMolOP database~\citep{20ChRoAl.exo} to run a retrieval and compared the results to the model input spectra. In this way we were able to make some predictions of how suitable JWST will be for detecting the hydrocarbon species C$_2$H$_2$~\citep{ref:20ChTeYu}, CH$_4$~\citep{jt698}, and C$_2$H$_4$~\citep{ref:18MaYaTe}, and which instruments we consider best to observe their spectral signatures.

In general, we find that the combination of NIRSpec G395M--MIRI LRS and NIRCam F322W2--MIRI LRS are best for detecting hydrocarbons in the most favourable hot Jupiter targets (those modelled on HD~189733b and HD~209458b). NIRSpec G395M--MIRI LRS is better than NIRCam F322W2--MIRI LRS in the HD~189733b case with clouds included, whereas NIRCam F322W2--MIRI LRS is the best combination for our selected target of the smaller super-Earth/sub-Neptune planet type (modelled on HD 97658b). The improvements using this combination in comparison to NIRCam alone are much greater in the latter smaller target. For the warm-Jupiter-type planet orbiting a faint star (modelled on Kepler-30c) the best results for an individual instrument are found using NIRSpec Prism, with potential improvements when including MIRI LRS slit (in combination), or possibly with the combination of NIRISS SOSS and NIRSpec G395M. Our conclusions for faint planets, however, remain inconclusive and could benefit from further investigation on a still faint but slightly brighter target.

Various factors can affect the ability to accurately retrieve molecular abundances, such as target selection. Faint small targets with dense and less extensive atmospheres are in general more difficult to characterise; in this study we base models on Kepler-30c,  which  is a warm Jupiter but also has a faint host star, and on the  smaller super-Earth target  HD~97658b. Our models based on two other targets, HD~189733b and HD~209458b, are brighter and therefore easier in general to characterise.
Combining instruments increases the observation time, but can greatly improve the certainty of detections, as described above. 
Combinations that cover a larger wavelength region can be favourable for hazy atmospheres where absorption features at shorter wavelengths may become muted.

The detectability limits for our most favourable targets, the hot Jupiters orbiting bright stars such as HD 189733b and HD 209458b, are between 10$^{-6}$ and 10$^{-7}$ for NIRCam F322W2 and NIRspec G395M, while the limit for MIRI LRS is closer to 10$^{-6}$.
We do not trust any retrievals that suggest the ability to detect anything past the 10$^{-7}$ mark in these compositions, since there is no visual change in the modelled spectra after this abundance. This suggests that abundances near a visual change in the spectrum can be detected, and generally those at least a factor of 10 higher, at least for the brighter targets. We note, however, that our conclusions on detectability will also be affected by the base compositions of our atmospheres. A lower abundance of H$_2$O in our base atmospheres, for example, could mean that we may be able to detect the hydrocarbons at lower abundances.

Smaller targets, when still orbiting relatively bright stars (in this study HD 97658b), have similar limits to those of the highly favourable targets mentioned previously, although the error on the observation is increased, and the certainty of detection decreased. Furthermore,  there are more problems with hydrocarbons being mistaken for one another. Generally, not all hydrocarbons may be detected when considering single-instrument single-transit observations. These smaller objects suffering from more noise can benefit from being observed over multiple transits since this reduces the noise and causes the measurements to be closer to the model input spectrum. Figure~\ref{fig:transits} and Table~\ref{t:stats_HD976} demonstrate that observing multiple transits (25) yields improvements for all single instruments 
with HD~97658b.
However, as  improvements are also found by combining different instruments together, it would be difficult to justify increasing the number of transits, and therefore observation time, using a single instrument over combining observations from two instruments.

Objects orbiting faint stars, even when they are relatively large warm Jupiters like Kepler-30c, are much more difficult to characterise. NIRSpec Prism is the best  single instrument mode. It can  weakly detect CH$_4$ down to VMR~=~10$^{-4}$; however, it struggles to detect down to the VMR~=~10$^{-4}$ region for C$_2$H$_2$ and C$_2$H$_4$, which is not helped by the large equilibrium abundance of CH$_4$ in a planet of this temperature. It is unable to detect down to VMR~=~10$^{-6}$ for any of the three species. This situation is marginally improved when NIRSpec Prism is combined with MIRI LRS in slit mode, and similar results are found with NIRISS SOSS and NIRSpec G395M (combined). All other single instruments tested are not able to significantly retrieve any of the three species we are looking for.

Whether or not the correct hydrocarbon is retrieved depends on the distribution of the noise around the model input spectrum. This noise, due to its random nature, may sometimes push the best fit to the wrong model spectrum, especially when the difference between the two spectra is minimal. This is described in Sect.~\ref{sec:ab_feat}.
NIRISS SOSS is affected by this the most, due to the limited hydrocarbon signatures in its wavelength range meaning that the difference between a spectra with and without the species included is small. NIRSpec G395M and NIRCam F322W2 perform best (see Fig.~\ref{fig:noise_comp}). The performance of MIRI LRS falls somewhere in between.

\subsection{Future work}

The main aim of this study was to give an idea of what hydrocarbon species could be detectable under different circumstances, and what instruments of JWST could be used to best characterise their abundances. The analysis was largely performed by simulating hypothetical clear atmospheres based on equilibrium chemistry calculations, and artificially adding hydrocarbons, with a moderately cloudy atmosphere tested in the case of HD~189733b. As such, they may not be representative of realistic atmospheres. 
We demonstrated that, in some cases, higher abundances of CH$_4$ in the input spectra in combination with C$_2$H$_2$ or C$_2$H$_4$ could lead to one or more spectral features being mistaken for those of the other species. Increased abundances of other non-hydrocarbon species could also cause similar issues (see e.g. \citealt{ref:19SoPeSe}).  
\citet{ref:20ZiMiMo} found that when looking at more nitrogen-dominated rather than hydrogen-dominated atmospheres, in particular when investigating cooler super-Earths and sub-Neptunes, HCN can become a major carbon-bearing species and can prevent larger hydrocarbons from becoming abundant. Increased abundances of HCN in particular could inhibit the accurate abundance detection of C$_2$H$_2$, as many of their spectral features, particularly at the resolutions we are investigating, are in very similar wavelength regions~\citep{ref:18TeYu}. A future expansion of the current work could be to investigate the inclusion of such nitrogen-bearing species. A synergy with high-resolution studies could be a good way to proceed  in order to distinguish between these two species in particular~\citep{17BrLiBe.exo,20GaBrYu.exo,21GiBrGa}. Although we do not consider MIRI MRS here,  it could be an interesting instrument to test in addition to those considered in this study~\citep{21PaNaCu}.

We would expect disequilibrium chemistry processes to potentially have some effect at the temperatures of the planets focused on in this study, which all have  equilibrium temperature under 1500~K.
Investigating all possible combinations is beyond the scope of the current work, however it could be useful to expand on this study in the future to include other species and different C/O ratios, and to explicitly include disequilibrium effects. ARCiS has recently been updated to explicitly handle disequilibrium effects~\citep{21KaMi.arcis}, which would facilitate such a follow-up study. 

We assume clear atmospheres when computing models for all planets, apart from HD~189733b, although in reality we do expect clouds and/or hazes to be present in the atmospheres of potentially all of the planets on which our models were based  (see e.g. \citealt{16LeDoHe,16HeLeDo,18HaHeDu,18LiMaBo,20Barstow}). HD~189733b in particular is assumed to be relatively cloudy due to the observed Rayleigh scattering slope~\citep{16SiFoNi.exo}. We therefore ran a series of retrievals for HD 189733b for a case where we added a simple cloud deck at 0.01~bar uniformly across the atmosphere, with a scattering slope of $f_{\rm mix}$=1$\times$10$^{2}$~$\times$~H$_2$ scattering.
It would be beneficial in the future to expand on the current study to investigate more comprehensively the effects of cloudy and hazy atmospheres on detecting the hydrocarbon species of interest to the present study in greater detail, including planets other than HD~189733b.

Furthermore, a handful of transiting gas giant planets have been identified as being potentially carbon-rich (C/O$\geq$~1),  the most promising being XO-1b, WASP-12b, CoRoT-2b~\citep{12MoMaVi.exo,ref:12Ma}, and HD~209458b \citep{21GiBrGa}. Of these, we only include a planet modelled on HD~209458b in our study. The high equilibrium temperature of WASP-12b~\citep{11MaErRa.wasp12b} could make the detection of the species we are interested in challenging. CoRoT-12b has been observed to have a featureless emission spectrum, indicating the presence of thick clouds, which could potentially mute some the features of the hydrocarbon species we consider here, particularly at lower wavelengths~\citep{18DaCoSc.corot2b}. Nevertheless, it would be of interest to expand upon the current set of planets on which we base our models   to include those planets assumed to be carbon-rich. 

\section{Supplementary material}

The supplementary information document associated with this work is available from \url{https://doi.org/10.5281/zenodo.5918500}.

\section{Acknowledgements}

This project has received funding from the European Union's Horizon 2020 Research and Innovation Programme, under Grant Agreement 776403, ExoplANETS-A. We would like to thank various people including Yui Kawashima, Christiane Helling, and Fabienne Schiefenender for some very useful discussions in relation to this work. We also thank the developers of the PandExo package for a very useful community tool. We are grateful to Frank Uittenbosch and the IT team at SRON for our use of the computing grid system where we undertook the calculations necessary for this study. Finally, we thank the anonymous reviewer for their useful feedback, which has helped us greatly improve the manuscript.

\bibliographystyle{aa}
\bibliography{hydrocarbons_bib_file}

\begin{thebibliography}{125}
\expandafter\ifx\csname natexlab\endcsname\relax\def\natexlab#1{#1}\fi

\bibitem[{Baines {et~al.}(2008)Baines, McAlister, ten Brummelaar, Turner,
  Sturmann, Sturmann, Goldfinger, \& Ridgway}]{08BaMcBr.exo}
Baines, E.~K., McAlister, H.~A., ten Brummelaar, T.~A., {et~al.} 2008, ApJ,
  680, 728

\bibitem[{Barman(2007)}]{ref:07Ba}
Barman, T.~S. 2007, ApJ, 661

\bibitem[{Barstow(2020)}]{20Barstow}
Barstow, J.~K. 2020, MNRAS, 497, 4183

\bibitem[{Barstow {et~al.}(2015)Barstow, Aigrain, Irwin, Kendrew, \&
  Fletcher}]{ref:15BaAiIr}
Barstow, J.~K., Aigrain, S., Irwin, P. G.~J., Kendrew, S., \& Fletcher, L.~N.
  2015, MNRAS, 448, 2546–2561

\bibitem[{Barstow {et~al.}(2016)Barstow, Aigrain, Irwin, Kendrew, \&
  Fletcher}]{ref:16BaAiIr}
Barstow, J.~K., Aigrain, S., Irwin, P. G.~J., Kendrew, S., \& Fletcher, L.~N.
  2016, MNRAS, 458, 2657–2666

\bibitem[{Batalha {et~al.}(2015)Batalha, Kalirai, Lunine, Clampin, \&
  Lindler}]{ref:15BaKaLu}
Batalha, N., Kalirai, J., Lunine, J., Clampin, M., \& Lindler, D. 2015,
  arXiv:1507.02655

\bibitem[{Batalha \& Line(2017)}]{ref:17BaLi}
Batalha, N. \& Line, M. 2017, AJ, 153, 151

\bibitem[{Batalha {et~al.}(2017)Batalha, Mandell, Pontoppidan, Stevenson,
  Lewis, Kalirai, Greene, Albert, Nielsen, \& Earl}]{ref:17BaMaKl}
Batalha, N., Mandell, A., Pontoppidan, K., {et~al.} 2017, PASP, 129, 064501

\bibitem[{Benneke \& Seager(2013)}]{13BeSe}
Benneke, B. \& Seager, S. 2013, ApJ, 778, 153

\bibitem[{Bernath(2020)}]{MOLLIST}
Bernath, P.~F. 2020, JQSRT, 240, 106687

\bibitem[{Bilger {et~al.}(2013)Bilger, Rimmer, \& Helling}]{ref:13BiRiHe}
Bilger, C., Rimmer, P., \& Helling, C. 2013, MNRAS, 435, 1888

\bibitem[{Birkby {et~al.}(2013)Birkby, de~Kok, Brogi, de~Mooij, Schwarz,
  Albrecht, \& Snellen}]{13BiKoBr}
Birkby, J., de~Kok, R., Brogi, M., {et~al.} 2013, MNRAS Letters, 436, L35

\bibitem[{Bouchy {et~al.}(2005)Bouchy, Udry, Mayor, Moutou, Pont, Iribarne,
  Da~Silva, Ilovaisky, Queloz, Santos, S\'egransan, \& Zucker}]{05BoUdMa}
Bouchy, F., Udry, S., Mayor, M., {et~al.} 2005, A\&A, 444, L15

\bibitem[{Boyajian {et~al.}(2014)Boyajian, von Braun, Feiden, Huber, Basu,
  Demarque, Fischer, Schaefer, Mann, White, Maestro, Brewer, Lamell, Spada,
  López-Morales, Ireland, Farrington, van Belle, Kane, Jones, ten Brummelaar,
  Ciardi, McAlister, Ridgway, Goldfinger, Turner, \& Sturmann}]{14BoBrFe.exo}
Boyajian, T., von Braun, K., Feiden, G.~A., {et~al.} 2014, MNRAS, 447, 846

\bibitem[{Braam {et~al.}(2021)Braam, van~der Tak, Chubb, \& Min}]{20Br.exo}
Braam, M., van~der Tak, F. F.~S., Chubb, K.~L., \& Min, M. 2021, A\&A, 646, A17

\bibitem[{Brewer {et~al.}(2017)Brewer, Fischer, \& Madhusudhan}]{17BrFiMa.exo}
Brewer, J.~M., Fischer, D.~A., \& Madhusudhan, N. 2017, The Astronomical
  Journal, 153, 83

\bibitem[{Brogi {et~al.}(2017)Brogi, Line, Bean, D{\'{e}}sert, \&
  Schwarz}]{17BrLiBe.exo}
Brogi, M., Line, M., Bean, J., D{\'{e}}sert, J.-M., \& Schwarz, H. 2017, ApJ,
  839, L2

\bibitem[{Burgdorf {et~al.}(2006)Burgdorf, Orton, van Cleve, Meadows, \&
  Houck}]{06BuOrCl.uranus}
Burgdorf, M., Orton, G., van Cleve, J., Meadows, V., \& Houck, J. 2006, Icarus,
  184, 634

\bibitem[{Cabot {et~al.}(2018)Cabot, Madhusudhan, Hawker, \& Gandhi}]{18CaMaHa}
Cabot, S. H.~C., Madhusudhan, N., Hawker, G.~A., \& Gandhi, S. 2018, MNRAS,
  482, 4422

\bibitem[{Chubb {et~al.}(2021)Chubb, Rocchetto, Al-Refaie, Waldmann, Min,
  Barstow, Molli{\'e}re, Phillips, Tennyson, \& Yurchenko}]{20ChRoAl.exo}
Chubb, K.~L., Rocchetto, M., Al-Refaie, A.~F., {et~al.} 2021, A\&A, 646

\bibitem[{Chubb {et~al.}(2020)Chubb, Tennyson, \& Yurchenko}]{ref:20ChTeYu}
Chubb, K.~L., Tennyson, J., \& Yurchenko, S.~N. 2020, MNRAS, 493, 1531

\bibitem[{Clark {et~al.}(2010)Clark, Curchin, Barnes, Jaumann, Soderblom,
  Cruikshank, Brown, Rodriguez, Lunine, Stephan, Hoefen, Le~Mouélic, Sotin,
  Baines, Buratti, \& Nicholson}]{10ClCuBa.titan}
Clark, R.~N., Curchin, J.~M., Barnes, J.~W., {et~al.} 2010, J. Geophys. Res.:
  Planets, 115

\bibitem[{Coles {et~al.}(2019)Coles, Yurchenko, \& Tennyson}]{jt771}
Coles, P.~A., Yurchenko, S.~N., \& Tennyson, J. 2019, MNRAS, 490, 4638

\bibitem[{{Cutri} {et~al.}(2003){Cutri}, {Skrutskie}, {van Dyk}, {Beichman},
  {Carpenter}, {Chester}, {Cambresy}, {Evans}, {Fowler}, {Gizis}, {Howard},
  {Huchra}, {Jarrett}, {Kopan}, {Kirkpatrick}, {Light}, {Marsh}, {McCallon},
  {Schneider}, {Stiening}, {Sykes}, {Weinberg}, {Wheaton}, {Wheelock}, \&
  {Zacarias}}]{ref:star_flux}
{Cutri}, R.~M., {Skrutskie}, M.~F., {van Dyk}, S., {et~al.} 2003, VizieR Online
  Data Catalog, II/246

\bibitem[{Dang {et~al.}(2018)Dang, Cowan, Schwartz, Rauscher, Zhang, Knutson,
  Line, Dobbs-Dixon, Deming, Sundararajan, Fortney, \& Zhao}]{18DaCoSc.corot2b}
Dang, L., Cowan, N.~B., Schwartz, J.~C., {et~al.} 2018, Nature Astronomy, 2,
  220

\bibitem[{de~Kok {et~al.}(2013)de~Kok, Brogi, Snellen, Birkby, Albrecht, \&
  de~Mooij}]{13KoBrSn}
de~Kok, R., Brogi, M., Snellen, I., {et~al.} 2013, A\&A, 554, A82

\bibitem[{Deming \& Seager(2017)}]{ref:17DeSe}
Deming, D. \& Seager, S. 2017, J. Geophys. Res. Planets, 122, 53

\bibitem[{Deming {et~al.}(2009)Deming, Seager, Winn, Miller-Ricci, Clampin,
  Lindler, Greene, Charbonneau, Laughlin, Ricker, Latham, \&
  Ennico}]{ref:09DeSeWi}
Deming, D., Seager, S., Winn, J., {et~al.} 2009, PASP, 121, 952

\bibitem[{Deming {et~al.}(2013)Deming, Wilkins, McCullough, Burrows, Fortney,
  Agol, Dobbs-Dixon, Madhusudhan, Crouzet, Desert, \& et~al.}]{ref:13DeWiMc}
Deming, D., Wilkins, A., McCullough, P., {et~al.} 2013, ApJ, 774, 95

\bibitem[{Drummond {et~al.}(2019)Drummond, Carter, Hébrard, Mayne, Sing,
  Evans, \& Goyal}]{19DrCaHe.exo}
Drummond, B., Carter, A.~L., Hébrard, E., {et~al.} 2019, MNRAS, 486, 1123

\bibitem[{{Eistrup, Christian} {et~al.}(2018){Eistrup, Christian}, {Walsh,
  Catherine}, \& {van Dishoeck, Ewine F.}}]{18EiWaDi}
{Eistrup, Christian}, {Walsh, Catherine}, \& {van Dishoeck, Ewine F.} 2018,
  A\&A, 613, A14

\bibitem[{Fabrycky {et~al.}(2012)Fabrycky, Ford, Steffen, Rowe, Carter,
  Moorhead, Batalha, Borucki, Bryson, Buchhave, Christiansen, Ciardi, Cochran,
  Endl, Fanelli, Fischer, Fressin, Geary, Haas, Hall, Holman, Jenkins, Koch,
  Latham, Li, Lissauer, Lucas, Marcy, Mazeh, McCauliff, Quinn, Ragozzine,
  Sasselov, \& Shporer}]{12FaFoSt.exo}
Fabrycky, D.~C., Ford, E.~B., Steffen, J.~H., {et~al.} 2012, ApJ, 750, 114

\bibitem[{Feng {et~al.}(2016)Feng, Line, Fortney, Stevenson, Bean, Kreidberg,
  \& Parmentier}]{16FeLiFo.exo}
Feng, Y.~K., Line, M.~R., Fortney, J.~J., {et~al.} 2016, Astrophys. J., 829, 52

\bibitem[{Feroz {et~al.}(2009)Feroz, Gair, Hobson, \& Porter}]{09FeGaHo.multi}
Feroz, F., Gair, J.~R., Hobson, M.~P., \& Porter, E.~K. 2009, Classical and
  Quantum Gravity, 26, 215003

\bibitem[{Feroz \& Hobson(2008)}]{08FeHo.multi}
Feroz, F. \& Hobson, M.~P. 2008, MNRAS, 384, 449

\bibitem[{Feroz {et~al.}(2013)Feroz, Hobson, Cameron, \&
  Pettitt}]{13FeHoCa.multi}
Feroz, F., Hobson, M.~P., Cameron, E., \& Pettitt, A.~N. 2013,
  arXiv:1306.2144v2

\bibitem[{Fortenbach \& Dressing(2020)}]{ref:20FoDr}
Fortenbach, C. \& Dressing, C. 2020, PASP, 132, 054501

\bibitem[{Gandhi {et~al.}(2020)Gandhi, Brogi, Yurchenko, Tennyson, Coles, Webb,
  Birkby, Guilluy, Hawker, Madhusudhan, Bonomo, \& Sozzetti}]{20GaBrYu.exo}
Gandhi, S., Brogi, M., Yurchenko, S.~N., {et~al.} 2020, MNRAS, 495, 224

\bibitem[{Gao {et~al.}(2020)Gao, Thorngren, Lee, Fortney, Morley, Wakeford,
  Powell, Stevenson, \& Zhang}]{ref:20GaThLe}
Gao, P., Thorngren, D.~P., Lee, E. K.~H., {et~al.} 2020, Nat Astron, 4, 951

\bibitem[{Giacobbe {et~al.}(2021)Giacobbe, Brogi, Gandhi, Cubillos, Bonomo,
  Sozzetti, Fossati, Guilluy, Carleo, Rainer, Harutyunyan, Borsa, Pino,
  Nascimbeni, Benatti, Biazzo, Bignamini, Chubb, Claudi, Cosentino, Covino,
  Damasso, Desidera, Fiorenzano, Ghedina, Lanza, Leto, Maggio, Malavolta,
  Maldonado, Micela, Molinari, Pagano, Pedani, Piotto, Poretti, Scandariato,
  Yurchenko, Fantinel, Galli, Lodi, Sanna, \& Tozzi}]{21GiBrGa}
Giacobbe, P., Brogi, M., Gandhi, S., {et~al.} 2021, Nature, 592, 205

\bibitem[{Gordon {et~al.}(2017)}]{ref:17Go}
Gordon, I.~E. {et~al.} 2017, JQSRT, 203, 3

\bibitem[{Greene {et~al.}(2007)Greene, Beichman, Eisenstein, Horner, Kelly,
  Mao, Meyer, Rieke, \& Shi}]{ref:07GrBeEi}
Greene, T., Beichman, C., Eisenstein, D., {et~al.} 2007, Proceedings of SPIE -
  The International Society for Optical Engineering, 6693, 153

\bibitem[{Greene {et~al.}(2016)Greene, Line, Montero, Fortney, Lustig-Yaeger,
  \& Luther}]{ref:16GrLiMo}
Greene, T., Line, M.~R., Montero, C., {et~al.} 2016, AAS, 817, 17

\bibitem[{Grootel {et~al.}(2014)Grootel, Gillon, Valencia, Madhusudhan,
  Dragomir, Howe, Burrows, Demory, Deming, Ehrenreich, Lovis, Mayor, Pepe,
  Queloz, Scuflaire, Seager, Segransan, \& Udry}]{14VaGiVa.exo}
Grootel, V.~V., Gillon, M., Valencia, D., {et~al.} 2014, ApJ, 786, 2

\bibitem[{Guilluy {et~al.}(2019)Guilluy, Sozzetti, Brogi, Bonomo, Giacobbe,
  Claudi, \& Benatti}]{ref:19GuSoBr}
Guilluy, G., Sozzetti, A., Brogi, M., {et~al.} 2019, A\&A, 625, A107

\bibitem[{Guo {et~al.}(2020)Guo, Crossfield, Dragomir, Kosiarek, Lothringer,
  Mikal-Evans, Rosenthal, Benneke, Knutson, Dalba, Kempton, Henry, McCullough,
  Barman, Blunt, Chontos, Fortney, Fulton, Hirsch, Howard, Isaacson, Matthews,
  Mocnik, Morley, Petigura, \& Weiss}]{20GuCoDr.exo}
Guo, X., Crossfield, I. J.~M., Dragomir, D., {et~al.} 2020, The Astronomical
  Journal, 159, 239

\bibitem[{Guzmán-Mesa {et~al.}(2020)Guzmán-Mesa, Kitzmann, Fisher, Burgasser,
  Hoeijmakers, Márquez-Neila, Grimm, Mandell, Sznitman, \&
  Heng}]{ref:20GuKiFi}
Guzmán-Mesa, A., Kitzmann, D., Fisher, C., {et~al.} 2020, AJ, 160, 15

\bibitem[{Harper {et~al.}(2018)Harper, Helling, \& Dufek}]{18HaHeDu}
Harper, J.~M., Helling, C., \& Dufek, J. 2018, ApJ, 867, 123

\bibitem[{Hawker {et~al.}(2018)Hawker, Madhusudhan, Cabot, \&
  Gandhi}]{18HaMaCa}
Hawker, G.~A., Madhusudhan, N., Cabot, S. H.~C., \& Gandhi, S. 2018, ApJ, 863,
  L11

\bibitem[{Helling {et~al.}(2016)Helling, Lee, Dobbs-Dixon, Mayne, Amundsen,
  Khaimova, Unger, Manners, Acreman, \& Smith}]{16HeLeDo}
Helling, C., Lee, E., Dobbs-Dixon, I., {et~al.} 2016, MNRAS, 460, 855

\bibitem[{Howard {et~al.}(2011)Howard, Johnson, \& et~al.}]{11HoAsMa.exo}
Howard, A.~W., Johnson, J.~A., \& et~al. 2011, ApJ, 730, 10

\bibitem[{Howe {et~al.}(2017)Howe, Burrows, \& Deming}]{ref:17HoBuDe}
Howe, A., Burrows, A., \& Deming, D. 2017, ApJ, 835, 96

\bibitem[{Hu \& Seager(2014)}]{ref:14HuSe}
Hu, R. \& Seager, S. 2014, ApJ, 784, 63

\bibitem[{Hörst {et~al.}(2018)Hörst, Yoon, Ugelow, Parker, Li, [de Gouw], \&
  Tolbert}]{18HoYoUg.exo}
Hörst, S.~M., Yoon, Y.~H., Ugelow, M.~S., {et~al.} 2018, Icarus, 301, 136

\bibitem[{Kaltenegger \& Traub(2009)}]{ref:09KaTr}
Kaltenegger, L. \& Traub, W. 2009, ApJ, 698, 519

\bibitem[{Kawashima {et~al.}(2019)Kawashima, Hu, \& Ikoma}]{19KaReIk.exo}
Kawashima, Y., Hu, R., \& Ikoma, M. 2019, ApJ, 876, L5

\bibitem[{Kawashima \& Ikoma(2019)}]{ref:19KaIk}
Kawashima, Y. \& Ikoma, M. 2019, ApJ, 877, 109

\bibitem[{{Kawashima, Yui} \& {Min, Michiel}(2021)}]{21KaMi.arcis}
{Kawashima, Yui} \& {Min, Michiel}. 2021, A\&A, 656, A90

\bibitem[{Keles {et~al.}(2019)Keles, Mallonn, von Essen, Carroll, Alexoudi,
  Pino, Ilyin, Poppenhäger, Kitzmann, Nascimbeni, Turner, \&
  Strassmeier}]{19KeMaEs}
Keles, E., Mallonn, M., von Essen, C., {et~al.} 2019, MNRAS Letters, 489, L37

\bibitem[{Knutson {et~al.}(2009)Knutson, Charbonneau, Cowan, Fortney, Showman,
  Agol, \& Henry}]{ref:09KnChCo}
Knutson, H.~A., Charbonneau, D., Cowan, N.~B., {et~al.} 2009, ApJ, 703

\bibitem[{Knutson {et~al.}(2014)Knutson, Dragomir, Kreidberg, Kempton,
  McCullough, Fortney, Bean, Gillon, Homeier, \& Howard}]{14KnDrKr}
Knutson, H.~A., Dragomir, D., Kreidberg, L., {et~al.} 2014, ApJ, 794, 155

\bibitem[{Kreidberg {et~al.}(2014)Kreidberg, Bean, Désert, Line, Fortney,
  Madhusudhan, Stevenson, Showman, Charbonneau, McCullough, Seager, Burrows,
  Henry, Williamson, Kataria, \& Homeier}]{ref:14KrBeDe}
Kreidberg, L., Bean, J.~L., Désert, J.-M., {et~al.} 2014, ApJ, 793

\bibitem[{Lee {et~al.}(2016)Lee, Dobbs-Dixon, Helling, Bognar, \&
  Woitke}]{16LeDoHe}
Lee, E., Dobbs-Dixon, I., Helling, C., Bognar, K., \& Woitke, P. 2016, A\&A,
  594, A48

\bibitem[{Li {et~al.}({2015})Li, Gordon, Rothman, Tan, Hu, Kassi, Campargue, \&
  Medvedev}]{15LiGoRo.CO}
Li, G., Gordon, I.~E., Rothman, L.~S., {et~al.} {2015}, ApJ Suppl., {216}, 15

\bibitem[{Lin {et~al.}(2021)Lin, MacDonald, Kaltenegger, \& Wilson}]{21LiMaKa}
Lin, Z., MacDonald, R.~J., Kaltenegger, L., \& Wilson, D.~J. 2021, MNRAS, 505,
  3562

\bibitem[{Line {et~al.}(2016)Line, Stevenson, Bean, Desert, Fortney, Kreidberg,
  Madhusudhan, Showman, \& Diamond-Lowe}]{16LiStBe}
Line, M.~R., Stevenson, K.~B., Bean, J., {et~al.} 2016, The Astronomical
  Journal, 152, 203

\bibitem[{Lines {et~al.}(2018)Lines, Mayne, Boutle, Manners, Lee, Helling,
  Drummond, Amundsen, Goyal, Acreman, Tremblin, \& Kerslake}]{18LiMaBo}
Lines, S., Mayne, N.~J., Boutle, I.~A., {et~al.} 2018, A\&A, 615, A97

\bibitem[{Lombardo {et~al.}(2019)Lombardo, Nixon, Achterberg, Jolly, Sung,
  Irwin, \& Flasar}]{19LoNiAc.titan}
Lombardo, N.~A., Nixon, C.~A., Achterberg, R.~K., {et~al.} 2019, Icarus, 317,
  454

\bibitem[{MacDonald {et~al.}(2020)MacDonald, Goyal, \& Lewis}]{ref:20MaGoLe}
MacDonald, R.~J., Goyal, J.~M., \& Lewis, N.~K. 2020, ApJ, 893, L43

\bibitem[{MacDonald \& Madhusudhan(2017)}]{17MaMa}
MacDonald, R.~J. \& Madhusudhan, N. 2017, MNRAS, 469, 1979

\bibitem[{Maciejewski {et~al.}(2011)Maciejewski, Errmann, Raetz, Seeliger,
  Spaleniak, \& Neuh\"auser}]{11MaErRa.wasp12b}
Maciejewski, G., Errmann, R., Raetz, S., {et~al.} 2011, A\&A, 528, A65

\bibitem[{Macintosh {et~al.}(2015)Macintosh, Graham, Barman, De~Rosa,
  Konopacky, Marley, Marois, Nielsen, Pueyo, Rajan, \& et~al.}]{ref:15MaGrBa}
Macintosh, B., Graham, J.~R., Barman, T., {et~al.} 2015, Science, 350, 64–67

\bibitem[{Madhusudhan(2012)}]{ref:12Ma}
Madhusudhan, N. 2012, ApJ, 758, 36

\bibitem[{Mai \& Line(2019)}]{ref:19MaLi}
Mai, C. \& Line, M.~R. 2019, ApJ, 883, 144

\bibitem[{Mant {et~al.}(2018)Mant, Yachmenev, Tennyson, \&
  Yurchenko}]{ref:18MaYaTe}
Mant, B.~P., Yachmenev, A., Tennyson, J., \& Yurchenko, S.~N. 2018, MNRAS, 478,
  3220

\bibitem[{McCullough {et~al.}(2006)McCullough, Stys, Valenti, Johns-Krull,
  Janes, Heasley, Bye, Dodd, Fleming, Pinnick, Bissinger, Gary, Howell, \&
  Vanmunster}]{06McStVa}
McCullough, P., Stys, J., Valenti, J.~A., {et~al.} 2006, ApJ, 648, 1228

\bibitem[{McKay(2016)}]{16Mc.titan}
McKay, C.~P. 2016, Life, 6, 8

\bibitem[{McKemmish {et~al.}(2019)McKemmish, Masseron, Hoeijmakers,
  P{\'e}rez-Mesa, Grimm, Yurchenko, \& Tennyson}]{ExoMol_TiO}
McKemmish, L.~K., Masseron, T., Hoeijmakers, H.~J., {et~al.} 2019, MNRAS, 488,
  2836

\bibitem[{McKemmish {et~al.}(2016)McKemmish, Yurchenko, \& Tennyson}]{jt644}
McKemmish, L.~K., Yurchenko, S.~N., \& Tennyson, J. 2016, MNRAS, 463, 771

\bibitem[{Miguel \& Kaltenegger(2014)}]{ref:14MiKa}
Miguel, Y. \& Kaltenegger, L. 2014, ApJ, 780

\bibitem[{Min {et~al.}(2005)Min, Hovenier, \& {de Koter}}]{05MiHoKo.arcis}
Min, M., Hovenier, J.~W., \& {de Koter}, A. 2005, A\&A, 432, 909

\bibitem[{Min {et~al.}(2020)Min, Ormel, Chubb, Helling, \&
  Kawashima}]{ref:20MiOrCh}
Min, M., Ormel, C.~W., Chubb, K., Helling, C., \& Kawashima, Y. 2020, A\&A,
  642, A28

\bibitem[{Molli{\'e}re {et~al.}(2017)Molli{\'e}re, van Boekel, Bouwman,
  Henning, Lagage, \& Min}]{ref:17MoBoBo}
Molli{\'e}re, P., van Boekel, R., Bouwman, J., {et~al.} 2017, A\&A, 600, A10

\bibitem[{Molli{\'e}re {et~al.}(2015)Molli{\'e}re, van Boekel, Dullemond,
  Henning, \& Mordasini}]{ref:15MoBoDu}
Molli{\'e}re, P., van Boekel, R., Dullemond, C., Henning, T., \& Mordasini, C.
  2015, ApJ, 813

\bibitem[{Moses {et~al.}(2000)Moses, Bézard, Lellouch, Gladstone,
  Feuchtgruber, \& Allen}]{00MoBeLe.saturn}
Moses, J., Bézard, B., Lellouch, E., {et~al.} 2000, Icarus, 143, 244

\bibitem[{Moses {et~al.}(2012)Moses, Madhusudhan, Visscher, \&
  Freedman}]{12MoMaVi.exo}
Moses, J.~I., Madhusudhan, N., Visscher, C., \& Freedman, R.~S. 2012, ApJ, 763,
  25

\bibitem[{Nielsen {et~al.}(2016)Nielsen, Ferruit, Giardino, Birkmann, Muñoz,
  Valenti, Isaak, de~Oliveira, Böker, Lützgendorf, Rawle, \&
  Sirianni}]{16NiFeGi}
Nielsen, L.~D., Ferruit, P., Giardino, G., {et~al.} 2016, 9904, 1218

\bibitem[{{\"O}berg {et~al.}(2011){\"O}berg, Murray-Clay, \&
  Bergin}]{11ObMuBe.exo}
{\"O}berg, K.~I., Murray-Clay, R., \& Bergin, E.~A. 2011, ApJ, 743, L16

\bibitem[{Ormel \& Min(2019)}]{ref:19OrMi}
Ormel, C.~W. \& Min, M. 2019, A\&A, 622, A121

\bibitem[{Pascale {et~al.}(2018)Pascale, Bezawada, Barstow, Beaulieu, Bowles,
  du~Foresto, Coustenis, Decin, Drossart, Eccleston, Encrenaz, Forget, Griffin,
  Güdel, Hartogh, Heske, Lagage, Leconte, Malaguti, Micela, Middleton, Min,
  Moneti, Morales, Mugnai, Ollivier, Pace, Papageorgiou, Pilbratt, Puig, Rataj,
  Ray, Ribas, Rocchetto, Sarkar, Selsis, Taylor, Tennyson, Tinetti, Turrini,
  Vandenbussche, Venot, Waldmann, Wolkenberg, Wright, Osorio, \&
  Zingales}]{18PaBeBa.ARIEL}
Pascale, E., Bezawada, N., Barstow, J., {et~al.} 2018, in Space Telescopes and
  Instrumentation 2018: Optical, Infrared, and Millimeter Wave, ed. M.~Lystrup,
  H.~A. MacEwen, G.~G. Fazio, N.~Batalha, N.~Siegler, \& E.~C. Tong, Vol.
  10698, International Society for Optics and Photonics (SPIE), 169 -- 178

\bibitem[{Patapis {et~al.}(2021)Patapis, Nasedkin, Cugno, Glauser, Argyriou,
  Whiteford, Molli{\'e}re, Glasse, \& Quanz}]{21PaNaCu}
Patapis, P., Nasedkin, E., Cugno, G., {et~al.} 2021, A\&A

\bibitem[{Polyansky {et~al.}(2018)Polyansky, Kyuberis, Zobov, Tennyson,
  Yurchenko, \& Lodi}]{jt734}
Polyansky, O.~L., Kyuberis, A.~A., Zobov, N.~F., {et~al.} 2018, MNRAS, 480,
  2597

\bibitem[{Rey {et~al.}(2016)Rey, Nikitin, Babikov, \& Tyuterev}]{TheoReTS}
Rey, M., Nikitin, A.~V., Babikov, Y.~L., \& Tyuterev, V.~G. 2016, JMS, 327, 138
  , new Visions of Spectroscopic Databases, Volume II

\bibitem[{Rimmer {et~al.}(2019)Rimmer, Ferus, Waldmann, Kn{\'{\i}}{\v{z}}ek,
  Kalvaitis, Ivanek, Kubel{\'{\i}}k, Yurchenko, Burian, Dost{\'{a}}l, Juha,
  Dud{\v{z}}{\'{a}}k, Kru{\r}s, Tennyson, Civi{\v{s}}, Archibald, \&
  Granville-Willett}]{19RiFeWa}
Rimmer, P.~B., Ferus, M., Waldmann, I.~P., {et~al.} 2019, ApJ, 888, 21

\bibitem[{Rimmer {et~al.}(2014)Rimmer, Helling, \& Bilger}]{14RiHeBi}
Rimmer, P.~B., Helling, C., \& Bilger, C. 2014, Int. J. Astrobiology, 13,
  173–181

\bibitem[{Rimmer \& Shorttle(2019)}]{19RiSh.exo}
Rimmer, P.~B. \& Shorttle, O. 2019, Life, 9, 12

\bibitem[{Rocchetto {et~al.}(2016)Rocchetto, Waldmann, Venot, Lagage, \&
  Tinetti}]{ref:16RoWaVe}
Rocchetto, M., Waldmann, I.~P., Venot, O., Lagage, P.-O., \& Tinetti, G. 2016,
  ApJ, 833, 120

\bibitem[{Rothman {et~al.}(2010)Rothman, Gordon, Barber, Dothe, Gamache,
  Goldman, Perevalov, Tashkun, \& Tennyson}]{HITEMP}
Rothman, L.~S., Gordon, I.~E., Barber, R.~J., {et~al.} 2010, JQSRT, 111, 2139

\bibitem[{Sanchis-Ojeda {et~al.}(2012)Sanchis-Ojeda, Fabrycky, Winn, Barclay,
  Clarke, Ford, Fortney, Geary, Holman, Howard, Jenkins, Koch, Lissauer, Marcy,
  Mullally, Ragozzine, Seader, Still, \& Thompson}]{12SaFaWi.exo}
Sanchis-Ojeda, R., Fabrycky, D.~C., Winn, J.~N., {et~al.} 2012, Nature, 487,
  449

\bibitem[{Schuler {et~al.}(2011)Schuler, Flateau, Cunha, King, Ghezzi, \&
  Smith}]{11ScFlCu.exo}
Schuler, S.~C., Flateau, D., Cunha, K., {et~al.} 2011, ApJ, 732, 55

\bibitem[{Sharpe {et~al.}(2002)Sharpe, Sams, \& Johnson}]{ref:PNNL}
Sharpe, S., Sams, R., \& Johnson, T. 2002, in Applied Imagery Pattern
  Recognition Workshop, 2002. Proceedings., 45--48

\bibitem[{Sinclair {et~al.}(2019)Sinclair, Moses, Hue, Greathouse, Orton,
  Fletcher, \& Irwin}]{19SiMoHu.jupiter}
Sinclair, J., Moses, J., Hue, V., {et~al.} 2019, Icarus, 328, 176

\bibitem[{Sing {et~al.}(2016)Sing, Fortney, Nikolov, Wakeford, Kataria, Evans,
  Aigrain, Ballester, Burrows, Deming, D{\'e}sert, Gibson, Henry, Huitson,
  Knutson, Etangs, Pont, Showman, Vidal-Madjar, Williamson, \&
  Wilson}]{16SiFoNi.exo}
Sing, D.~K., Fortney, J.~J., Nikolov, N., {et~al.} 2016, Nature, 529, 59

\bibitem[{Sousa-Silva {et~al.}(2019)Sousa-Silva, Petkowski, \&
  Seager}]{ref:19SoPeSe}
Sousa-Silva, C., Petkowski, J., \& Seager, S. 2019, PCCP, 21, 18970

\bibitem[{Southworth(2010)}]{10Sou.exo}
Southworth, J. 2010, MNRAS, 408, 1689

\bibitem[{Southworth {et~al.}(2018)Southworth, Tregloan-Reed, Pinhas,
  Madhusudhan, Mancini, \& Smith}]{18SoTrPi.exo}
Southworth, J., Tregloan-Reed, J., Pinhas, A., {et~al.} 2018, MNRAS, 481, 4261

\bibitem[{Swain {et~al.}(2008)Swain, Vasisht, \& Tinetti}]{08SwVaTi.exo}
Swain, M.~R., Vasisht, G., \& Tinetti, G. 2008, Nature, 452, 329

\bibitem[{Tarafdar \& Wickramasinghe(1975)}]{75TaWi.exo}
Tarafdar, S.~P. \& Wickramasinghe, N.~C. 1975, Ap\&SS, 35, L41

\bibitem[{Taylor {et~al.}(2020)Taylor, Parmentier, Irwin, Aigrain, Lee, \&
  Krissansen-Totton}]{20TaPaIr.exo}
Taylor, J., Parmentier, V., Irwin, P. G.~J., {et~al.} 2020, MNRAS, 493, 4342

\bibitem[{Tennyson \& Yurchenko(2018)}]{ref:18TeYu}
Tennyson, J. \& Yurchenko, S. 2018, Atoms, 6, 26

\bibitem[{Tennyson {et~al.}(2020)Tennyson, Yurchenko, Al-Refaie, Clark, Chubb,
  Conway, Dewan, Gorman, Hill, Lynas-Gray, Mellor, McKemmish, Owens, Polyansky,
  Semenov, Somogyi, Tinetti, Upadhyay, Waldmann, Wang, Wright, \&
  Yurchenko}]{ref:20TeYu}
Tennyson, J., Yurchenko, S.~N., Al-Refaie, A.~F., {et~al.} 2020, JQSRT, 255,
  107228

\bibitem[{Tielens(2008)}]{08Ti.exo}
Tielens, A. 2008, Annu. Rev. Astron. Astrophys., 46, 289

\bibitem[{Torres {et~al.}(2008)Torres, Winn, \& Holman}]{08ToWiHo.exo}
Torres, G., Winn, J.~N., \& Holman, M.~J. 2008, ApJ, 677, 1324

\bibitem[{Trotta(2008)}]{08Trotta.stats}
Trotta, R. 2008, Contemp Phys, 49, 71

\bibitem[{Tsai {et~al.}(2017)Tsai, Lyons, Grosheintz, Rimmer, Kitzmann, \&
  Heng}]{17TsLyGr}
Tsai, S.-M., Lyons, J.~R., Grosheintz, L., {et~al.} 2017, ApJ Supplement
  Series, 228, 20

\bibitem[{van Belle(2008)}]{08Belle.exo}
van Belle, G.~T. 2008, Publications of the Astronomical Society of the Pacific,
  120, 617

\bibitem[{Venot {et~al.}(2015)Venot, Hébrard, Agúndez, Decin, \&
  Bounaceur}]{ref:15VeHeAg}
Venot, O., Hébrard, E., Agúndez, M., Decin, L., \& Bounaceur, R. 2015, A\&A,
  577

\bibitem[{Waldmann {et~al.}(2015)Waldmann, Tinetti, Rocchetto, Barton,
  Yurchenko, \& Tennyson}]{15WaTiRo.taurex}
Waldmann, I.~P., Tinetti, G., Rocchetto, M., {et~al.} 2015, Astrophys. J., 802,
  107

\bibitem[{Woitke {et~al.}(2018)Woitke, Helling, Hunter, Millard, Turner,
  Worters, Blecic, \& Stock}]{ref:18WoHeHu}
Woitke, P., Helling, C., Hunter, G.~H., {et~al.} 2018, A\&A, 614, A1

\bibitem[{Wunderlich {et~al.}(2020)Wunderlich, Scheucher, Godolt, Grenfell,
  Schreier, Schneider, Wilson, S{\'{a}}nchez-L{\'{o}}pez, L{\'{o}}pez-Puertas,
  \& Rauer}]{20WuScGo}
Wunderlich, F., Scheucher, M., Godolt, M., {et~al.} 2020, The Astrophysical
  Journal, 901, 126

\bibitem[{Wyttenbach {et~al.}(2015)Wyttenbach, Ehrenreich, Lovis, Udry, \&
  Pepe}]{15WyEhLo}
Wyttenbach, A., Ehrenreich, D., Lovis, C., Udry, S., \& Pepe, F. 2015, A\&A,
  577, A62

\bibitem[{Yurchenko {et~al.}(2017)Yurchenko, Amundsen, Tennyson, \&
  Waldmann}]{jt698}
Yurchenko, S.~N., Amundsen, D.~S., Tennyson, J., \& Waldmann, I.~P. 2017, A\&A,
  605, A95

\bibitem[{Yurchenko {et~al.}(2020)Yurchenko, Mellor, Freedman, \&
  Tennyson}]{20YuMeFr.co2}
Yurchenko, S.~N., Mellor, T.~M., Freedman, R.~S., \& Tennyson, J. 2020, MNRAS,
  496, 5282

\bibitem[{Zhang {et~al.}(2019)Zhang, Chachan, Kempton, \&
  Knutson}]{ref:19ZhChKe}
Zhang, M., Chachan, Y., Kempton, E. M.-R., \& Knutson, H.~A. 2019, PASP, 131

\bibitem[{Zilinskas {et~al.}(2020)Zilinskas, Miguel, Molli{\'e}re, \&
  Tsai}]{ref:20ZiMiMo}
Zilinskas, M., Miguel, Y., Molli{\'e}re, P., \& Tsai, S.-M. 2020, MNRAS, 494,
  1490

\end{thebibliography}

\begin{appendix}

\onecolumn
\section{Additional figures}\label{sec:append1}

\begin{figure}[h!]
        \centering
        \includegraphics[width=0.45\textwidth]{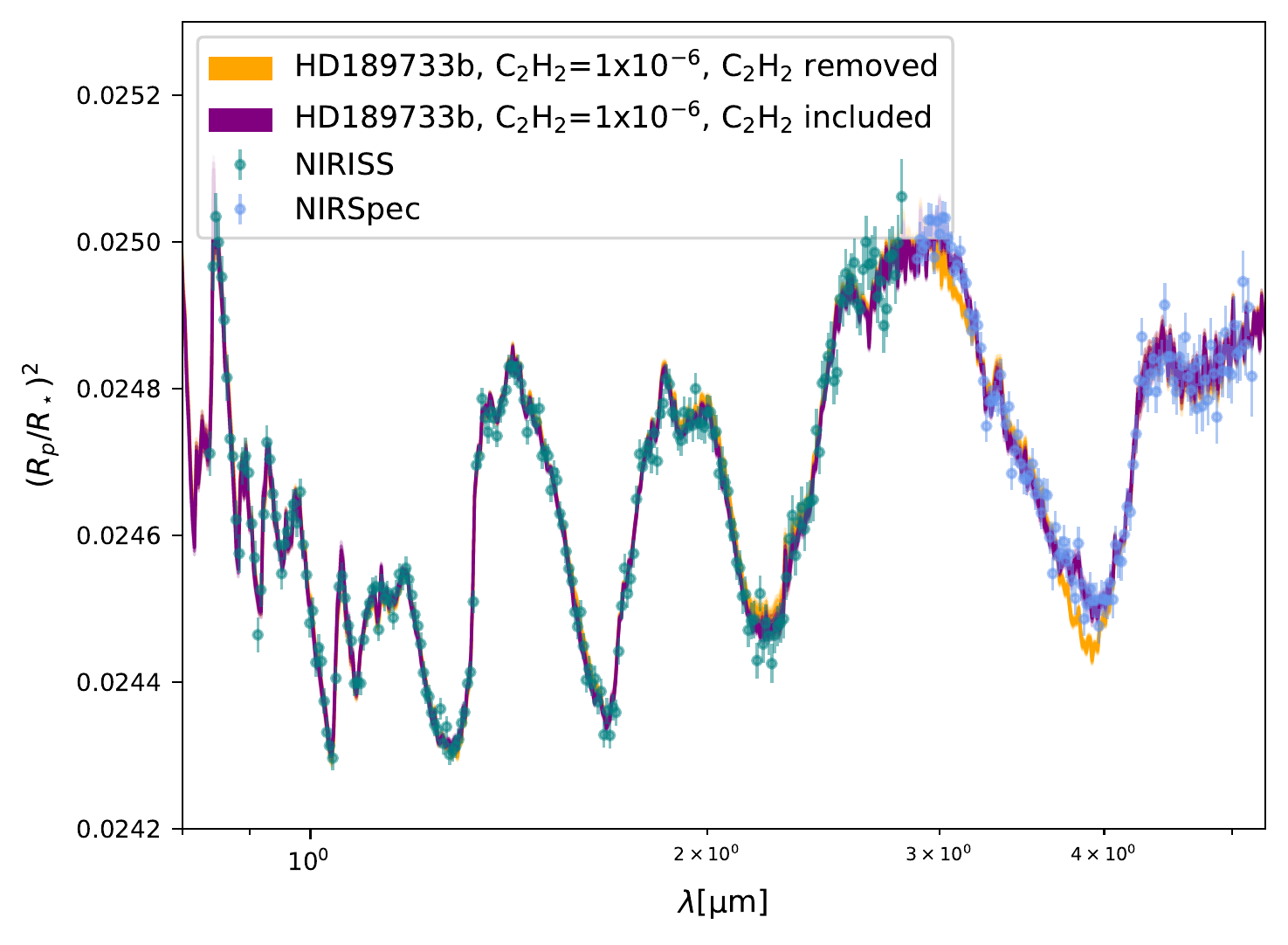}
        \includegraphics[width=0.45\textwidth]{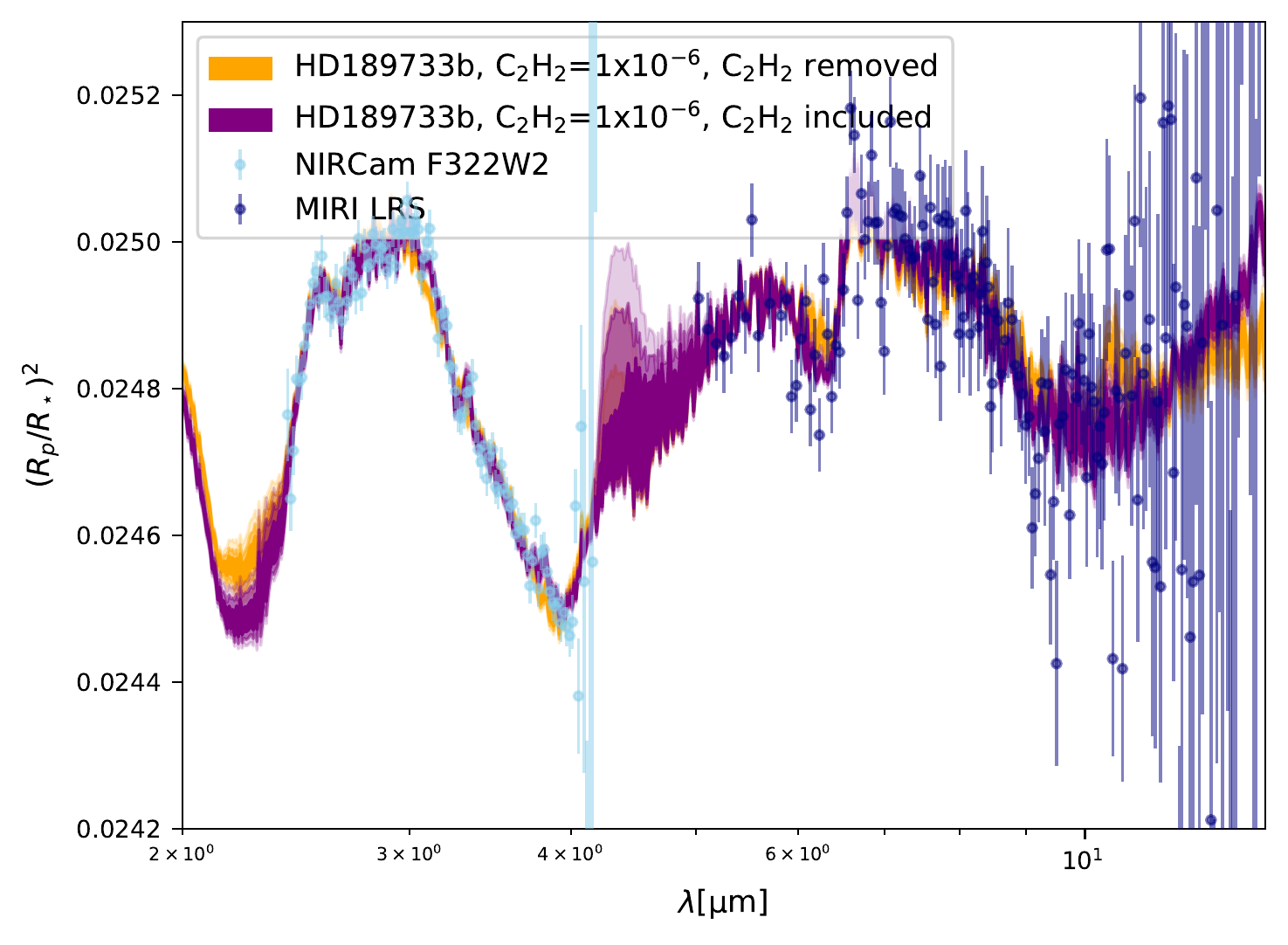}
        \includegraphics[width=0.45\textwidth]{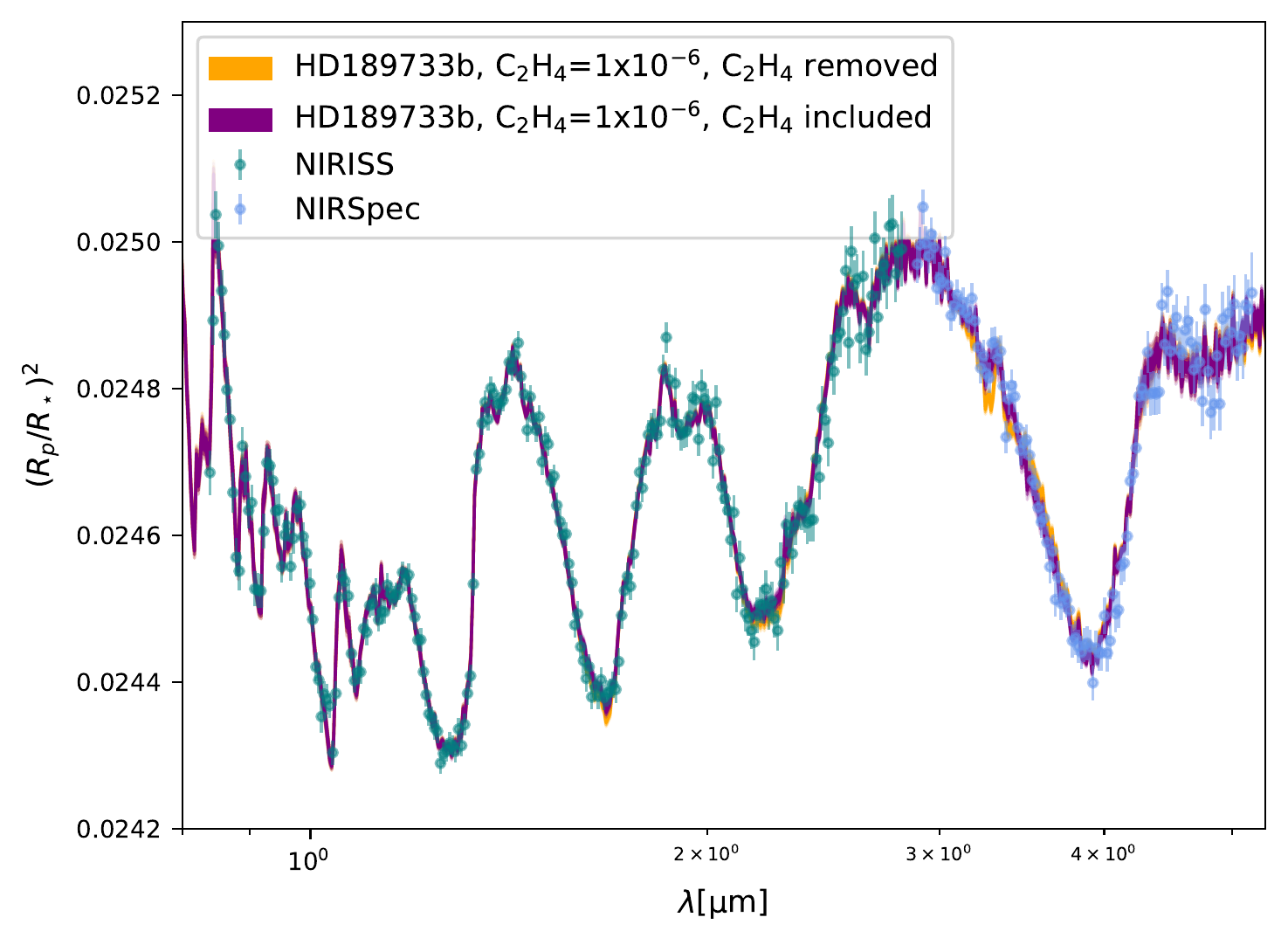}
        \includegraphics[width=0.45\textwidth]{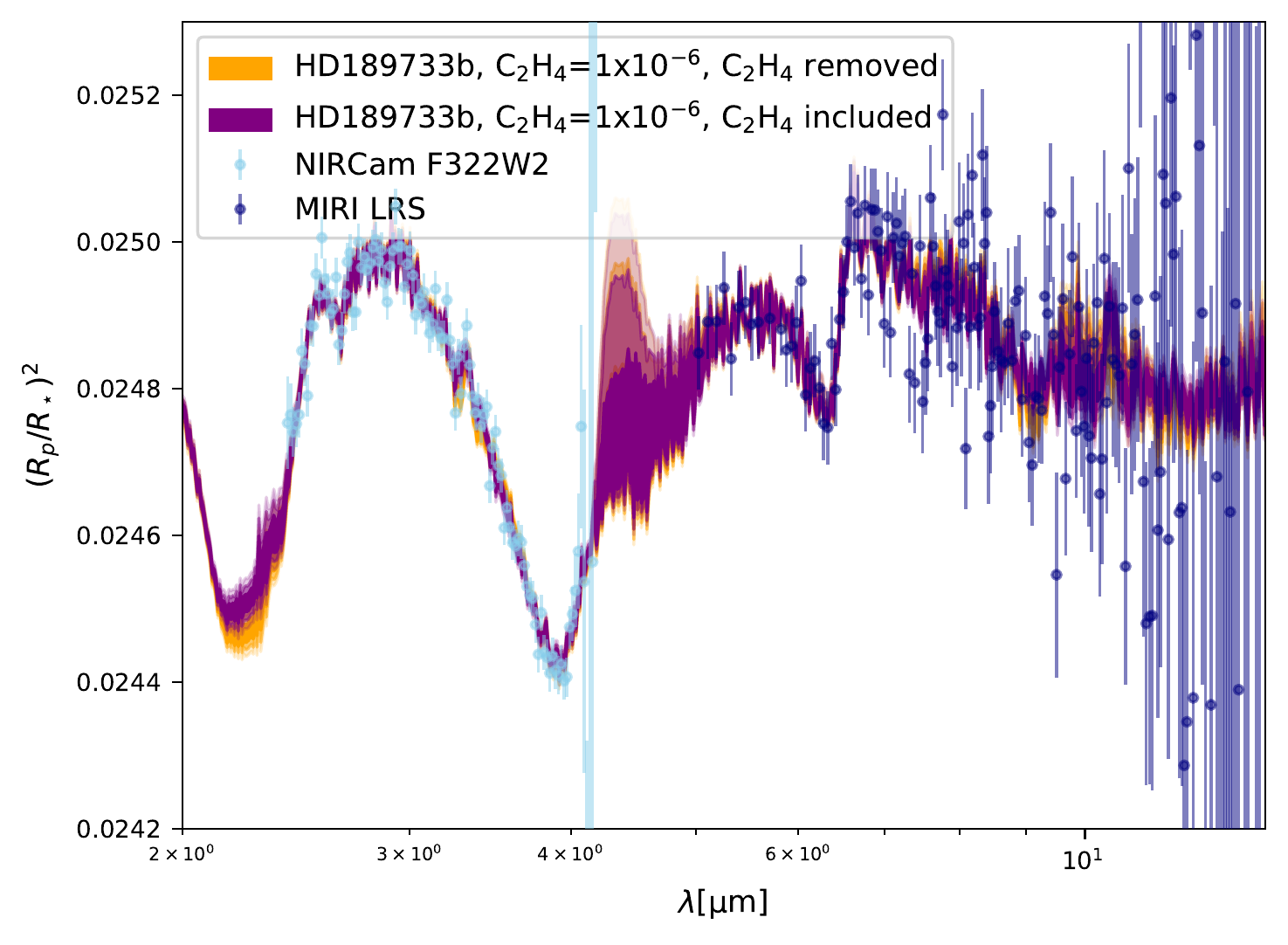}
        \includegraphics[width=0.45\textwidth]{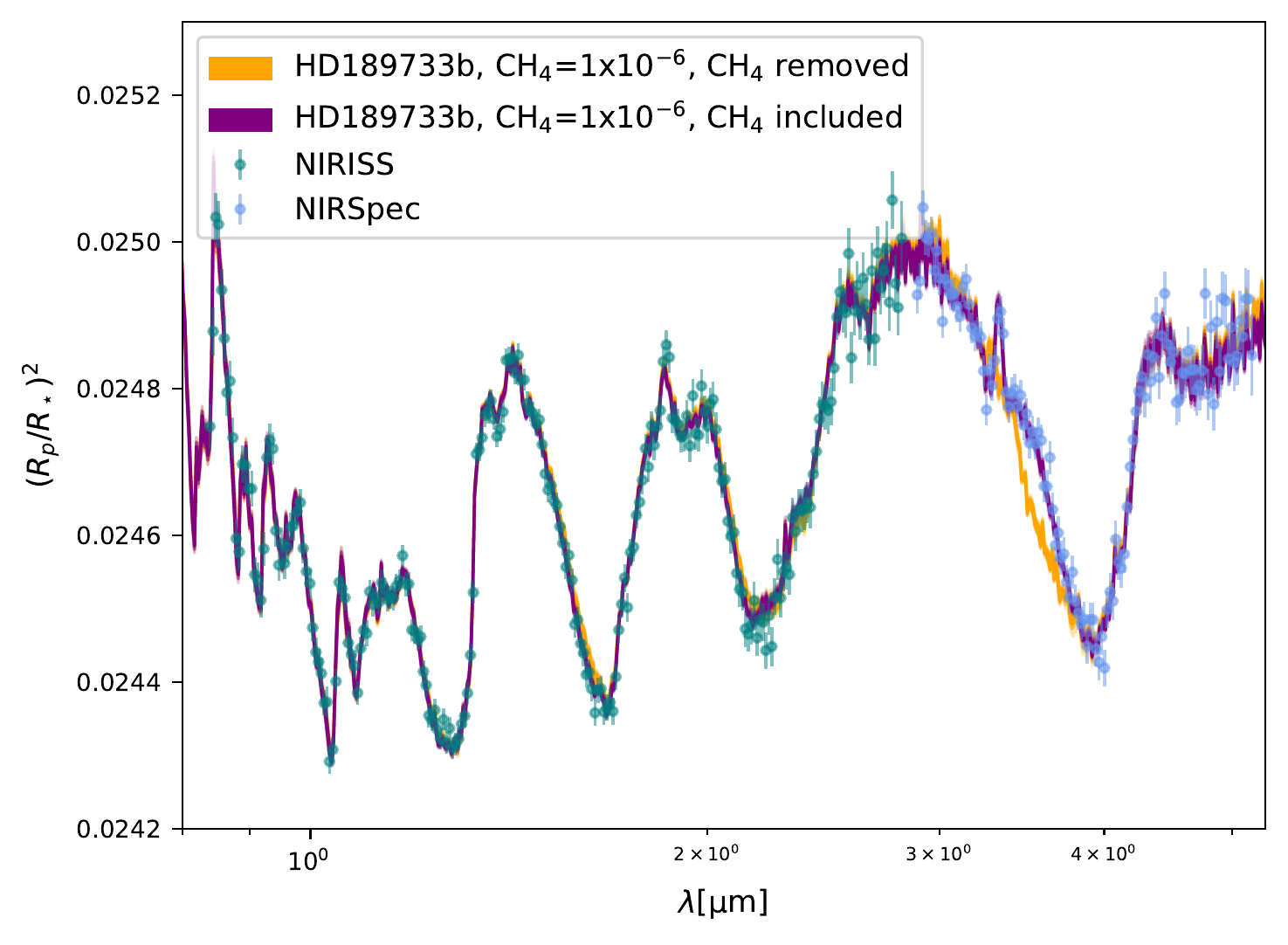}
        \includegraphics[width=0.45\textwidth]{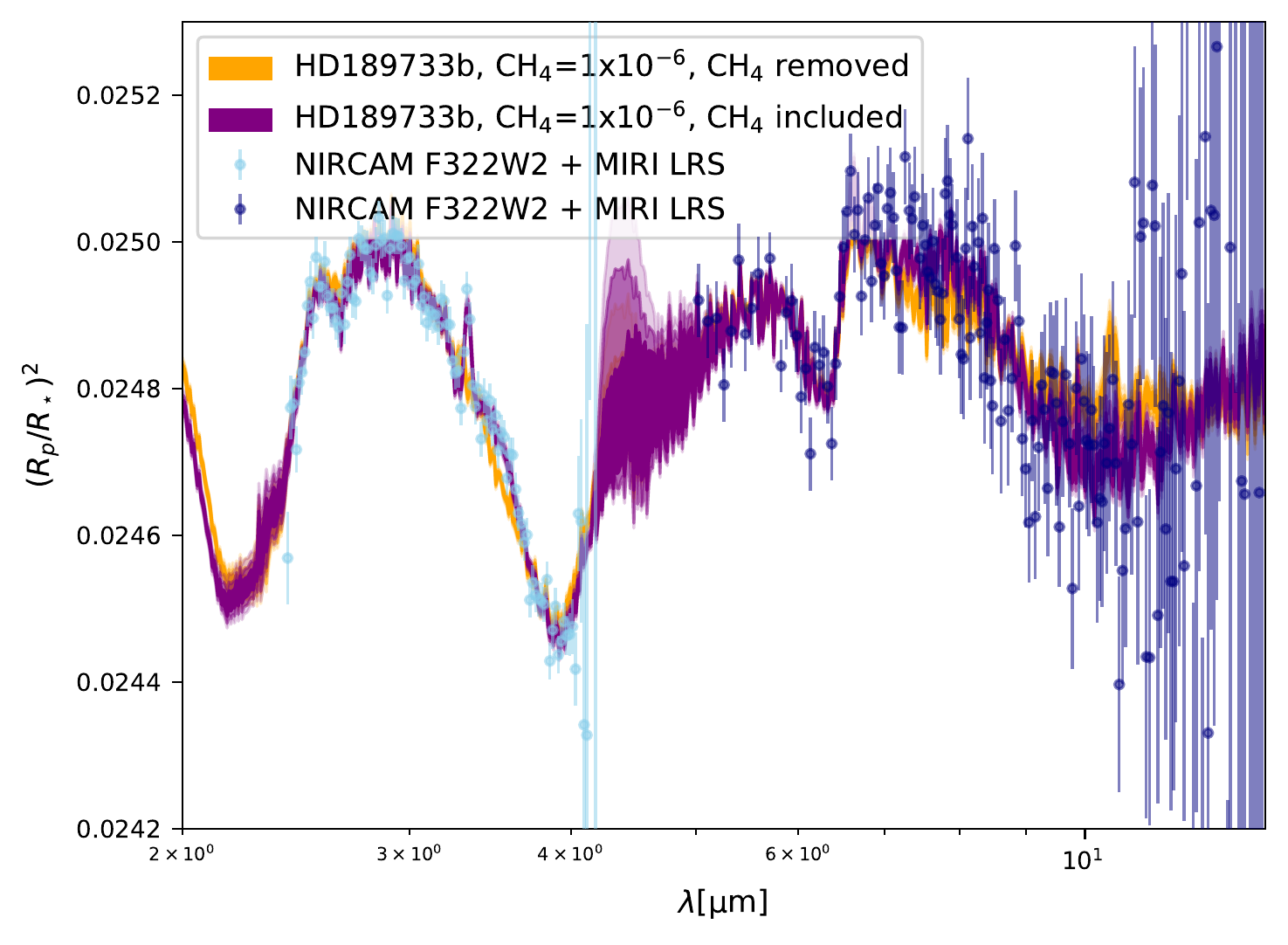}
        \caption{Simulated spectra (re-binned to R=100) for the atmospheres of a HD189733b-like planet observed with NIRISS SOSS and NIRSpec G395M (left column) and with NIRCam F322W2 and MIRI LRS (right column). The VMR was set to $10^{-6}$ for the hydrocarbon molecule specified in each legend for the simulated observed spectra.  The best-fit retrieved spectra are given in orange for the cases where the molecule was removed, and in purple   when it was included in the retrieval. The shading of the best-fit spectra corresponds to 1, 2, and 3~$\sigma$ regions. }
        \label{fig:comp1_HD189}
\end{figure}

\begin{figure}[h!]
        \centering
        \includegraphics[width=0.45\textwidth]{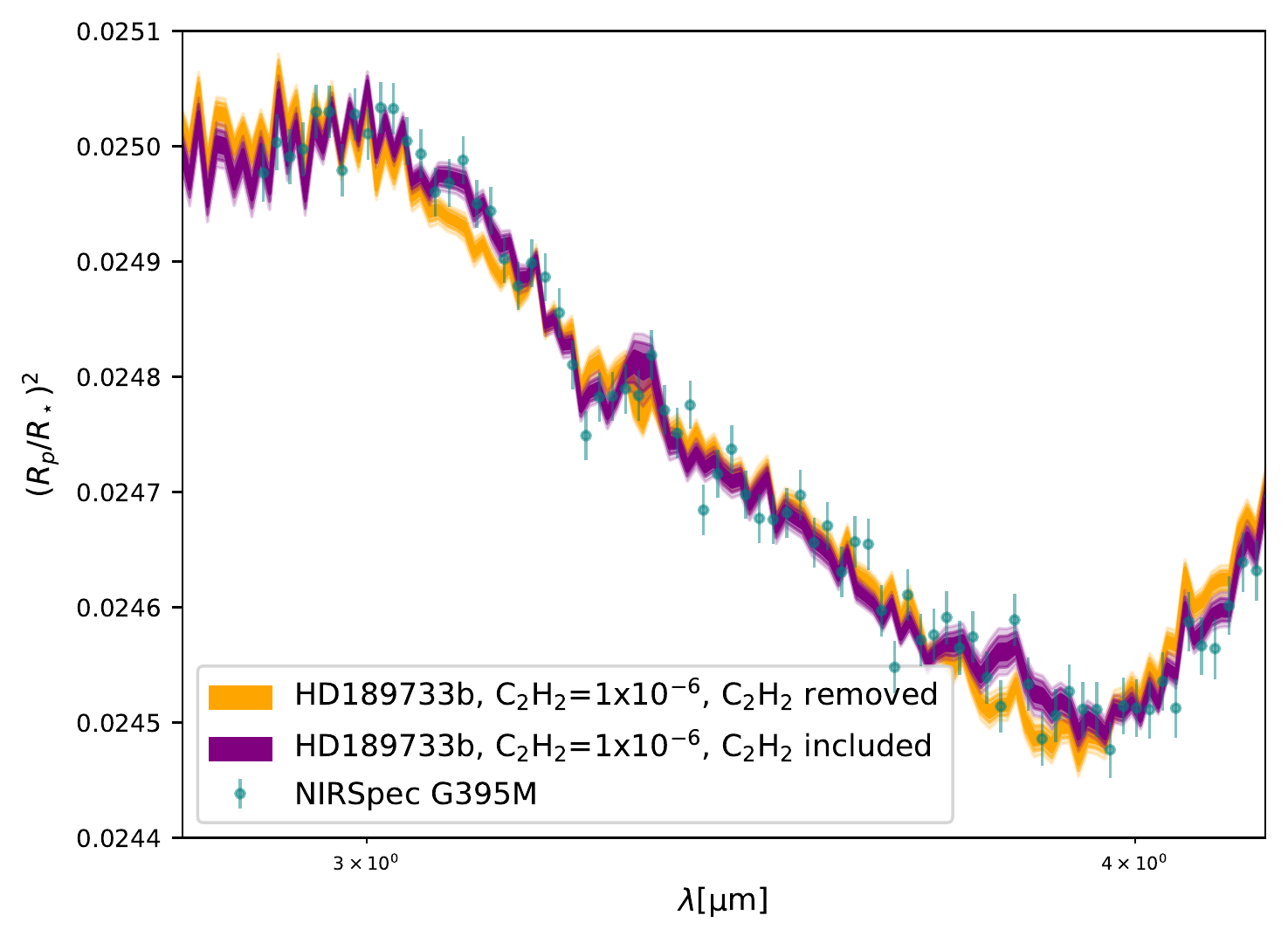}
        \includegraphics[width=0.45\textwidth]{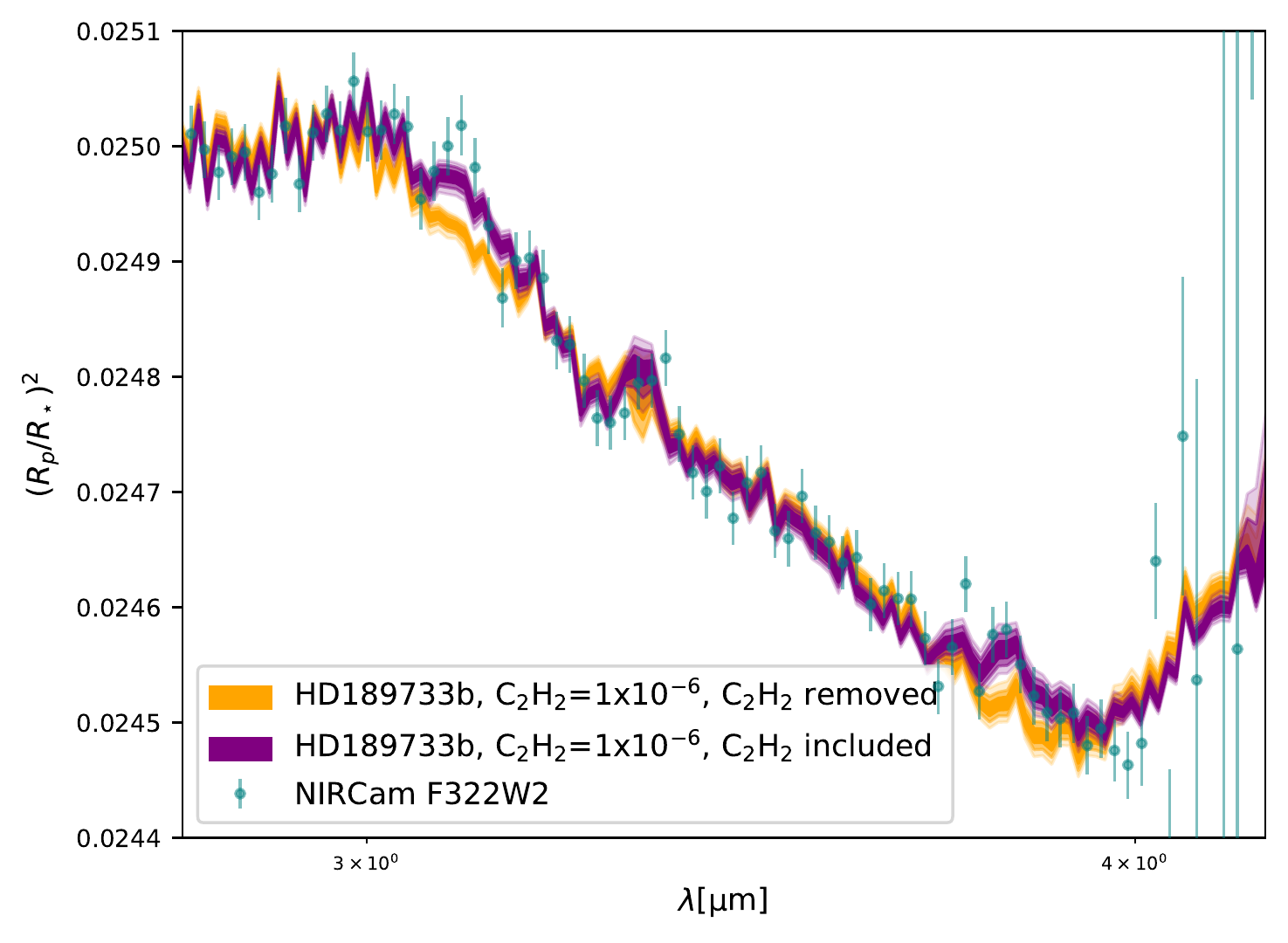}
        \includegraphics[width=0.45\textwidth]{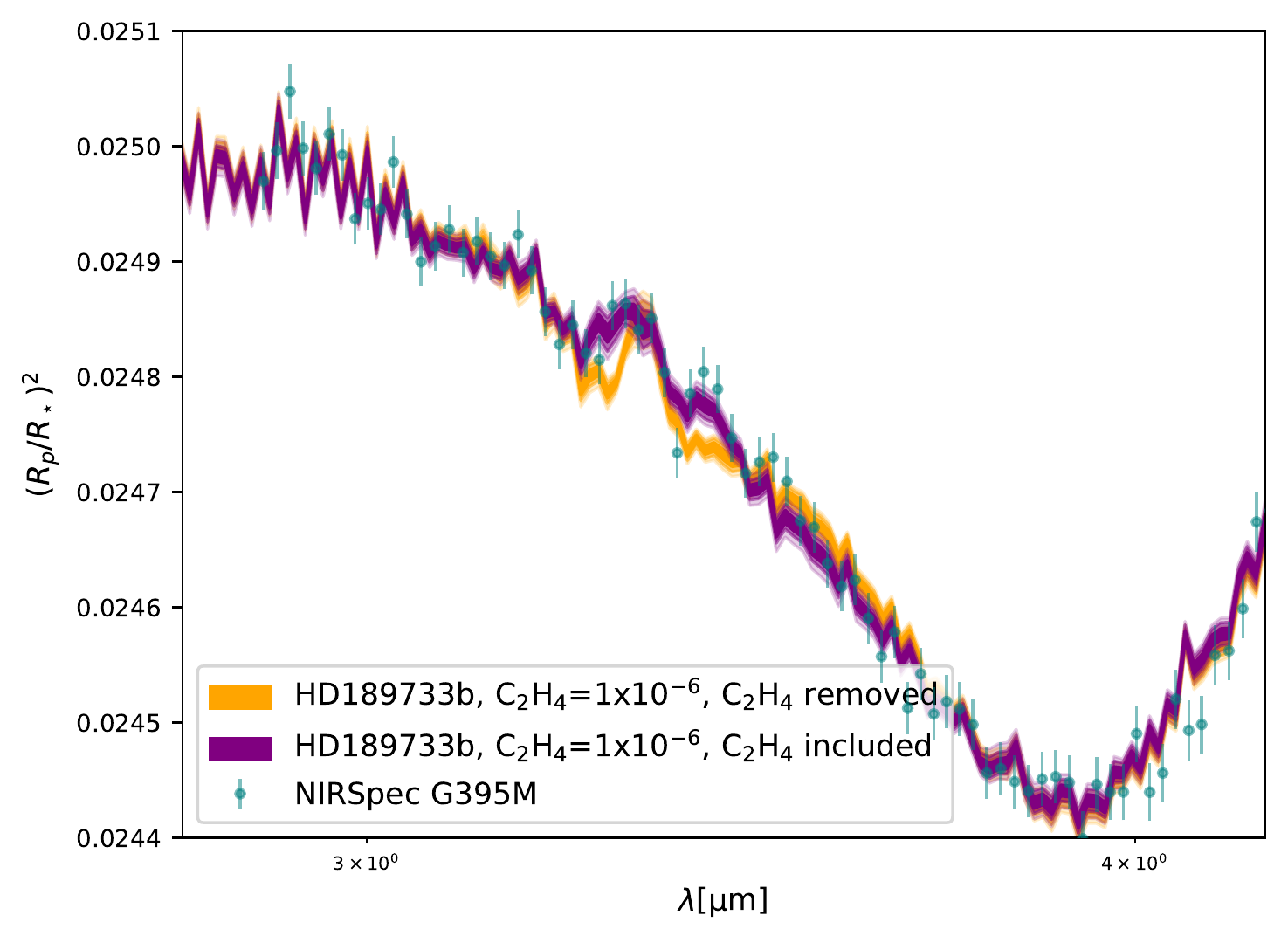}
        \includegraphics[width=0.45\textwidth]{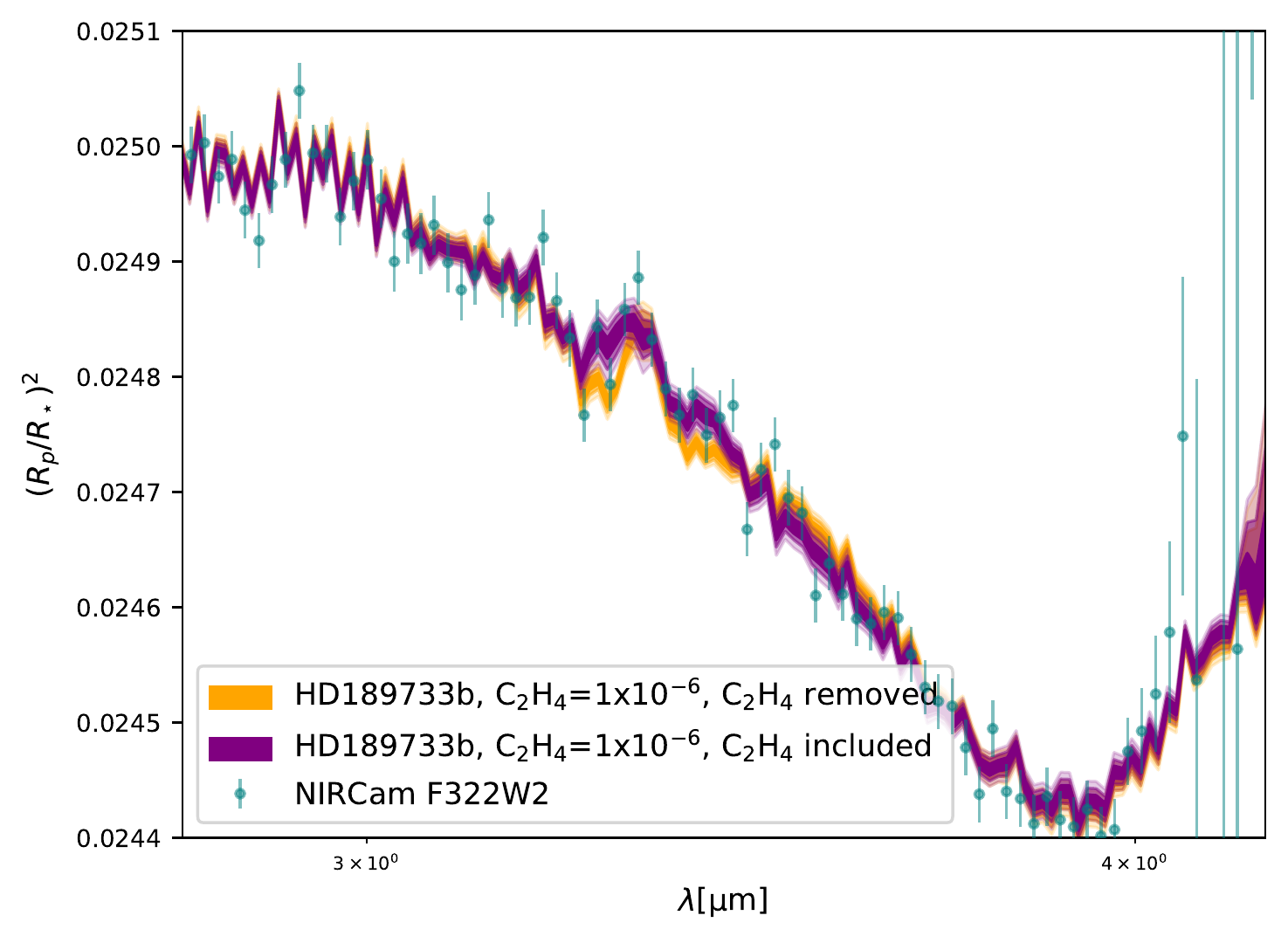}
        \includegraphics[width=0.45\textwidth]{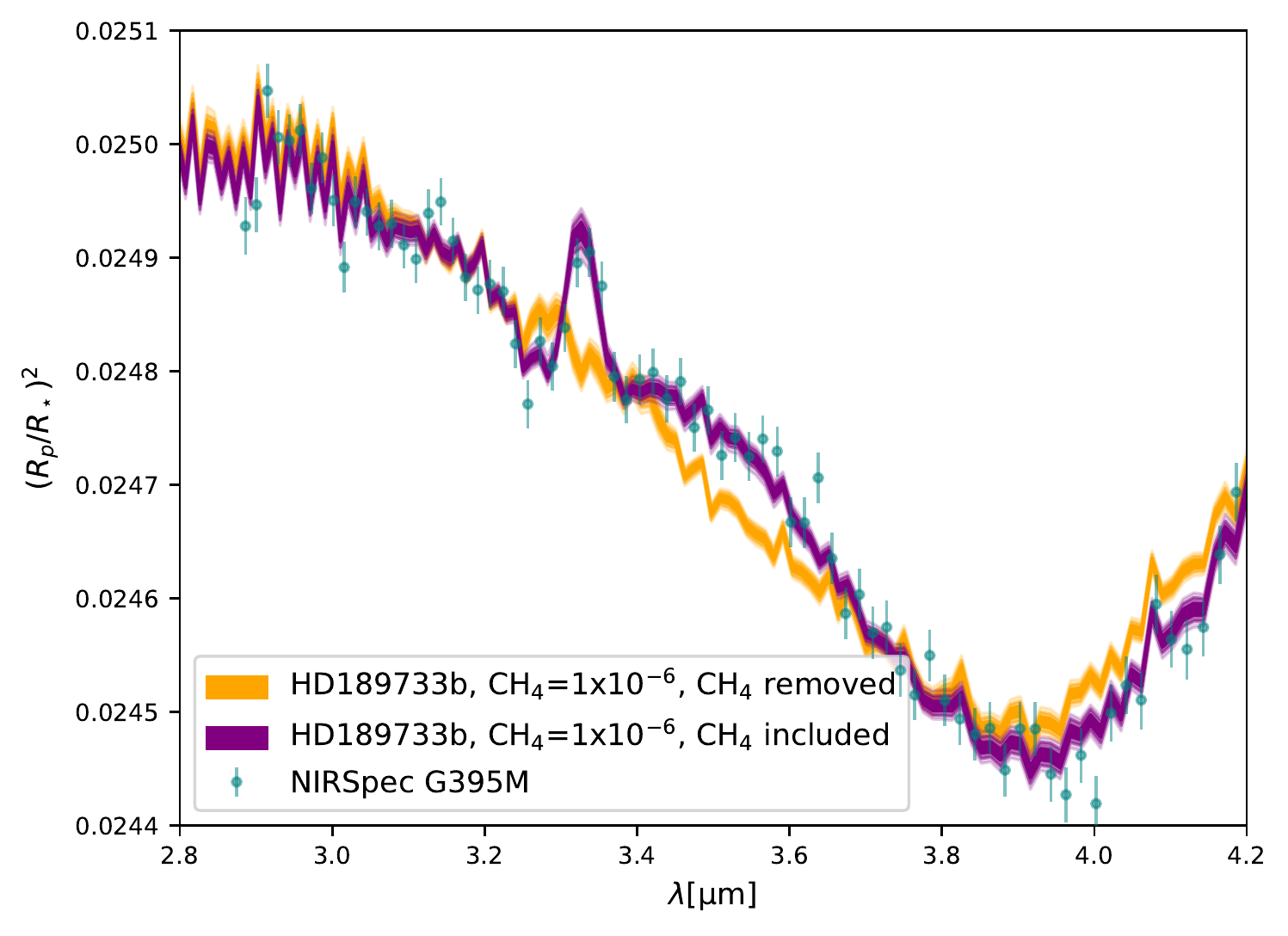}
        \includegraphics[width=0.45\textwidth]{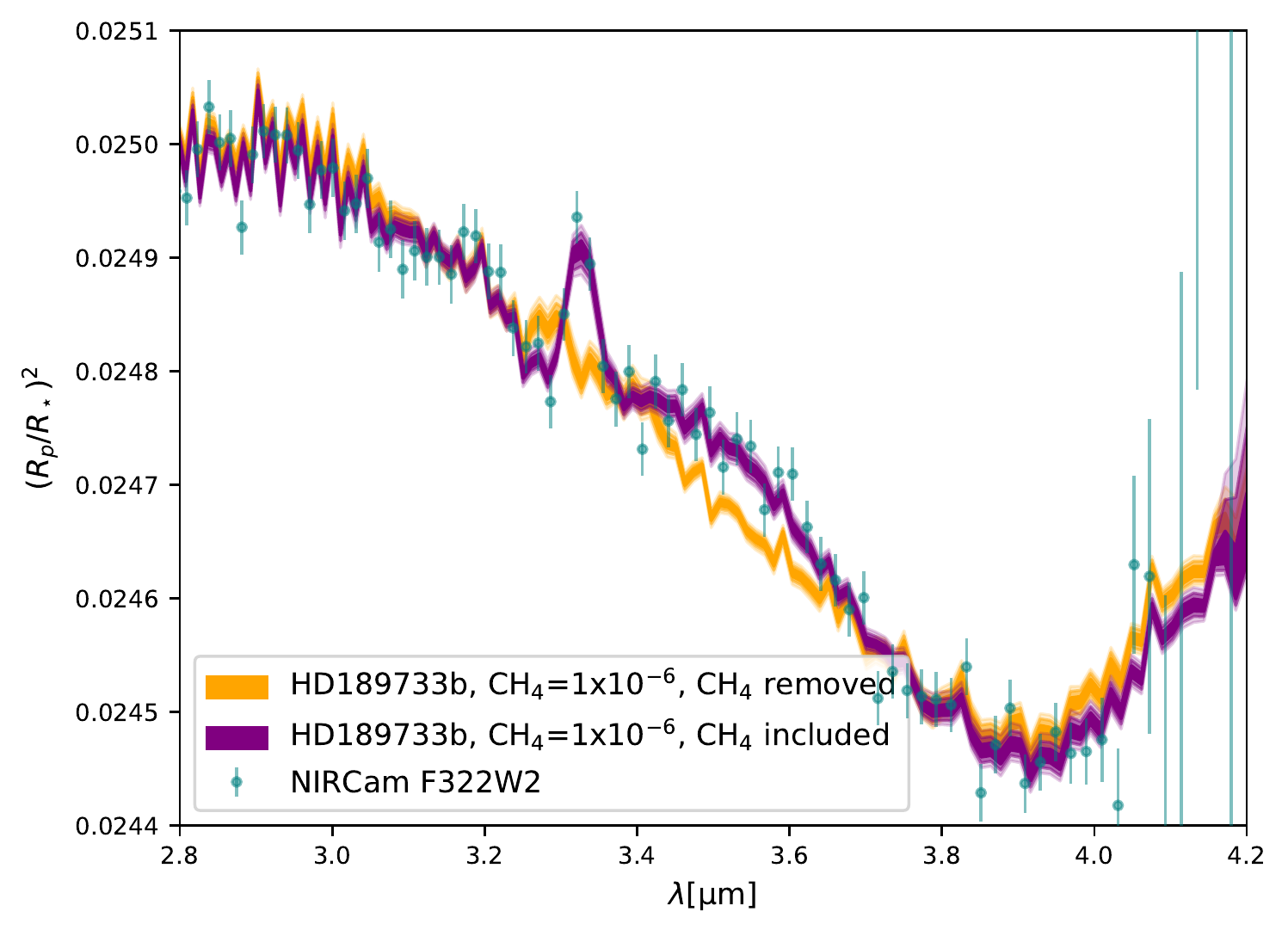}
        \caption{Zoomed-in section of simulated spectra (re-binned to R=100) for the atmospheres of a HD189733b-like planet observed with  NIRISS SOSS and NIRSpec G395M (left column) and with NIRCam F322W2 and MIRI LRS (right column). The region illustrated is covered by both the NIRSpec G395M and  NIRCam F322W2 instruments. The VMR was set to $10^{-6}$ for the hydrocarbon molecule specified in each legend for the simulated observed spectra.  The best-fit retrieved spectra are given in orange for the cases where the molecule was removed, and in purple   when it was included in the retrieval. The shading of the best-fit spectra corresponds to 1, 2, and 3~$\sigma$ regions. }
        \label{fig:HD189_zoomed_NIRSpec_NIRCam}
\end{figure}

\begin{figure}[h!]
        \centering
 \includegraphics[width=0.45\textwidth]{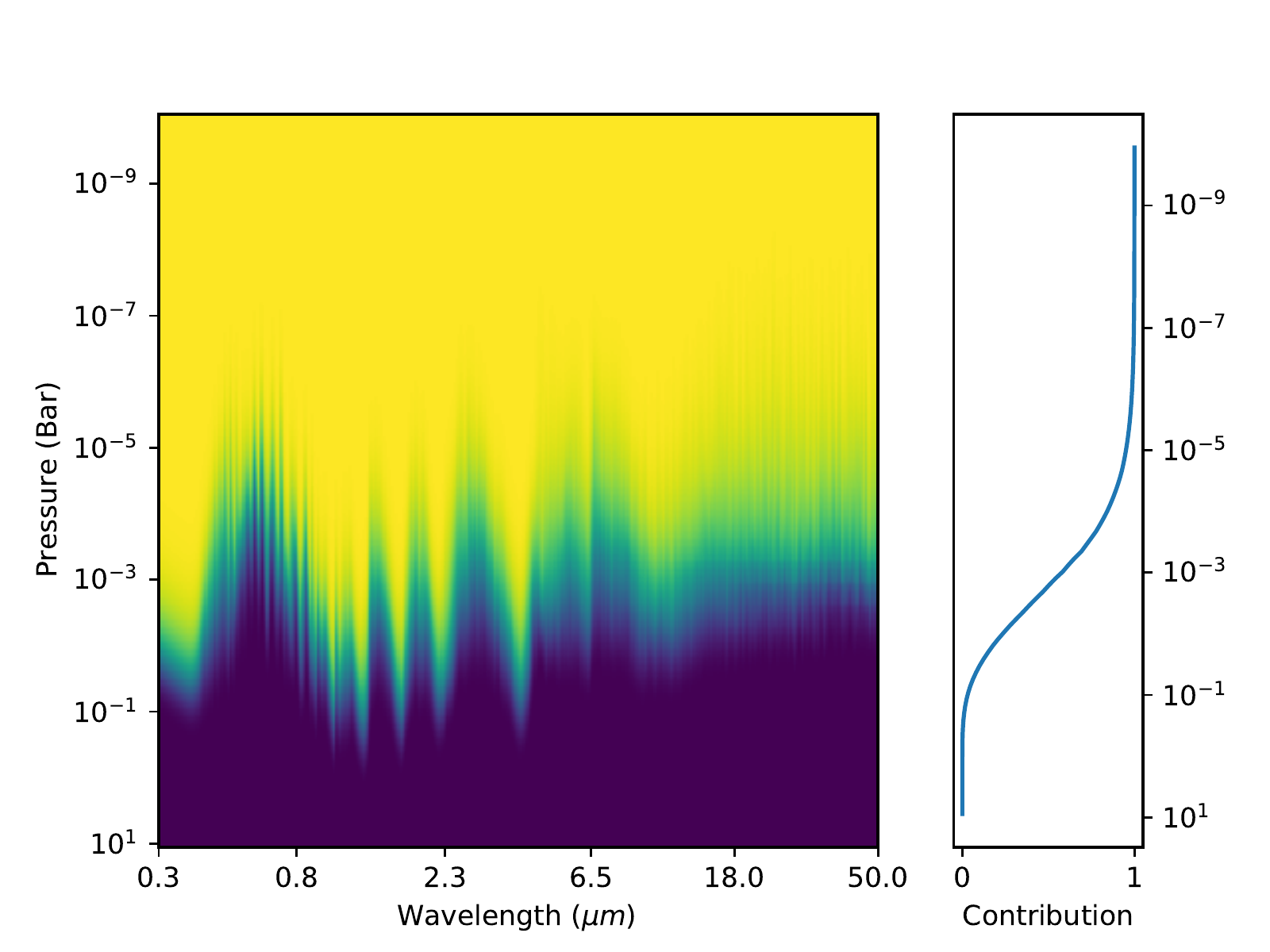}
  \includegraphics[width=0.45\textwidth]{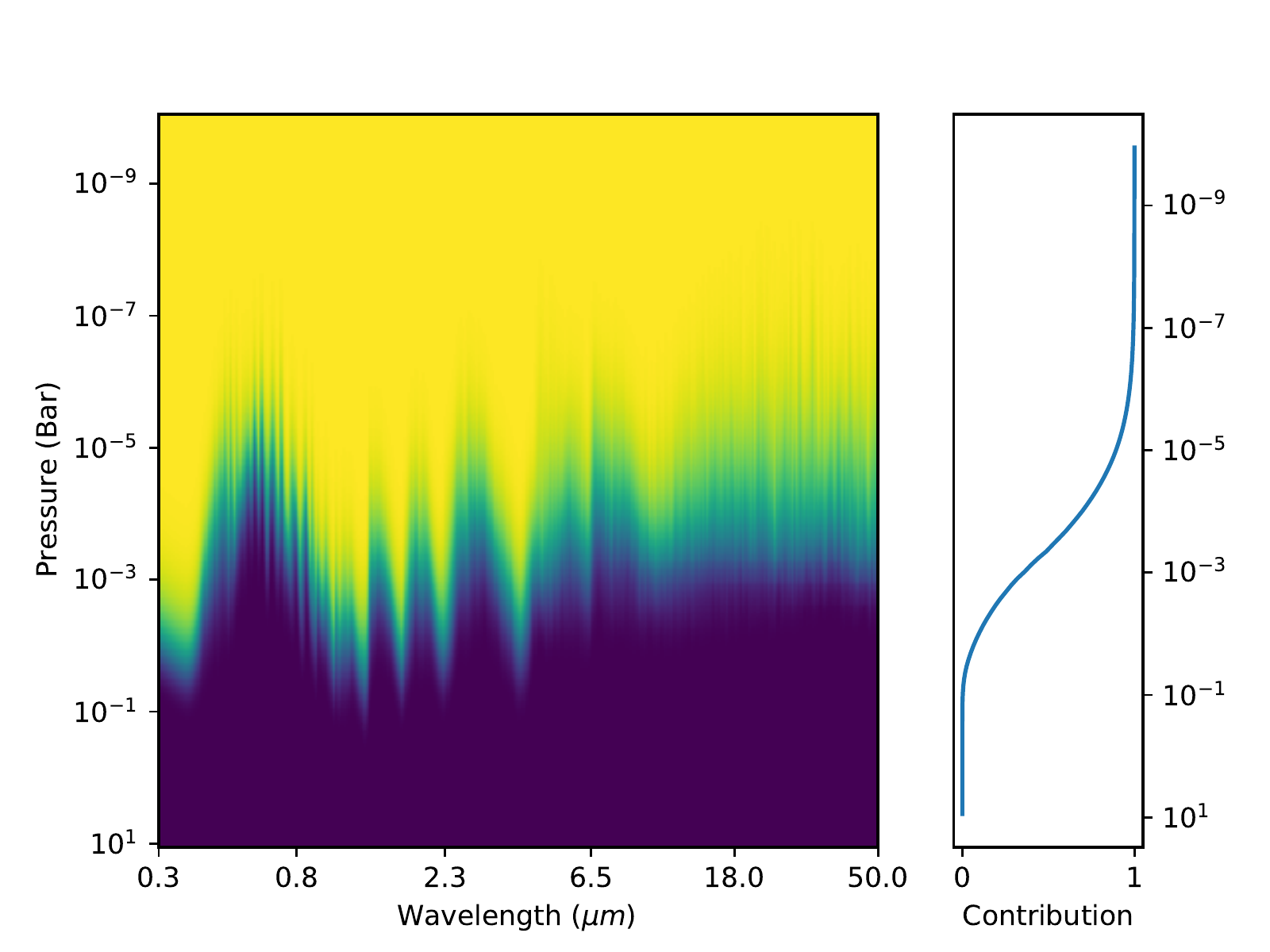}
   \includegraphics[width=0.45\textwidth]{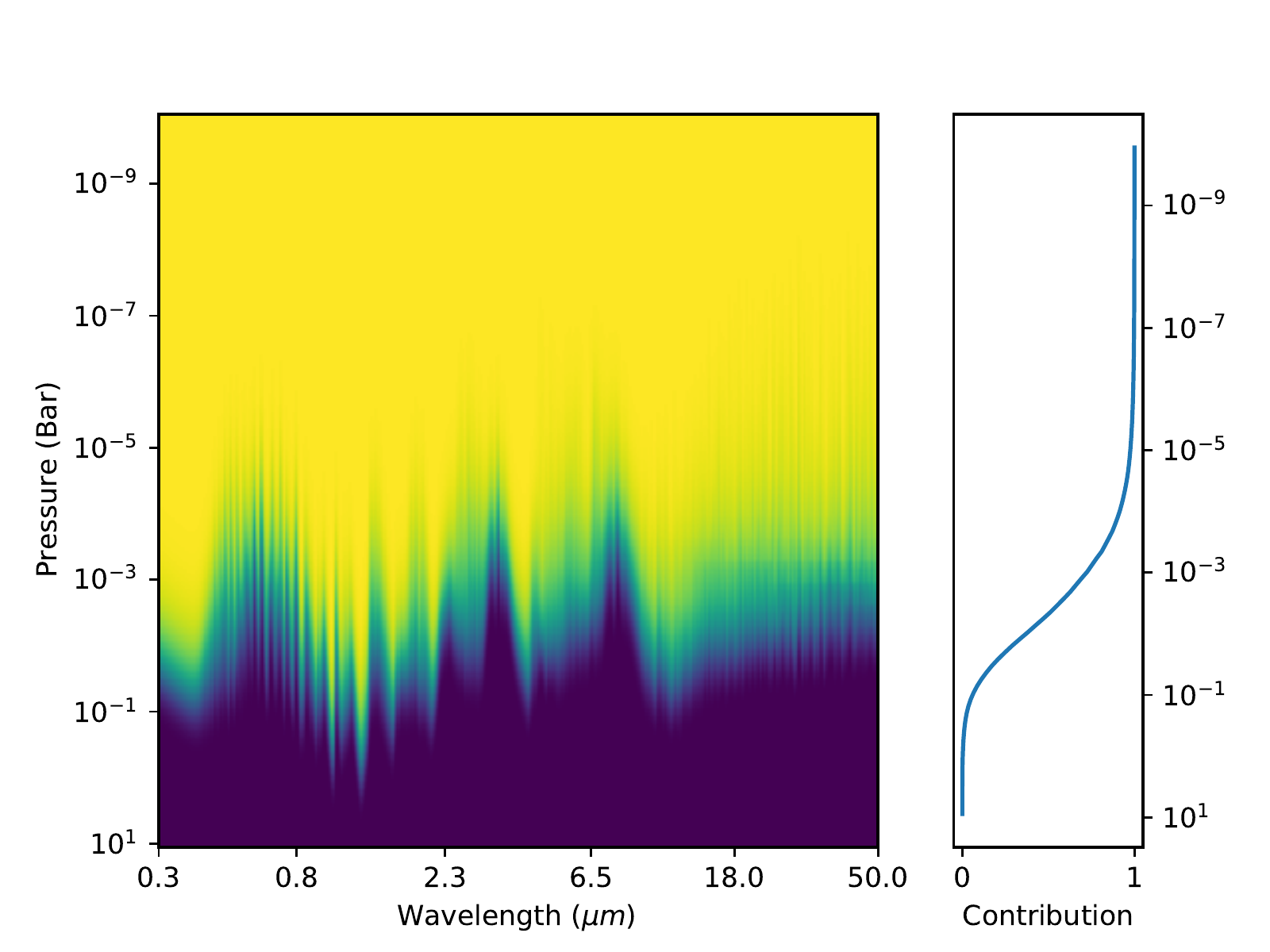}
    \includegraphics[width=0.45\textwidth]{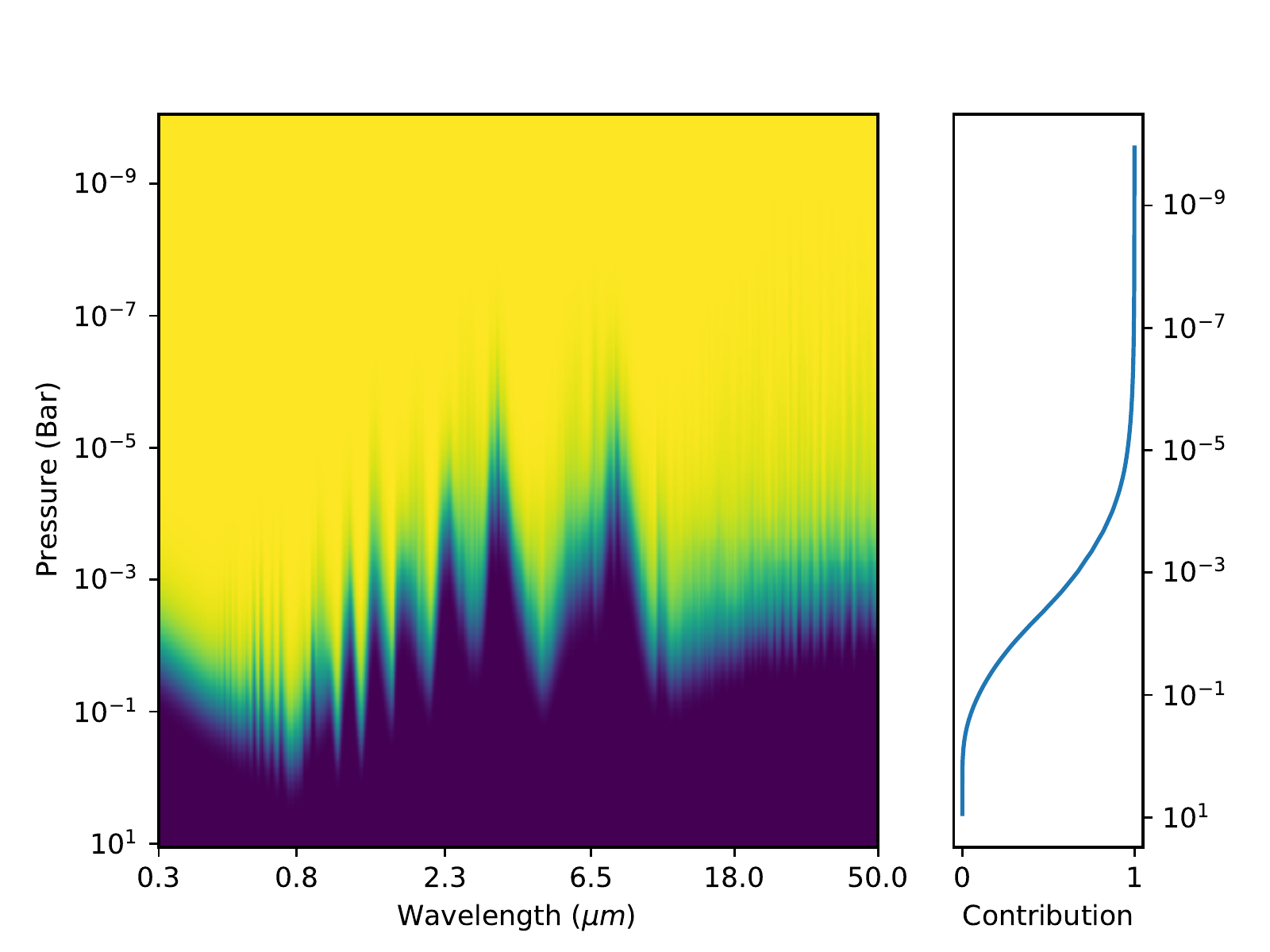}
        \caption{Contribution functions for the base spectra of each of the four planets considered in this study; HD 189733b (upper left), HD 209458b (upper right), HD 97658b (lower left), and Kepler-30c (lower right). }
        \label{fig:contrib_funcs}
\end{figure}

\begin{figure}[h!]
        \centering
 \includegraphics[width=0.6\textwidth]{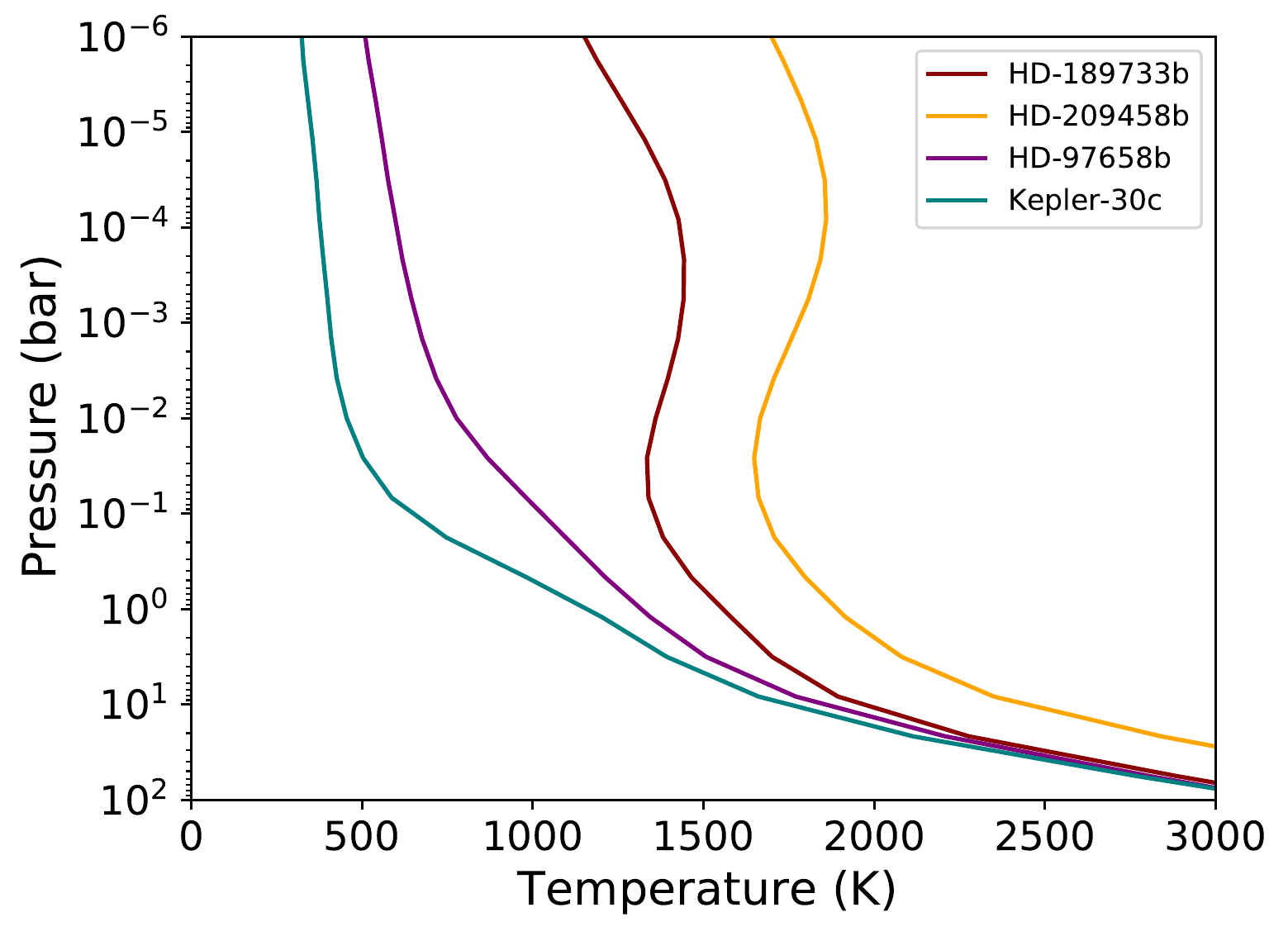}
        \caption{Pressure--temperature profiles of the four plants considered in this study, as determined from self-consistent temperature structure calculations using ARCiS~\citep{ref:20MiOrCh}. }
        \label{fig:PT_profs}
\end{figure}

\end{appendix}

\end{document}